\documentclass[useAMS,usenatbib]{mnras}
\pdfoutput=1

\usepackage{ifpdf}
\usepackage{graphicx}
\usepackage{amsmath}
\usepackage{amsfonts}
\usepackage{amssymb}
\usepackage{array}
\usepackage{supertabular}
\usepackage{url} 
\usepackage[figuresright]{rotating}
\usepackage{array}
\usepackage{afterpage}
\usepackage{fixltx2e} 
\usepackage{lastpage}
\usepackage{hyperref}
\usepackage{tikz}
\usetikzlibrary{calc,arrows,decorations.pathmorphing,intersections}
\usepackage[font={small,sf},labelfont={bf},labelsep=endash]{caption}
\usepackage{sansmath}
\usepackage{colortbl}  

\def \da{$\Delta\alpha / \alpha$}

\def \la{Lyman alpha}

\newcolumntype{:}{>{\global\let\currentrowstyle\relax}}
\newcolumntype{;}{>{\currentrowstyle}}

\newcommand\oldtabcolsep{\tabcolsep}
\newcommand{\be}{\begin{equation}}
\newcommand{\ee}{\end{equation}}
\newcommand{\zabs}{z_{abs}}

\title[Artificial intelligence and spectroscopic analysis]{Artificial
intelligence applied to the automatic analysis of absorption spectra. 
Objective measurement of the fine structure constant.}

\author[M. Bainbridge et al.]{
Matthew B. Bainbridge$^{1,2}$\thanks{E-mail: mbb8@le.ac.uk},
John K. Webb$^{2}$\thanks{E-mail: jkw@phys.unsw.edu.au}\\
$^{1}$College of Science and Engineering, University of Leicester, University Road, Leicester, LE17RH, UK.\\
$^{2}$School of Physics, University of New South Wales, Sydney, NSW, 2052, Australia.\\
}

\begin{document}

\date{\today}

\pagerange{\pageref{firstpage}--\pageref{lastpage}} \pubyear{2011}

\maketitle

\label{firstpage}

\begin{abstract}

A new and automated method is presented for the analysis of 
high-resolution absorption spectra.  Three established numerical 
methods are unified into one ``artificial intelligence'' process: 
a genetic algorithm (GVPFIT); non-linear least-squares with parameter 
constraints (VPFIT); and Bayesian Model Averaging (BMA). 

The method has broad application but here we apply it specifically 
to the problem of measuring the fine structure constant at high 
redshift.  For this we need objectivity and reproducibility. 
GVPFIT is also motivated by the importance of obtaining a large 
statistical sample of measurements of \da.  Interactive analyses 
are both time consuming and complex and automation makes obtaining 
a large sample feasible.

In contrast to previous methodologies, we use {\sc BMA} to 
derive results using a large set of models and show that this 
procedure is more robust than a human picking a single preferred 
model since BMA avoids the systematic uncertainties associated 
with model choice.

Numerical simulations provide stringent 
tests of the whole process and we show using both real and simulated 
spectra that the unified automated fitting procedure out-performs 
a human interactive analysis.  The method should be invaluable 
in the context of future instrumentation 
like ESPRESSO on the VLT and indeed future ELTs.

We apply the method to the $z_{abs} = 1.8389$ absorber towards 
the $z_{em} = 2.145$ quasar J110325-264515. The derived constraint 
of $\Delta\alpha/\alpha = 3.3 \pm 2.9 \times 10^{-6}$ 
is consistent with no variation and also consistent with the 
tentative spatial variation reported in \citet{webb2011} and 
\citet{king2012}.

\vspace{5mm}

\end{abstract}


\section{INTRODUCTION}

Interactive methods for analysing high-resolution quasar spectra of
heavy element absorption systems are complex and require considerable
expertise.   Modelling even just one absorption system can be very
time-consuming if the velocity structure is complex, as is the case for
a typical damped Lyman-$\alpha$ absorption system for example  
\citep{riemer2015}.

The literature contains many examples of fits to absorption systems
which are clearly inadequate, highlighting the complexity. However, if
the numerical methods have been developed and applied properly, whilst
the end product will not be a unique solution, it will at least be a
statistically acceptable fit to the data. Parameter errors are derived
 from the diagonal terms of the covariance matrix, which has been shown to be reliable \citep{king2009}.  However,
the derived parameter errors are strictly only valid if the model is
correct.

Much of our recent work has focused on a search for space-time variation
of the fine structure constant $\alpha$, and we have presented tentative
evidence for spatial variations over cosmological scales, modelled by a
spatial dipole \citep{webb2011,king2012}. Such a potentially important effect
obviously needs critical checking.  Any possible subjectivities
associated with the analysis process must be identified and if present
accounted for. The ``traditional'' method used to analyse the data
requires significant human input and human decisions. Given the
complexity of the analysis and given the methodology used, it is
unlikely that the human element could introduce a systematic bias on
$\alpha$, especially causing a spatial dipole. Nevertheless, it is
imperative to check.

One would also like to know more about the individual parameter
estimates and uncertainties associated with ambiguities in the
``manual'' fitting of the absorption complexes used. For example, given
that one cannot uniquely determine the absorbing cloud model, how robust
is the $\alpha$ estimate to changes in the velocity structure? To
address these issues, we would like the ability to explore a large set
of models of the absorption line complex, mapping out the corresponding
probability space for $\alpha$ and any other parameters of particular
interest. This is the main motivation for the work described in this
paper. However, the methods developed in this paper are of wider
applicability.

A secondary motivation is one of practicality. A considerable amount of
time is devoted to echelle spectroscopy of quasars on large optical
telescopes. Whilst some analyses of such data are based on computing
statistical properties of the data directly from the spectrum, e.g.
power spectra to measure baryon acoustic oscillations, others require a
detailed Voigt profile analysis to extract parameters associated with
specific absorption lines/systems e.g. measurements of metallicity,
$\alpha$, T(CMB), D/H, abundance measurements. Interactive measurements
with such data can be very time consuming and telescope archives (e.g.
HIRES/Keck, UVES/VLT) generally contain a considerable amount of echelle
spectra which either remain unpublished, or which have only been
partially analysed and hence from which a great deal of valuable
scientific information has yet to be extracted.

To address all these shortcomings, we have developed a new genetic
algorithm that is likely to have broad application in observational
cosmology, applied here specifically to the problem of modelling quasar
absorption systems.  It unifies a memetic``artificially intelligent''
process with Bayesian Model Averaging (BMA), providing a more 
robust and more
automated methodology, hence bringing a far greater degree of
objectivity and reproducibility to the problem.  The new method is used
to derive a measurement of \da\ from thousands of potential models and
has allowed an exploration of the stability of \da\ to model complexity.
Our new method enables almost complete removal of human interaction and
hence subjectivity.

The structure of this paper is as follows. In Section \ref{prelim} we
discuss the considerations in designing algorithms that aim to replace
human decision making in the context of automated spectral fitting and
explain why one in particular is highly suited. A genetic algorithm is
used to apply a large set of trial models to the data and this is
combined with local non-linear least-squares minimisation for each
candidate model to select the fittest. In Section \ref{memetic} we
develop in detail this hybrid approach. Section \ref{simplesim} outlines
a simple simulation to test the methods. Section \ref{prefit} explains
how we incorporate additional free parameters and parameter constraints.
Section \ref{postfit} introduces our method of BMA 
and develops the mathematical formalism. Section \ref{applic} presents
the results of applying our new methods to an absorption complex at
$z_{abs} = 1.839$ towards the quasar J110325-264515, deriving a
measurement of the fine structure constant in that gas cloud. A
comprehensive discussion on robustness is also given. Finally in Section
\ref{conclusions} we summarise our conclusions.

\section{PRELIMINARY CONSIDERATIONS}\label{prelim}

Interactive analyses of high-resolution spectra are currently done using
programs such as {\sc VPFIT} \citep{vpfit}. Recent papers summarising
the application of {\sc VPFIT} to quasar spectroscopy include
\citet{king2012}, \citet{riemer2015} and \citet{wilczynska2015}. Small
regions of the spectrum are selected by eye such that each is flanked by
a sufficiently long continuum region. First guesses for the absorption
line parameters are generated through trial-and-error, the efficiency
depending heavily on the user expertise. Usually there is considerable
blending of adjacent absorption features and the user decides on the
first guesses. The human brain is good at visually filtering out random
noise, picking out stationary points, inflection points, and small
asymmetries, such that a reasonable guess can be made as to the velocity
structure.  By simultaneously visually inspecting different species
aligned in velocity space, contaminating lines from different redshift
systems can be identified and the set of first guesses further refined.

Nevertheless, experience shows that two people, working independently,
can often produce a {\it different} first guess model.  Their end
results after least-squares minimisation can therefore also end up
differently \citep{wilczynska2015}.  In the absence of any further,
comprehensive, exploration of many models, there thus remains
uncertainty as to the robustness of conclusions drawn from individual
models.

Emulating what the human brain does so well poses interesting
challenges. The eye is able to filter the random noise contribution,
differentiate the observed data, and spot stationary and inflection
points. The eye assesses the underlying model by simultaneously
inspecting an entire spectral segment of the data, identifying the
simplest underlying element, the ``building block'' from which the
complex is constructed (i.e. a single absorption line in our case), and
then reconstructing the whole spectral segment by assigning it a set of
individual, varying-strength, absorption components.

Clearly a simple computer algorithm can be used to differentiate the
spectrum. However this increases noise, requiring prior Weiner (or
similar) filtering to help determine stationary and inflection points
numerically, but then small-scale information can be lost in the
smoothing process. An additional complication in converting fixed points
into first guesses for parameters is the instrumental resolution, so
deconvolution methods are needed.  Further complications are caused by
possible breaks in the data --- one cannot then determine the derivatives
continuously and the data has to be dealt with as fragments.
Non-Gaussian noise (cosmic rays, bad CCD pixels) can also corrupt
small-scale information content.

However, all the problems above can in principle be dealt with and one
can automate obtaining the first-guess parameters. Nevertheless, such a
method still lacks some important aspects of the eye's more
sophisticated pattern-matching ability and therefore the end result is
not as good.  Nor does a deterministic method such as this provide any
mechanism for exploring robustness of certain ``embedded'' fundamental
parameters of interest, for example $\alpha$, to the choice of model. 
To do that we need to incorporate additional search methods in order to
explore a large set of candidate models.  With these things in mind, we
therefore initially assessed the applicability of various statistical
methods to this problem, the first important decision being how best to
generate candidate model databases, i.e. randomly or intelligently. 

\subsection{Different optimisation approaches}

Parameter estimation methods can generally fall into one of the
categories, deterministic, stochastic and heuristic. A deterministic
algorithm will always take the same path through parameter space,
producing the same result from a given starting point \citep{bos2007}. A
non-linear least-squares method such as
\textsc{VPFIT} \citep{vpfit}
is an example of a purely deterministic method. A stochastic algorithm
relies on a random sampling of parameter space to reach a solution.
Markov Chain Monte Carlo is an example of this type of method.  This has
been applied to quasar spectroscopy \citep{KingMCMC} and was found
useful primarily to check on the validity of covariance matrix diagonals
as parameter error estimates. A heuristic algorithm conducts an
``experienced-based'' search of parameter space, i.e. the current
search direction depends on the previous search history.  Genetic
algorithms involving such machine-learning or ``artificial intelligence"
are examples of this type of method.

Stochastic methods \citep{spall2003} are insensitive to noise, unlike
deterministic algorithms. Also, stochastic methods make no assumption of
the objective function.  The exact solution can in principle be reached
although requires an infinite number of trials.  Many calculations are
needed and the approach is inefficient and may not be feasible for
problems with a large number of free parameters.

Heuristic methods \citep{winker2004} require far fewer trial models
compared to stochastic methods, hence are more efficient, and like
stochastic methods, are insensitive to noise. However, because of the
directed nature of the search, regions of parameter space can be left
unexplored and there is no guarantee of reaching an acceptable solution. 

A genetic algorithm (one type of heuristic method) offers the best
features of deterministic and stochastic \citep{holland1975}. It reduces
the number of required calculations, compared to stochastic, and because
the search direction is ``artificially intelligent'' the optimal
solution is reached more efficiently.

In some sense the genetic method reflects a manual procedure. Manually,
one would initially fit the data with all the obvious features
and then use some goodness-of-fit measure to make decisions about
including extra parameters. The genetic method makes decisions in a
similar way.  Nevertheless, by itself, a genetic algorithm cannot provide
the combination of efficiency and precision required for our problem.

\section{A HYBRID APPROACH} \label{memetic}

In practical applications, a hybrid approach can be very effective,
taking advantage of the best characteristics of the fundamentally
different approaches described above e.g. \cite{Krasnogor2010toappear}.
The amalgamation of a deterministic and a heuristic technique, referred
to as a ``memetic algorithm'' by some authors e.g. \cite{Sudholt09},
has been widely applied and we have opted for this approach in this
paper.  We have chosen to use a genetic algorithm which sits ``above''
{\sc VPFIT}, controlling a high level or global parameter space search,
within which {\sc VPFIT} is used for local optimisation. In this paper
we refer to {\sc GVPFIT} as the whole process, i.e. genetic algorithm + {\sc VPFIT}.

\begin{figure*}
\begin{center}
\includegraphics[width=18.5cm]{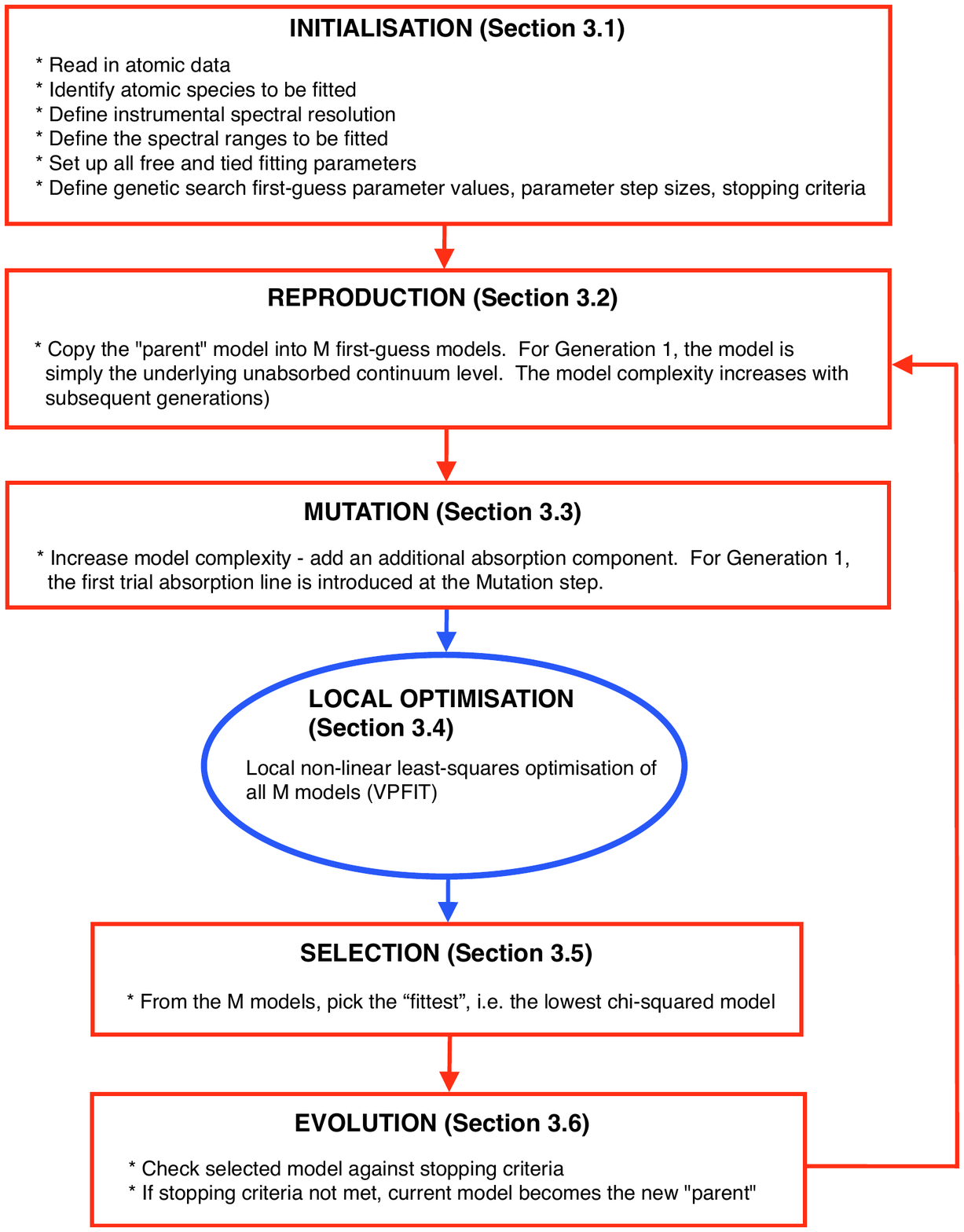}
\vspace{-5cm}
\caption{\label{fig:flowchartv4} 
The whole process is the combination of a genetic algorithm and
non-linear least-squares minimisation, sometimes referred to collectively 
as a ``memetic algorithm''.  Red boxes indicate the genetic part of the
process that constitutes {\sc GVPFIT}. The blue oval indicates
standard VPFIT calls. $M$ is the ``population size'' i.e. it is the 
total number of trial or competing models at some generation, 
$M = n_N \times n_b \times n_z$ i.e. the product of the number of 
trial values for each of the 3 variables, $N, b, z$.
}
\end{center}
\end{figure*}

Whilst genetic algorithms come in different varieties, the Darwinian
analogies can be generalised and put in context: an iterative procedure
is set up to generate a population of ``individuals'' (theoretical
absorption cloud models in our case) and subject them to ``environmental
pressure'' (goodness-of-fit testing) and select using ``survival of the
fittest'' (minimum-$\chi^2$). This results in natural selection (keeping
only the best candidate model) and increasingly stronger ``offspring''
(a new theoretical model, slightly better than the previous). The
algorithm can thus be broken down into an initialisation procedure,
followed by an iterative 5-step process: Reproduction, Mutation, Local
Optimisation, Selection and Evolution, illustrated in Figure
\ref{fig:flowchartv4}.

The remainder of this section describes the algorithm details.

\subsection{Initialisation}\label{init}

The setup for using {\sc GVPFIT} is very similar to the {\sc VPFIT}
setup. The spectral data format is the same as for {\sc VPFIT} (FITS or
ascii). The atomic data used in the analysis in this paper is given in
Tables \ref{tab:app_ato_ide}, \ref{tab:app_ato_dat},
\ref{tab:app_ato_wav}, and \ref{tab:app_ato_ref}. Model absorption
profiles are produced within {\sc VPFIT}.  However, additional input
information is needed for GVPFIT:

\begin{enumerate}
\item  Initialisation requires a ``parent'' model. The parent can be
simply an ``empty model'' i.e. an input files containing no absorption
line parameters (but it must identify which species are going to be
modelled). The parent can also be more complex if there are known
absorption features which ``contaminate'' the system being fitted. Other
examples of additional parameters which may be included at this stage,
before the actual absorption model has been developed, are floating 
zero and continuum levels;
\item {\sc GVPFIT} search parameters may be altered from their default
values (see Section \ref{ranges});
\item Stopping criteria for the genetic search may be altered from 
their default values (see Section \ref{stop}).
\end{enumerate}

\subsection{Reproduction}\label{repro}

The ``parent'' is now ``reproduced'' $M$ times to create the current
``generation'' of trial models (i.e. $M$ copies of the parent are made).
$M$ is defined by the number of steps to be taken through absorption
line parameter space (as described in the next Section).

\subsection{Mutation}\label{mutate}

Each of the models in the current generation are then ``mutated'' by
adding one additional velocity component.  
At the mutation step components are only added and not rejected.  
Component rejection only occurs during the local-optimisation stage 
(see Section \ref{lopt}).  
The parameters for the
additional component are determined by the global parameter ranges and
parameter step-sizes. The choice of these ranges and step-sizes
determine the overall efficiency and success of the procedure
(see Section \ref{ranges}).

Each absorption component is specified by three parameters, $N, z, b$.
$N$ is the column density of absorbing atoms along the line of sight,
$z$ is the absorption system redshift and $b$ is the velocity dispersion
parameter. The total number of trial models is $M = n_N \times n_b
\times n_z$ ($n_N$, $n_b$, $n_z$ are the number of trial values for
column density, b-parameter and redshift). 

We wish to minimise the number of calculations, but at the same time,
the initial guesses must cover a range sufficiently close to the
solution for a descent direction to be found by the local non-linear
minimisation (Section \ref{lopt}).  We carried out extensive tests,
using both real data and synthetic spectra, to find suitable ranges and
step sizes. The next section summarises reasonable search ranges and
stopping criteria.

\subsubsection{Parameter first guesses and step-sizes for the genetic
algorithm}\label{ranges}

The genetic part of {\sc GVPFIT} trials a range of first guess
parameters which are then subject to local non-linear optimisation.
Obtaining values outside the first guess range in the final fit is not
precluded. The ranges in parameters discussed here are only to set the
initial guesses. If the default ranges and step-sizes fail in some cases
(the failure becomes obvious very quickly), these can be reset.

A reasonable lower limit for the first column density guesses is
suggested by the data quality (i.e. the detection threshold). A
reasonable upper limit for first guess column density can be derived
from an empirical knowledge of the column density probability
distribution.  Absorption components for heavy elements are generally
concentrated in the range $11 < \log N < 14$ but in practice we found
that using $12 < \log N < 14$ for the first guess range for heavy
element lines is adequate to find a non-linear minimisation search
direction.  For {\sc HI} lines, the column density range is larger,
ranging from the weaker Lyman forest lines at $\log N \approx 11$ up to
the stronger damped Lyman alpha lines at $\log N \approx 21$. The
genetic algorithm has also been applied to Lyman forest data but at this
stage we constrained the fits to segments without strong lines, and for
this we found that using $12 < \log N < 14$ for the first guess range
also worked well.

Although crude, a column density step-size of $\Delta log N = 1$ 
permits a reasonable computation time and does not result in the local 
non-linear least squares search getting stuck in local false minima.

The permitted redshift search range $z_{start}$ -- $z_{end}$ is constrained
to lie within the observed wavelength range of the data. Since in
general we fit multiple species simultaneously, we transform the
spectral segment or segments being fitted to velocity space, $\Delta v =
c \Delta \lambda/\lambda$. The starting and end points of the velocity
range are defined by the smallest and largest values amongst the input
spectral segments.

Before the mutation stage, one must decide how finely to step through
the redshift range,
striking a compromise between computational efficiency and robustness.
One would expect results to be robust within plausible step-size limits
suggested by the data. For the analysis reported in Section
\ref{applic}, heavy elements have typical line widths of 3 or 4 km/s. 
The instrumental resolution (FWHM) is $\sim 4$ km/s. Empirically, we tried 3 step-sizes for the
data in Section \ref{applic}: 0.6, 1.6, and 14.5 km/s.  Surprisingly, we
obtained very good results even with the largest step-size (and in fact
the detailed results in Tables \ref{tab:preferred_model} and
\ref{tab:j11_sta} were obtained using that large step-size).  The reason
such a large step-size produces robust results is because local
optimisation, i.e. VPFIT, is able to identify a reliable
search-direction even when first-guesses are not particularly close to
the local minimum.

For heavy element systems, the empirical spread in $b$-parameters is of
the order of the typical parameter error estimates.  In practice we
found that taking a constant first-guesses of $b = 4$ km/s for heavy
element lines worked well. For {\sc HI} lines, a first guess of $b = 21$
km/s also worked well in practice.

\subsection{Local optimisation (VPFIT)}\label{lopt}

Once the ``local'' first guesses are established, these are then
subjected to local non-linear least-squares optimisation using
{\sc VPFIT}, \citep{vpfit}.

{\sc VPFIT} implements optimisation methods (Gauss-Newton or
Levenberg-Marqhuardt) with optional parameter constraints.  The method
makes use of first-order derivatives of $\chi^2$ with respect to the
free parameters, and estimates of the second-order derivatives.
Parameter search directions are derived by solving matrix equations
involving the Hessian matrix and gradient vector. Solutions thus depend
on the $\chi^2$ derivatives being reasonably accurate. However, when the
model parameters are far from the best-fit values (as can be the case in
our genetic process), the derivatives will be noisy, and search
directions can be poorly determined.  We get around this problem by
stabilising {\sc VPFIT} i.e. minimising or eliminating ill-conditioning,
by adding a constant value to the diagonal term of Hessian, following
the procedure described in \citet{GMW81}.

Related to this is the way in which model complexity, i.e. the number of
free parameters, evolves with increasing generation number.  As the
genetic component of {\sc GVPFIT} increases complexity, trial models are
sent to {\sc VPFIT} for refinement.  In situations where the added
absorption components are such that the associated $\chi^2$ derivatives
are close to zero, and hence dominated by the finite precision of the
calculation, {\sc VPFIT} may fail to identify a search direction. 
Depending on the parameters specified in the {\sc VPFIT} setup parameter
file, the newly added
component may then be rejected and the model complexity at that
iteration remains the same.  The effect of this is to cause fluctuations
in the total number of free parameters, and hence a jagged appearance in
the absolute $\chi^2$ values.  This can be seen in the continuous grey
lines plotted Figure \ref{fig:da_fig1} and is a natural consequence of
the overall procedure.

\subsection{Selection}\label{selection}

Following optimisation, we have a set of $M$ trial models. In this
Selection step, we select the minimum-$\chi^2$ model which proceeds to
the next step and which subsequently becomes the parent for the next
generation (provided stopping criteria are not met).

\subsection{Evolution}\label{evol}

This step has 2 simple functions: (i) to check that the entire set of
models so far generated comfortably spans the range of acceptable models
and (ii) if not, to make the current model the new parent and then to
continue iterating from the Reproduction step.

\subsubsection{Stopping criteria for the genetic part of {\sc GVPFIT}}\label{stop}

As {\sc GVPFIT} loops through the steps illustrated in Figure
\ref{fig:flowchartv4}, the model complexity generally increases as
generation number increases. To avoid this happening indefinitely, we
must establish suitable stopping criteria.

The end-goal is to calculate an entire suite of potential models, such
that at one extreme the models have far too few parameters to
satisfactorily describe the data, and at the other extreme, have too
many, i.e. the data are ``over-fitted''. In other words, it does not
matter when the calculations are stopped (other than economy of
computing time) provided that we go well into the ``over-fitting''
region. After we have calculated the entire suite of models, we can then
analyse the entire set to optimally extract the science data of interest.

The termination of the calculations therefore does not correspond to
obtaining the ``correct'' or ``preferred'' model. The stopping criteria
are required only to force the models generated to lie in a relevant
range.  A number of stopping criteria were explored and the following
adopted.

One possibility is to simply pre-set a maximum number of generations,
based on experience. If the pre-set maximum number of generations fails to reach a
statistically acceptable fit, the user can simply increase the number of
generations and re-start from that point. 
Alternatively we can establish the stopping point
using a goodness-of-fit measure (for example, once $\chi^2_\nu$ falls
below some reasonable threshold, e.g. 0.9 or less, where $\nu$ is the
number of degrees of freedom). One could define analogous conditions
using other statistical measures such as the Akaike Information
Criterion or the Bayesian Information Criterion (Section \ref{3stats}).

If it proves impossible to reach $\chi^2_\nu \approx 1$ irrespective of
the number of pre-set generations, the input model never becomes
satisfactory, one likely inference is the presence of interloping
absorption lines at some other redshift.  We discuss this situation
separately in Section \ref{prefit}.

\section{Testing GVPFIT using simulated spectra with relatively simple
velocity structure}\label{simplesim}

We have carried out extensive tests of GVPFIT using a broad range of
simulated spectra.  We avoid discussing simulations with complex
velocity structure at this point in the paper and instead describe 
simulations with relatively simple velocity structure that serve both 
as an illustration of how GVPFIT works and as a simple test.  

\begin{figure*}
\begin{center}
\includegraphics[trim=0.00cm 6.50cm 0.00cm 2.25cm, clip=true, 
width=17.0cm]{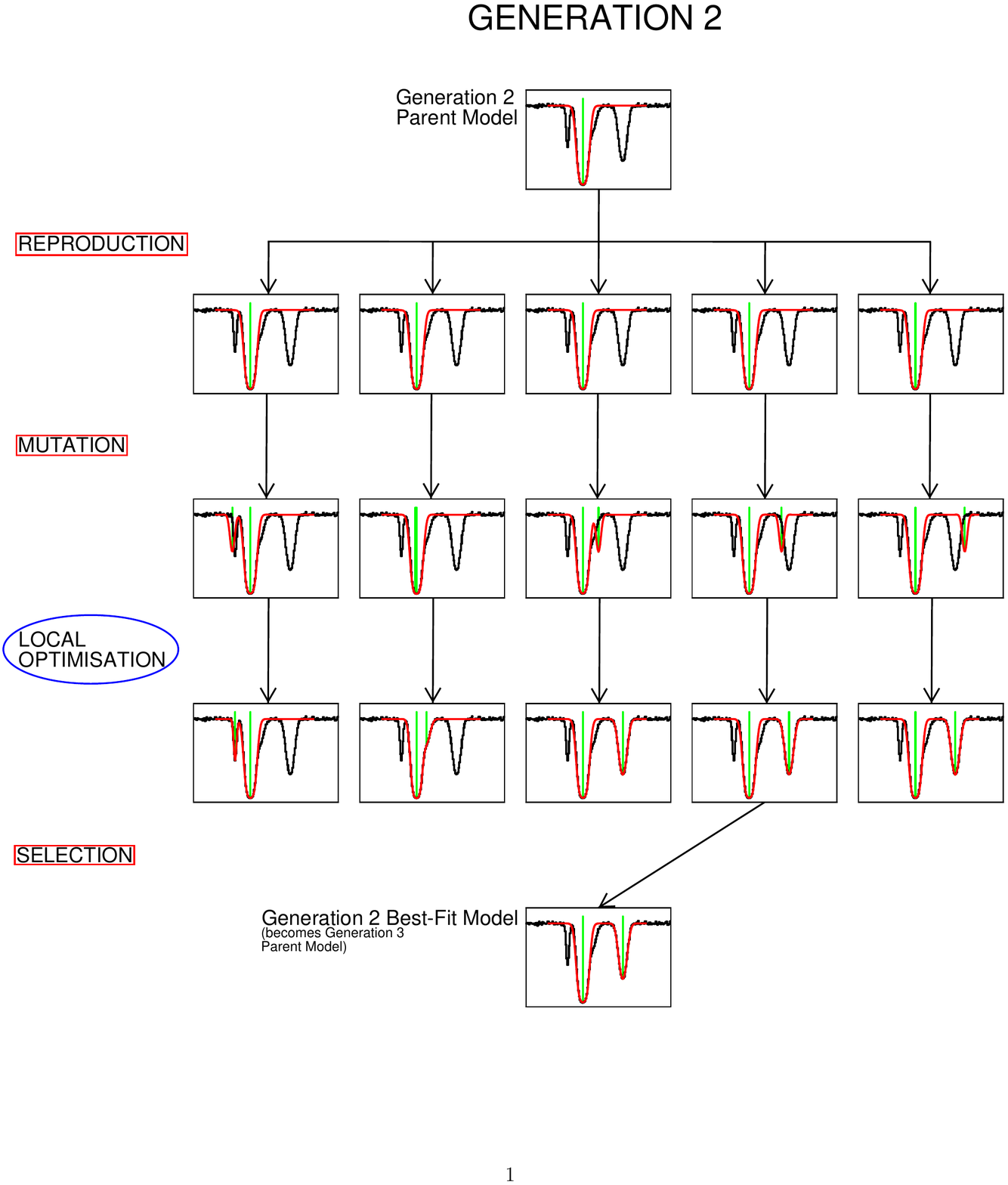}
\end{center}
\caption[
A simplified example to illustrate how the algorithm emulates
Darwinian evolution, as portrayed in Figure \ref{fig:flowchartv4}.
]{
\label{fig:met_gen}
\textbf{
A simplified example to illustrate how the algorithm emulates
Darwinian evolution, as portrayed in Figure \ref{fig:flowchartv4}.
}
Shown is the second generation or iteration of {\sc GVPFIT} when applied to a simple simulated spectra as part of the test described in Section \ref{simplesim}.
The data (black
line) is a simple 4-component synthetic spectrum created using a
spectral resolution of FWHM$= 3.2$ kms$^{-1}$, a pixel size of 1.25 kms$^{-1}$, and
the mean signal-to-noise is approximately 100. These observational parameters
correspond to the real values for the 2006 data for the quasar
J110325-264515 (Tables \ref{tab:dat_spe_obs} and \ref{tab:app_sys_tra}).
Two transitions were simulated and then fitted simultaneously, FeII2600 
(shown in this Figure) and 1608 (not illustrated).
In this simplified example, the first-guess column density N and
b-parameter remain the same each time a new trial component is
introduced. 5 trial redshift positions were used.  Therefore,
specifically for this example, $M=5$ (see Figure \ref{fig:flowchartv4})
and this happens to be the same as the number of trial redshifts, but
this would not normally be the case. In general, $M = n_N \times n_b
\times n_z$ i.e. the product of the number of trial values for each of
the 3 variables, $N, b, z$.
In the 1st row of this figure, the model (the smooth red line), is the result from
Generation 1, i.e. it is the model chosen after local optimisation (VPFIT).
In the 2nd row, the model has been reproduced, ready for mutation.
In the 3rd row, the successful parent model has been copied from the 
previous step, modified by adding one extra velocity component with
different redshifts. $b$ and $N$ are the same for the added velocity component
in each case, as these are first guesses that are subsequently optimised 
in the next step.  However, the example illustrated here has been deliberately
kept simple and in general all 3 parameters are varied.
The 4th row shows the result after local optimisation of each model from 
the 3rd row. In the example shown, the right-hand three models are visually 
indistinguishable.  The redshift step-size is relatively coarse in this 
simple example, so VPFIT local optimisation produced 3 almost identical models.
Finally, in the 5th row, the model shown is the best (minimum-$\chi^2$)
fit for this Generation.
}
\end{figure*}

\begin{figure*}
\begin{center}
\includegraphics[trim=0.00cm 6.95cm 2.60cm 2.00cm, clip=true, width=14.0cm]
{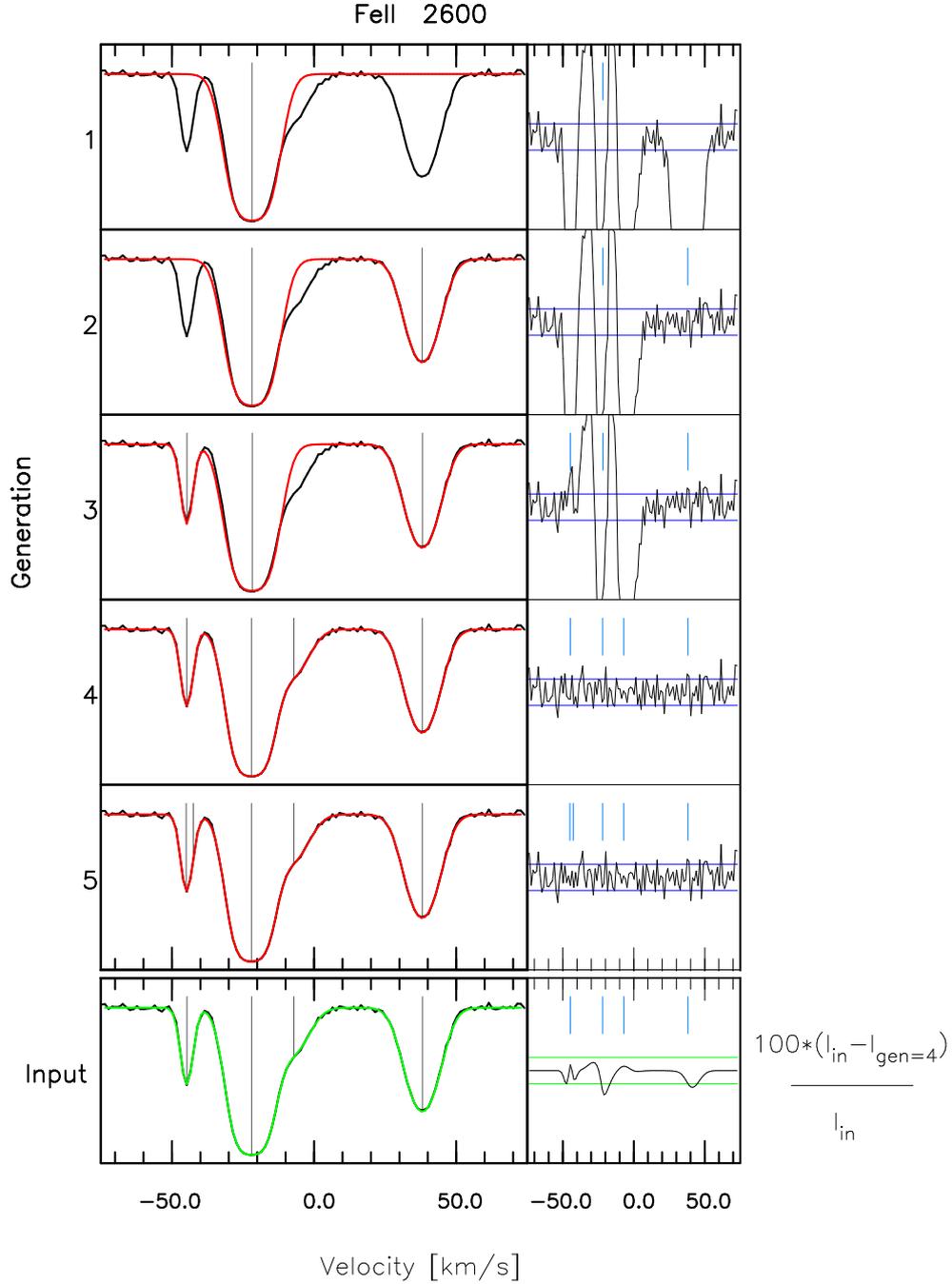}\\
\vspace{0.50mm}
\includegraphics[trim=0.00cm 0.50cm 2.60cm 20.95cm, clip=true, width=14.0cm]
{method_growth_figure.pdf}
\end{center}
\caption{\label{fig:example_growth} 
This illustrates the application of the genetic process to an simulated 
spectrum with known parameters. 
We simulated 2 FeII transitions (2600 and 
1608\AA) with 4 velocity components.  Only FeII 2600 is illustrated here. 
The absorption line parameters used to create the model are given in Table
\ref{tab:example_comparison}. The signal-to-noise used is approximately 100 
and the spectral resolution is $\sigma = 1.6$ km/s (as per the 2006 data). In the upper 5 left-hand panels, the spectral data are shown in
black and the best-fit model at each generation in red, $I_{gen}$. Grey
vertical ticks show the position of absorption components at each
generation. The upper 5 right-hand panels illustrate the normalised
residuals and the horizontal blue lines show 1$\sigma$ bounds. Each of
generations 1 through 5 in this Figure corresponds to 5 iterations of the ``EVOLUTION'' step in Figure \ref{fig:flowchartv4}. The genetic process 
found a satisfactory solution in 4 generations, plotted here on a velocity 
scale. In practice, we ran GVPFIT for 10 generations. Generations 6-10 appear 
visually almost identical so we  only plot the first 5. The lowest (6th) 
left-hand panel illustrates the ``true'' spectrum, $I_{in}$ and the 
corresponding right-hand panel plots the blown-up fractional difference 
$(I_{in} - I_{gen=4})/I_{in} \times 100$. In the same panel, the horizontal 
green lines show a fractional flux difference of 1\%. A comparison between 
the input and output parameters is given in Table \ref{tab:example_comparison}.  
The minimum-AICc model was found within the minimum possible number of
iterations, since there are 4 components in the model. 
}
\end{figure*}

The parameters used to generate the simulations described in this Section 
are given in Table \ref{tab:example_comparison} and in the caption to 
Figure \ref{fig:example_growth}, which shows the best fit at each
generation. 
Figure \ref{fig:met_gen} illustrates one iteration of {\sc GVPFIT}, demonstrating the application of our genetic algorithm to this simple simulated spectra. 

The most important conclusions of this initial simple test are: the 
minimum-AICc agrees extremely well with the ``true'' input spectrum, 
and this model was found within the minimum possible number of
iterations, since there are 4 components in the model.  Out of the 13
free parameters in the model, Table \ref{tab:example_comparison} shows
that 11 agree to better than 1$\sigma$ and the other 2 free parameters
are within or close to 2$\sigma$, as expected.

In Section \ref{synthetic} we describe more sophisticated tests based 
on simulated spectra with velocity structures both simpler than and 
more complicated than the $\protect\zabs$ = 1.839 towards J110325-264515.

\setlength{\tabcolsep}{0.75em}
\begin{table*}
\caption[Comparison of Input and Fitted models. ]{
\label{tab:example_comparison}
Here we show the results of applying {\sc GVPFIT} to the simple
simulated spectrum shown in Figure \ref{fig:example_growth}
and discussed in Section \ref{simplesim}.
The absorption line parameters for the input model and the
best-fit (minimum AICc) parameters derived using GVPFIT are 
tabulated.  The statistical uncertainties, as determined by 
{\sc VPFIT}, are also given. 
Input model: Four-component absorption system comprising only 
two transitions: FeII 1608 and FeII 2600. In the input model,
$\Delta \alpha /\alpha$ was set to $2.6 \times 10^{-6}$.
The total number of free parameters in the model $= 13$.
The number of data points $= 239$ and the number of degrees 
of freedom is therefore $= 226$. The best-fit $\alpha$ 
parameters are: $\Delta \alpha / \alpha = (0.7 \pm 1.8) 
\times 10^{-6}$, $\chi^{2} = 211.8$.  The first column gives 
the velocity component number. Columns 2 and 3: redshift
parameters; Columns 4 and 5: Doppler parameters; Columns 6 
and 7: FeII column density. Out of the 13 free parameters 
in the model, 11 agree to better than 1$\sigma$, and the 
other 2 free parameters are within or close to 2$\sigma$, 
as expected from Gaussian statistics.
}

\begin{center}
\begin{tabular}{lrrrrrr}
\hline
\# & $z_{input}$ \hspace{0.8mm} &   $z_{fitted}$ \hspace{9.0mm} &    $b_{input}$ \hspace{0.4mm} &    $b_{fitted}$ \hspace{2.0mm} & $N_{input}$ \hspace{2.0mm} &            $N_{fitted}$ \hspace{4.0mm} \\
&                               &                               & $[$kms$^{-1}]$ \hspace{0.3mm} &  $[$kms$^{-1}]$ \hspace{1.5mm} &    $\log_{10} [$cm$^{-2}]$ & $\log_{10} [$cm$^{-2}]$ \hspace{2.0mm} \\
\hline
1  &                  1.8383248 &     1.8383245 $\pm$ 0.0000005 &           1.84 \hspace{2.5mm} &                1.77 $\pm$ 0.07 &     12.4166 \hspace{1.5mm} &                   12.4178 $\pm$ 0.0064 \\
2  &                  1.8385399 &     1.8385395 $\pm$ 0.0000004 &           6.62 \hspace{2.5mm} &                6.64 $\pm$ 0.03 &     13.5948 \hspace{1.5mm} &                   13.5937 $\pm$ 0.0014 \\
3  &                  1.8386810 &     1.8386810 $\pm$ 0.0000018 &           6.94 \hspace{2.5mm} &                6.85 $\pm$ 0.26 &     12.4404 \hspace{1.5mm} &                   12.4390 $\pm$ 0.0149 \\
4  &                  1.8391085 &     1.8391085 $\pm$ 0.0000005 &           7.03 \hspace{2.5mm} &                7.02 $\pm$ 0.04 &     12.9853 \hspace{1.5mm} &                   12.9811 $\pm$ 0.0020 \\
\hline
\end{tabular}
\end{center}
\end{table*}
\setlength{\tabcolsep}{\oldtabcolsep}

\section{Additional free parameters, parameter constraints, interlopers}\label{prefit}

Additional free parameters may arise depending on the data attributes
and on what one wants to solve for.  The same additional free parameters
that can be introduced in ``manual'' fits can be included in {\sc
GVPFIT}, e.g. continuum level, zero levels, velocity offsets
between separate fitting regions, cloud temperature, turbulent velocity
component, $\alpha$. Similarly, parameter constraints (e.g. tied
redshifts, tied $b$-parameters, summed column densities) are defined in
exactly the same way as in manual fits.  

\subsection{b-parameter constraints}

As with previous analyses, we require that the absorption line width
($b$-parameter, with $b=\sqrt{2}\sigma$) for a particular species in the
fit is related to the corresponding components for other species. The
two extremes are wholly thermal broadening ($b^2 = 2kT/M$ for a species
with atomic mass $M$) and wholly turbulent broadening ($b^2 =
b^2_{\mathrm{turb}}$). In general, there will be contributions from the
two mechanisms ($b^2 = 2kT/M + b^2_{\mathrm{turb}}$).  We discuss this
further in the context of our modelling of the $z_{abs} = 1.839$
absorption system towards J110325-264515 in Section
\ref{sect:broadening}.  See \citet{vpfit} for further
discussions of tied parameters. The fully-automated method described in
this paper, i.e. {\sc GVPFIT}, is therefore no less flexible than the
well-established manual version, i.e. {\sc VPFIT}.

\subsection{Interlopers}\label{interlopers}

Since each quasar sight-line generally contains many distinct 
absorption systems at different redshifts, blending between 
species at different redshifts occurs frequently.  One approach 
to solving this complication is to carefully identify all 
absorption redshifts, to examine where such blends can occur, 
and then to simultaneously fit all relevant parameters.  
Nevertheless, it may not always be possible to identify all
absorption systems present and an unidentified interloping 
absorption line or lines can blend with transitions of interest.  
This may be revealed as a poor goodness-of-fit for the blended 
spectral region. Manually, in this circumstance, one would 
then guess at the positions, strengths and number of 
interloping velocity components, and allow {\sc VPFIT} 
to iterate to a best-fit solution (\citet{vpfit}).

In some cases, a statistically acceptable fit may be obtained, 
yet a weak interloper may nevertheless be present. 
A conservative approach when using {\sc GVPFIT} is therefore 
to {\it assume} interlopers are present (i.e. without 
needing to identify them beforehand) and allow GVPFIT to 
search parameter space for interlopers. 

The application of {\sc GVPFIT} is then as follows.  First we 
carry out a complete {\sc GVPFIT} assuming no interlopers are 
present.  Then we select a preferred model using AICc.  
We note that selecting a preferred model does not follow the 
spirit of using BMA (Section \ref{postfit}) but it is 
nevertheless a useful thing to do in this context.  
That preferred model is then used as a parent (in the sense 
of Section \ref{memetic}) and {\sc GVPFIT} is applied again 
but is only allowed to add interlopers as unidentified species 
(i.e. it is prevented from adding further metals, MgII for example).

The result of this process may or may not end up identifying 
any interlopers (for the absorption system analysed in this paper, 
it did -- See Section \ref{consistentvel}).  

If an interloper is found and is blended with some transition of 
interest, it could potentially bias the measurement of \da.  It 
is then necessary to quantify this potential bias.  There are
several ways in which this could be done.  In this paper
we adopt the following very simple approach and avoid a general
discussion outlining other possible methods.

Having identified the interloper or interlopers and having
found best--fit parameters for them using the preferred
model, we could ``insert'' the interlopers into the underlying
continuum and re-run the entire GVPFIT process with the
interlopers as fixed parameters.  In practice, for the absorption
system analysed in this paper, that was unnecessary, as explained
in Section \ref{consistentvel}.

\subsection{Weak lines and ill-conditioning}

As described in Section \ref{stop}, as {\sc GVPFIT} evolves the fit into
the ``over-fitting'' regions the model may include one
or more very weak absorption components that have large parameter error
estimates.  When that happens, logarithmic error estimates on column
densities can be large. Finite-difference derivatives of $\chi^2$ with
respect to those parameters can then be close to zero and thus dominated
by machine noise (i.e. rounding errors).  The effect of that is to
produce ill-conditioning in the Hessian matrix, potentially making the
non-linear least-squares optimisation unreliable \citep{GMW81}.
This in turn can impact adversely on the ability to correctly determine the
parameter estimates (and uncertainties) for the ``good'' velocity
components. However, it is easy to check for this effect. When such
large parameter error estimates are spotted, one can re-run the model
but with those particular parameters fixed.  This then removes any
possibility of throwing noise into the Hessian and gradient vector.  If
the parameter estimates and associated error estimates for the
parameters of interest are unchanged by this process, ill-conditioning
is unimportant.

\section{Post-fitting analysis}\label{postfit}

\subsection{Bayesian Model Averaging}

The methodology outlined above produces a large set of possible models,
many of which are statistically acceptable representations of the data.
The question then naturally arises as to how to select {\it one}
parameter, or range of parameters, deemed to be ``correct'', or at least
the best, solution. Should we chose one preferred model or is it better
to select a set of models? The former is the current {\sc VPFIT}
methodology.  The latter represents a fundamentally different approach.  

It has been noted that averaging over all the candidate models provides
better predictive ability than using any single model
(\citet{leamer1978, draper1995, hodges1987}).  The improvement provided
by model averaging, as opposed to selecting one preferred model, has
been quantified using a logarithmic scoring rule \citep{raftery1997} ---
see \citet{hoeting1999} for a review describing empirical support for
this method. In this Section we describe how to average over a large set
of models to derive robust parameter estimates.

Let $\mathbf{x}$ be the set of parameters describing the data.  In our
case we are particularly interested in \da\ , one of the parameters
contained within the vector set $\mathbf{x}$.  Let {\bf D} be the
spectrum i.e. {\bf D} $ = \{d_i\}$ ($d_i$ being the intensity at the
$i^{th}$ pixel in the spectrum). Let $\mathcal{M}_j =
\{I(\mathbf{x})_i\}_j$, where $I(\mathbf{x})_i$ is 
the model intensity at the $i^{th}$ pixel, 
i.e. $\mathcal{M}_j$ is the $j^{th}$ model of the data.

Given the spectrum {\bf D} and some particular model $\mathcal{M}_j$,
the posterior weighted average of the parameter set $\mathbf{x}$ is
\citep{raftery1997,hoeting1999}
\be
\mbox{p}(\mathbf{x}|\mathbf{D}) = \displaystyle\sum_{i=1}^{S}
\mbox{p}(\mathbf{x}|\mathbf{D}, \mathcal{M}_j) \mbox{p}(\mathcal{M}_j|\mathbf{D})
\label{eq:pxd}
\ee
where the summation is taken over all $S$
models (i.e. we sum over all models within all generations),
$\mbox{p}(\mathbf{x}|\mathbf{D}, \mathcal{M}_j)$ is the probability of
obtaining the parameters $\mathbf{x}$ given $\mathbf{D}$ and
$\mathcal{M}_j$ and $\mbox{p}(\mathcal{M}_j|\mathbf{D})$ is the
probability that $\mathcal{M}_j$ is the correct model given the data
$\mathbf{D}$ (assuming that the correct model is in fact in the list of
models),
\be
\mbox{p}(\mathcal{M}_{j}|\mathbf{D}) = \frac{ \mbox{p}(\mathbf{D}|\mathcal{M}_{j})\mbox{p}(\mathcal{M}_{j}) }{ \sum^{S}_{l=1} \mbox{p}(\mathbf{D}|M_{l})\mbox{p}(\mathcal{M}_{l})}
\label{eq:pmd}
\ee
The expected value and variance of $\mathbf{x}$ is then (\citet{hoeting1999}),
\be
\mbox{E}[\mathbf{x}|\mathbf{D}] = \sum^{S}_{i=1} \mbox{E}[\mathbf{x}|\mathbf{D},\mathcal{M}_{j}]\mbox{p}(\mathcal{M}_{j}|\mathbf{D})
\ee 
and
\begin{multline}
\mbox{Var}[\mathbf{x}|\mathbf{D}] = \sum^{S}_{j=1} (\mbox{Var}[\mathbf{x}|\mathbf{D},\mathcal{M}_{j}] \\
+ \mbox{E}[\mathbf{x}|\mathbf{D},\mathcal{M}_{j}]^2)\mbox{p}(\mathcal{M}_{j}|\mathbf{D})
- E[\mathbf{x}|\mathbf{D}]^2
\end{multline}

\subsection{Three statistics}\label{3stats}

The above formulation requires us to compute the probability of
obtaining the $j^{th}$ model, given the data $\mathbf{D}$, i.e. we need
to compute $\mbox{p}(\mathbf{D}|\mathcal{M}_{j})$. We use three
different statistics in our analysis: chi-squared, $\chi^2$, Akaike
Information Criterion corrected for finite sample sizes, $AICc$, and the
Bayesian Information Criterion, $BIC$.  Whilst all three statistics
provide a measure of the goodness of fit to the data, $\chi^2$ decreases
with the addition of free parameters, requiring the user to decide when
a satisfactory representation of the data has been achieved.  $AICc$ and
$BIC$ both compensate for the addition of free parameters, but $BIC$ has a
stronger penalty and is thus more prone to under-fitting (i.e. too few
parameters) compared to $AICc$.\\

\noindent{\em (i) Chi-squared, $\chi^2$}\\

We use the absolute (not the normalised) quantity,
\be
\chi^2_j = \sum^n_i (\{I(\mathbf{x})_i\}_j - d_i)^2/\sigma_i^2 = 
-2 \ln \mathcal{L}(\chi^2_j)
\ee
where $I(\mathbf{x})_i$ is the intensity at the $i^{th}$ pixel in the
data, $\mathbf{x}$ is the vector set of model parameters, $n$ is the 
number of data points, $d_i$ is the
model intensity at the $i^{th}$ pixel, $\sigma_i$ is its estimated 
uncertainty, and $\mathcal{L}(\chi^2_j)$ is the associated likelihood.\\

\noindent{\em (ii) Akaike Information Criterion corrected for small sample size, $AICc$}\\

The standard $AIC$ \citep{akaike1973} is 
\be
AIC_j = \chi^2_j + 2k
\ee
where $k$ is the number of free parameters. The corrected $AIC$ allows for
finite sample size \citep{hurvich1989} and is
\be
AICc_j = AIC_j + \frac{2k(k+1)}{(n-k-1)} = -2 \ln \mathcal{L}(AICc_j)
\label{eq:aicc}
\ee
where $k$ is the number of free parameters, $n$ is the number of
data points, and $\mathcal{L}(AICc_j)$ is the associated likelihood.\\

\noindent{\em (iii) Bayesian Information Criterion, BIC}\\

The Bayesian Information Criterion \citep{bozdogan1987} is,
\be
BIC_j = \chi^2_j + k\ln(n) = -2 \ln \mathcal{L}(BIC_j)
\ee
where $k$ is the number of free parameters, $n$ is the number of
data points and $\mathcal{L}(BIC_j)$ is the associated likelihood.

\subsection{Implementation}

We assume that each model is equally likely, i.e.
before considering the data we adopt a uniform prior, such that
$\mbox{p}(\mathcal{M}_{j})$ and $\mbox{p}(\mathcal{M}_{l})$ cancel out
in equation (\ref{eq:pmd}),
\be
\mbox{p}(\mathcal{M}_{j}|\mathbf{D}) = \frac{ \mbox{p}(\mathbf{D}|\mathcal{M}_{j}) }{ \sum^{S}_{l=1} \mbox{p}(\mathbf{D}|\mathcal{M}_{l}) }
\ee 
so $\mbox{p}(\mathcal{M}_{j}|\mathbf{D})$ becomes a simple model
weighting function and the expectation value and variance become
\be
\mbox{E}[\mathbf{x}|\mathbf{D}] = \sum^{S}_{i=1} \mbox{E}[\mathbf{x}|\mathbf{D},\mathcal{M}_{j}] \omega_{j}
\ee 
and
\begin{multline}
\mbox{Var}[\mathbf{x}|\mathbf{D}] = \sum^{S}_{j=1} \Big(\mbox{Var}[\mathbf{x}|\mathbf{D},\mathcal{M}_{j}] \\
+ \mbox{E}[\mathbf{x}|\mathbf{D},\mathcal{M}_{j}]^2\Big) \omega_{j}  - E[\mathbf{x}|\mathbf{D}]^2
\end{multline}

The weighting function $\omega_{j}$ depends on the statistical criteria used. 
Relating the statistical criteria to the model likelihoods, taking
$\chi^2$ as an example,
\be
\mbox{p}(\mathbf{D}|\mathcal{M}_{i}) = \mathcal{L}(\chi^2_j) = e^{-\chi^2_j/2}
\ee

The weighting function (relative likelihood) for the $\chi^2$ statistic
then becomes

\be
\omega(\chi^2_j) = \frac{\mathcal{L}(\chi^2_j)}{\sum_{l=1}^{S}\mathcal{L}(\chi^2_l)} = 
\frac{e^{-\chi^2_{j}/2}}{\sum_{l=1}^{S} e^{-\chi^2_{l}/2}}
\label{eq:relatlik}
\ee 
with analogous equations for the $AICc$ and $BIC$ statistics.

\section{Measurement of the fine structure constant at $\protect\zabs$
= 1.839 towards J110325-264515} \label{applic}

\subsection{Background}

The $z_{abs} = 1.839$ absorption system towards J110325-264515 is an
ideal target to illustrate the method presented in Section
\ref{memetic}. Several analyses of \da\ in this absorption system have
been published (See
\citet{levshakov2005,levshakov2007,molaro2008,levshakov2009,king2012})
permitting a direct comparison between the previous methodologies and
the genetic algorithm presented in this work (See Section
\ref{tab:dis_com}). The spectral data are high signal-to-noise and high
spectral resolution (\ref{tab:app_sys_tra}) and the absorption system is
relatively complex and thus presents a good challenge to our new method.

In addition, the combination of MgII and FeII absorption, including the
particularly important FeII1608 transition (because of its opposite-sign
sensitivity to $\alpha$ compared to other FeII transitions --- see Table
\ref{tab:app_ato_dat}), makes this system unusually sensitive to \da,
giving added motivation to a detailed analysis of this absorption
system. 

\subsection{The observational data}

The quasar J110325-264515 has an emission redshift of 2.145 and an
apparent magnitude (in B, V and R) of approximately 16 \citep{osmer77}.
Early high-resolution spectroscopic studies of this quasar include
\citet{Carswell82,Carswell84,Carswell91}. The observational data used in
the analysis presented in this paper were obtained using the
Ultra-Violet Echelle Spectrograph on the Very Large Telescope
(UVES/VLT) \citep{dekker2000}.

There have been three separate observations runs towards J110325-264515,
2000a, 2000b, and 2006. The instrumental settings were different for
each. In Table \ref{tab:dat_spe_obs} we summarise the details by
grouping exposures according to instrumental setting. Exposure groupings
numbered 1--4 were obtained using a slit-width of 0.80 and 0.90
arcsecond and 1x2 CCD binning.  Groupings numbered 5--8 had a 1.00
arcsecond slit and 2x2 binning.  Groupings 9 and 10 had a 0.50 arcsecond
slit and 1x1 binning.  Therefore we do not combine these data into a
single spectrum and instead maintain 3 independent spectra which are
then modelled simultaneously in our analysis. 

There is a single data grouping with a slit-width of 0.90 (line 1 from Table \ref{tab:dat_spe_obs}). We have merged this data grouping with the 0.80 slit width exposures from the same year (line 2--4 from Table \ref{tab:dat_spe_obs}) instead of separating it into a fourth co-added spectra. 

\begin{figure*}
\begin{center}
\vspace{-15.0mm}
\includegraphics[trim=0cm 0cm 0cm 0cm, clip=true, width=23.0cm,angle=-270]{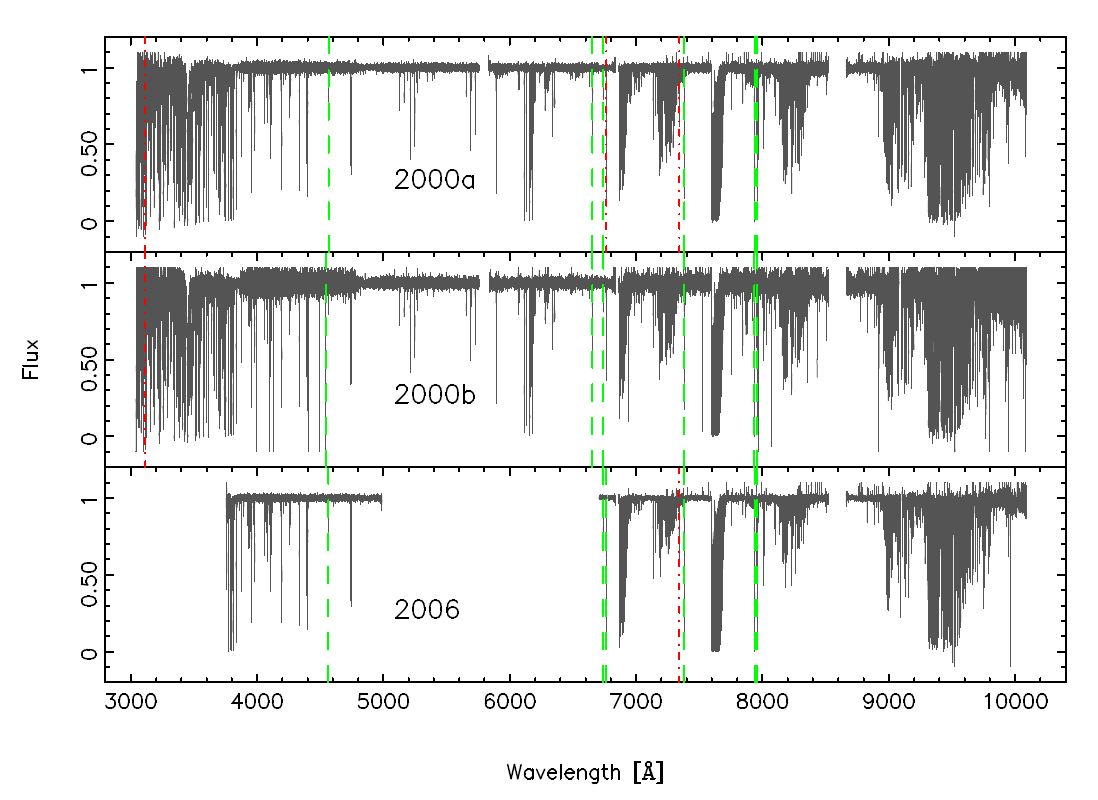}
\end{center}
\caption[
Illustration of the three combined spectra. 
]{
\label{fig:dat_spe_obs}
\textbf{
Illustration of the three combined spectra. 
}
Table \ref{tab:dat_spe_obs} gives the observing dates and parameters for
the quasar observations of J110325-264515 using UVES on the VLT. 
There are 10 groupings of exposures, co-added to make three spectra.  
Vertical green dashed lines illustrate the positions of the transitions used in
our analysis. Vertical red dot-dashed lines illustrate 5 transitions that are
present in the spectra but which have not been included in the analysis,
either because of suspected data problems or because no $q$-coefficient
is available (see Table \ref{tab:app_sys_tra}). 
}
\end{figure*}

Figure \ref{fig:dat_spe_obs} shows these three co-added spectra and illustrates
where the transitions used fall (green vertical lines) and where other
transitions were discarded because of suspected data problems (red
vertical lines).

\begin{table*}
\begin{center}
\caption[
Observational details for the quasar J110325-264515.
]{\textbf{
Observational details for the quasar J110325-264515.  
}
Column 1: data grouping. 
Column 2 and 3:  observation date range. 
Column 4: total exposure time. 
Column 5: number of exposures. 
Column 6: slit width. 
Column 7: UVES spectrograph grating setting (central $\lambda$ in nm). 
Column 8: on-chip binning (e.g. 1x2 = no binning in the spectral direction, 2 pixels
binning in the spatial direction). 
Column 9: astronomical seeing. 
Column 10: label for co-added spectrum used in Figure \ref{fig:dat_spe_obs} and Table \ref{tab:app_sys_tra}. 
The data groupings which contribute to each of the three co-added spectra are separated by horizontal lines and these co-added spectra are illustrated in Figure \ref{fig:dat_spe_obs}. 
}
\label{tab:dat_spe_obs}
\begin{tabular}{cccccccccc}
\hline
\# & Date from  & Date to    & Exp. Time & \# of Exp. & Slit Width & Setting & Bin. & Seeing   & Spectrum \\
&    yyyy-mm-dd & yyyy-mm-dd & [s]       &            & [arcsec]   &         &      & [arcsec] &          \\
\hline					   					   
1  & 2000-02-11 & 2000-02-12 & 16200.008 &          4 &       0.90 &   346   & 1x2  &   0.62   & 2000a    \\
2  & 2000-02-11 & 2000-02-16 & 18000.015 &          9 &       0.80 &   437   & 1x2  &   0.86   & 2000a    \\
3  & 2000-02-11 & 2000-02-12 & 16199.997 &          4 &       0.80 &   580   & 1x2  &   0.61   & 2000a    \\
4  & 2000-02-11 & 2000-02-16 & 17999.998 &          9 &       0.80 &   860   & 1x2  &   0.87   & 2000a    \\
\hline					   					   
5  & 2000-02-10 & 2000-02-12 &  7200.004 &          2 &       1.00 &   346   & 2x2  &   0.73   & 2000b    \\
6  & 2000-02-10 & 2000-02-10 &  3600.002 &          1 &       1.00 &   437   & 2x2  &   0.99   & 2000b    \\
7  & 2000-02-10 & 2000-02-12 &  7200.000 &          2 &       1.00 &   580   & 2x2  &   0.72   & 2000b    \\
8  & 2000-02-10 & 2000-02-10 &  3600.000 &          1 &       1.00 &   860   & 2x2  &   0.98   & 2000b    \\
\hline					   					   
9  & 2006-02-21 & 2006-02-23 & 55261.009 &          5 &       0.50 &   437   & 1x1  &   0.87   & 2006     \\
10 & 2006-02-21 & 2006-02-23 & 54271.992 &          5 &       0.50 &   860   & 1x1  &   0.88   & 2006     \\
\hline
\end{tabular}
\begin{tabular}{c}
\footnotesize{
}
\end{tabular}
\end{center}
\end{table*}

\subsection{Data reduction}

The spectra were extracted from 2d echelle format to 1d orders using the
ESO pipeline \citep{uves_pipeline}. Wavelength calibration made use of
the ThAr line list from \cite{murphy2007}. Order combination,
co-addition of multiple exposures to make the final 1d co-added spectra,
and continuum fitting, were done using {\sc UVES\_POPLER}
\citep{uves_popler}.

The MIDAS pipeline underestimates the 1-$\sigma$ spectral error array
near zero flux, impacting mostly at the base of saturated absorption
lines.  The effect is easily confirmed by fitting a straight (zero flux)
line within the saturated part of an absorption line and calculating the
normalised chi-squared, typically found to be around 2.  We correct the
spectral error arrays using the procedure described in Section 2.1 of
\citet{king2012}.

\subsection{Transitions used to measure $\alpha$}

We identified 32 transitions comprising the redshift 1.839 absorption
system: AlII, AlIII, CII, CIV, ClI, FeII, FeIII, HI, MgI, MgII, OI,
SiII, SiIII and SiIV. In this paper we focus on the FeII and
MgII transitions because of their similar ionisation potentials.

Between the three observation runs, there are 18 usable spectral 
regions containing FeII and MgII transitions at redshift 1.839
(see Table \ref{tab:app_sys_tra}).

The FeII 2586 transitions are present in our data in only the 2000a and
2006 data.  Unfortunately both of these spectral regions show apparent
observational glitches (perhaps due to cosmic rays or some other
corruption of the raw data) in the strongest velocity component and we
have not used either.  The same is true for the FeII 2382 transition in
2000a. The FeII 1096 transition is barely detectable and its sensitivity
to a variation on $\alpha$ (q-coefficient) has not yet been calculated.
The five data regions that were omitted are indicated in the Comment
column in Table \ref{tab:app_sys_tra}. 

The instrument profile was taken from the Ultra-Violet Echelle
Spectrograph user manual \citep{dekker2000}. 

\setlength{\tabcolsep}{0.20em}
\begin{table*}
\begin{center}
\caption[
FeII and MgII transitions in the $z=1.839$ absorption system towards J110325-264515. 
]{
\label{tab:app_sys_tra}
\textbf{
FeII and MgII transitions in the $z=1.839$ absorption system towards J110325-264515. 
}
Column 2: spectrum label. 
Column 3: species label. 
Column 4: transition label. 
Column 5: pixel size. 
Column 6: instrumental resolution. 
Column 7: observed central wavelength of absorption system. 
Column 8: average signal-to-noise ratio per pixel. 
Column 9: comments for data regions. 
}
\begin{tabular}{ccccccccl}
\hline
\# & Spectrum & Spec.        & Tran.        & Disp.         & Res. $\sigma$       & $\lambda_{\mbox{center}}$ & SNR        & Comment                                         \\
&             &              &              & [kms$^{-1}$]        & [kms$^{-1}$]              & [\AA]                     &            &                                                 \\
\hline                                                                                                        
1  & 2000a    & Fe II        &         1096 &          2.50 &                 2.3 &                    3113.7 &          13 & not used -- no $q$                              \\
2  & 2000a    & Fe II        &         1608 &          2.50 &                 2.3 &                    4565.9 &          74 &                                                 \\
3  & 2000a    & Fe II        &         2344 &          2.50 &                 2.3 &                    6654.6 &         101 &                                                 \\
4  & 2000a    & Fe II        &         2374 &          2.50 &                 2.3 &                    6740.4 &         136 &                                                 \\
5  & 2000a    & Fe II        &         2382 &          2.50 &                 2.3 &                    6764.0 &         128 & not used -- suspected data problem              \\
6  & 2000a    & Fe II        &         2586 &          2.50 &                 2.3 &                    7342.8 &          95 & not used -- suspected data problem              \\
7  & 2000a    & Fe II        &         2600 &          2.50 &                 2.3 &                    7381.1 &          91 &                                                 \\
8  & 2000a    & Mg II        &         2796 &          2.50 &                 2.3 &                    7937.2 &          71 &                                                 \\
9  & 2000a    & Mg II        &         2803 &          2.50 &                 2.3 &                    7957.6 &          60 &                                                 \\
\hline        									                              
10 & 2000b    & Fe II        &         1096 &          2.50 &                 2.7 &                    3113.7 &          12 & not used -- no $q$                              \\
11 & 2000b    & Fe II        &         1608 &          2.50 &                 2.7 &                    4565.9 &          36 &                                                 \\
12 & 2000b    & Fe II        &         2344 &          2.50 &                 2.7 &                    6654.6 &          68 &                                                 \\
13 & 2000b    & Fe II        &         2374 &          2.50 &                 2.7 &                    6740.4 &          87 &                                                 \\
14 & 2000b    & Fe II        &         2600 &          2.50 &                 2.7 &                    7381.1 &          46 &                                                 \\
15 & 2000b    & Mg II        &         2796 &          2.50 &                 2.7 &                    7936.8 &          37 &                                                 \\
16 & 2000b    & Mg II        &         2803 &          2.50 &                 2.7 &                    7957.6 &          29 &                                                 \\
\hline        									                              
17 & 2006     & Fe II        &         1608 &          1.25 &                 1.6 &                    4564.1 &         103 &                                                 \\
18 & 2006     & Fe II        &         2374 &          1.25 &                 1.6 &                    6737.7 &         153 &                                                 \\
19 & 2006     & Fe II        &         2382 &          1.25 &                 1.6 &                    6761.3 &         117 &                                                 \\
20 & 2006     & Fe II        &         2586 &          1.25 &                 1.6 &                    7339.8 &         145 & not used -- suspected data problem              \\
21 & 2006     & Fe II        &         2600 &          1.25 &                 1.6 &                    7378.2 &         110 &                                                 \\
22 & 2006     & Mg II        &         2796 &          1.25 &                 1.6 &                    7936.9 &         128 &                                                 \\
23 & 2006     & Mg II        &         2803 &          1.25 &                 1.6 &                    7957.2 &          99 &                                                 \\
\hline
\end{tabular}
\end{center}
\end{table*}
\setlength{\tabcolsep}{\oldtabcolsep}

\begin{figure*}
\begin{center}
\vspace{-0.50cm}
\includegraphics[trim=1cm 0.5cm 1cm 4cm, clip=true, width=15.5cm]{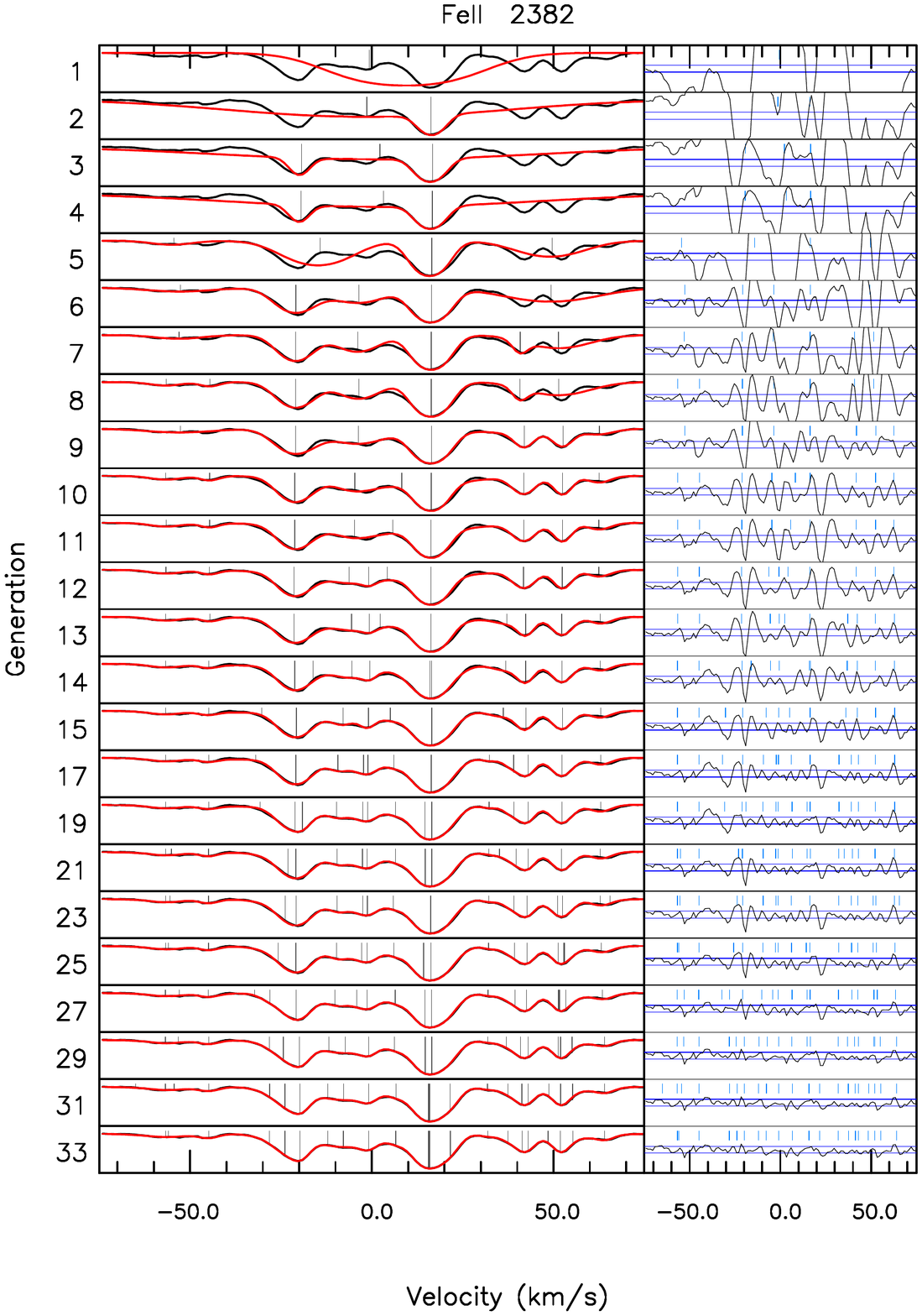}
\end{center}
\vspace{-0.25cm} 
\caption{
\label{fig:res_gro_one} 
\textbf{Evolution of minimum-$\chi^2$ models.}
This figure shows how the complexity of the
model absorption profile evolves with each successive generation.  
Shown are the best-fit model from generations 1--15 and then every odd numbered 
generation from 15--81. 
The minimum-$\chi^2$ model (selected from the whole set of turbulent and thermal models) 
is displayed for each generation until generation 81. 
Each model was fitted simultaneously to all 6 FeII and 2 MgII transitions listed in
Table \protect\ref{tab:app_sys_tra}.  
This figure contains a series of four plots. 
In the first two plots we illustrate how the fit evolved
over generations 1 to 81 for the strongest iron transition, FeII 2382. 
In the last two plots we illustrate the same thing for MgII 2796. 
In the left column, the spectral data are shown in black and the model
as the smooth red line. In the right column, the normalised residuals are
shown, plotted over the same velocity range (but on a smaller scale).
Blue horizontal lines illustrate the $\pm 1\sigma$ expectation values. 
Vertical ticks (grey in the left column, light blue in the right column) 
show the positions of absorption components at each generation. 
}
\end{figure*}

\begin{figure*}
\begin{center}
\includegraphics[trim=1cm 0.5cm 1cm 4cm, clip=true, width=17cm]{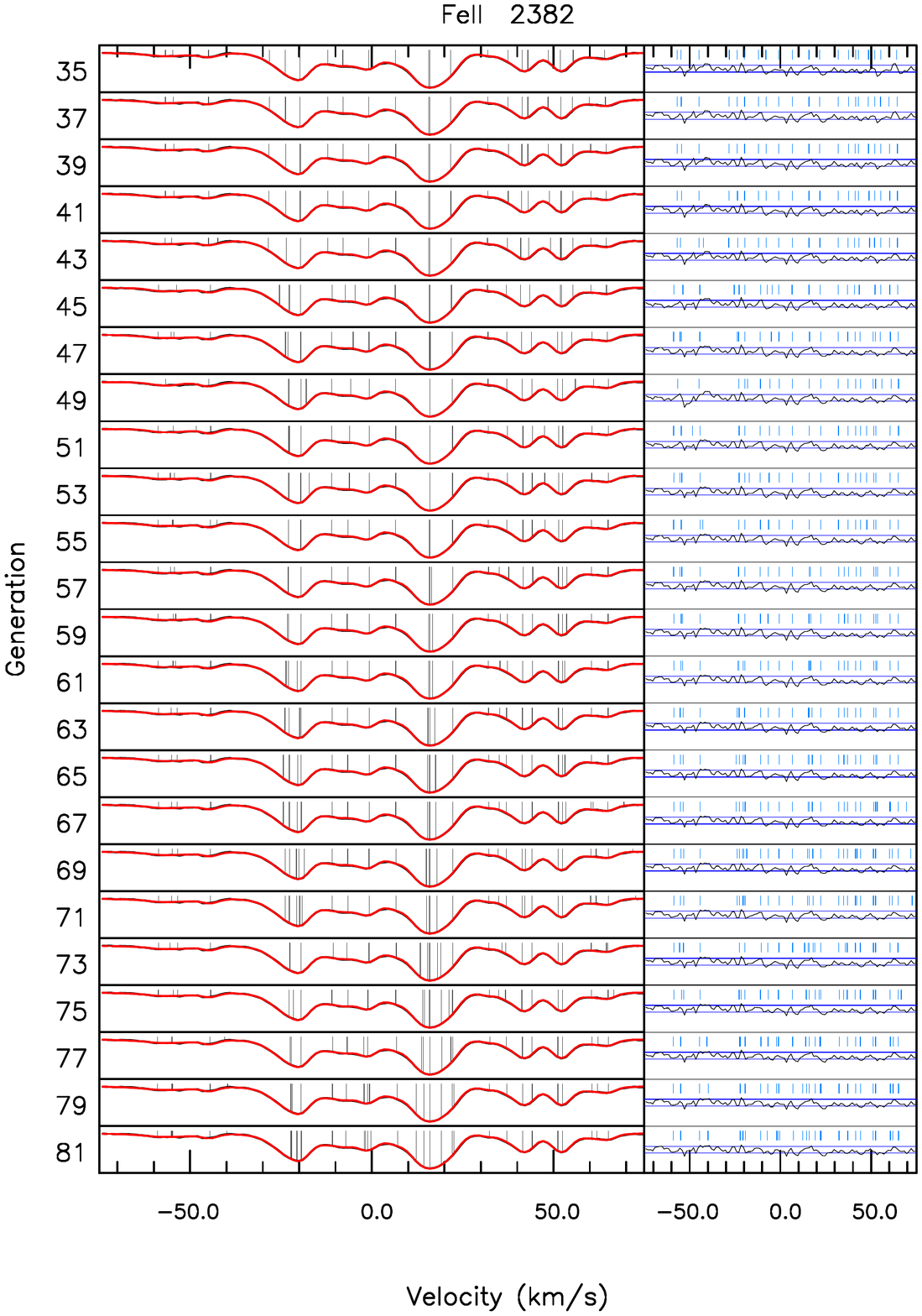}
\contcaption{
$\!\!.$ 
\textbf{Evolution of minimum-$\chi^2$ models.}
}
\end{center}
\end{figure*}

\begin{figure*}
\begin{center}
\includegraphics[trim=1cm 0.5cm 1cm 4cm, clip=true, width=17cm]{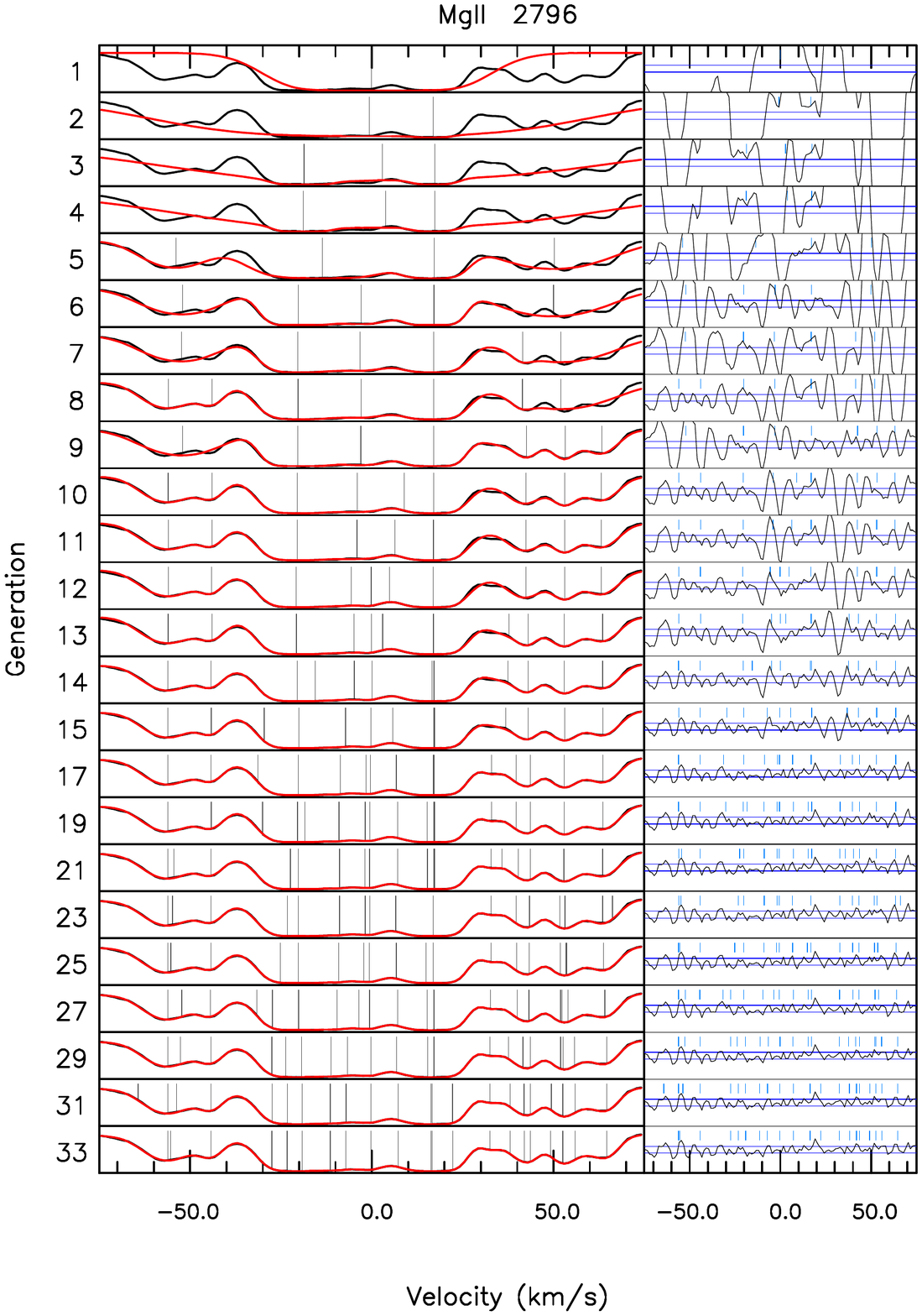}
\contcaption{
$\!\!.$ 
\textbf{Evolution of minimum-$\chi^2$ models.}
}
\end{center}
\end{figure*}

\begin{figure*}
\begin{center}
\includegraphics[trim=1cm 0.5cm 1cm 4cm, clip=true, width=17cm]{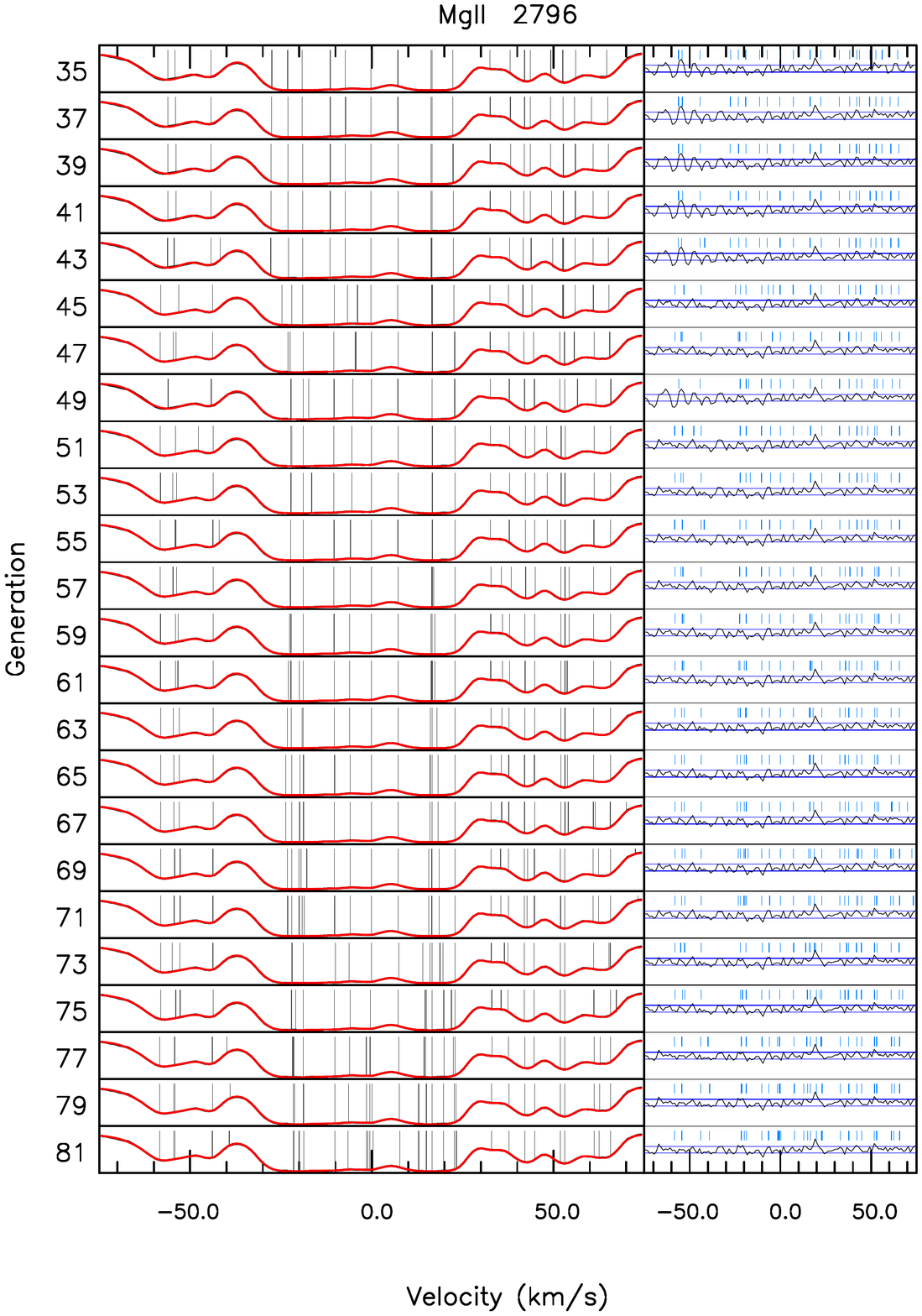}
\contcaption{
$\!\!.$ 
\textbf{Evolution of minimum-$\chi^2$ models.}
}
\end{center}
\end{figure*}

\subsection{Atomic data}

The laboratory wavelengths of the FeII and MgII transitions have
recently been re-measured. We use the most recent and most precise MgII
laboratory wavelengths \citep{batteiger2009}, and the most recent and
most precise FeII wavelengths \citep{nave2012}.  In generating model
absorption profiles, we use the $^{25}$MgII 3s to 3p hyperfine structures
(see Appendix \ref{app:atomic_data}).

\subsection{Choice of fitting regions}

It is important to ensure that absorption regions are flanked by
adequate continuum regions when fitting \citep{wilczynska2015}. Doing
this makes it less likely to miss weak lines on the system edges, more
likely to anchor the broader velocity components, and will result in a
more robust estimate of a floating continuum level, if included in the
fitting parameters. 

The final wavelength ranges for the fitting regions used in this paper
are tabulated in Table \ref{tab:preferred_model_data_regions} and
illustrated in Figure \ref{fig:j11_pre_one}.  The weaker FeII lines
include seemingly over-generous continuum regions (i.e. in the
approximate regions $-180 < v < -80$ km/s).  These regions provide extra
column density constraints (albeit weak) and the number of free
parameters (i.e. number of velocity components) is well-defined (using
the AICc) by the stronger FeII transitions.

\subsection{Line Broadening} \label{sect:broadening}

We do not know {\it a priori} whether this absorption system is best
represented by a thermal model, a turbulent model, or a hybrid of each
(Section \ref{prefit}).  In this analysis we are only considering 2
species, MgII and FeII.  In such a situation the degeneracy between
b-parameters and temperature is too severe to solve independently for
temperature (and hence solve for a hybrid model). In previous analyses,
the approach has been to first develop and solve for a turbulent model
and then to use that turbulent model as the starting point for
developing a thermal model (because of the human time required). 
However in the analysis presented here, both thermal and turbulent
models have been constructed independently. To our knowledge, this is
the first time that this has been done.  

\subsection{Obtaining a consistent velocity structure for all
atomic species at $z_{abs} = 1.839$ towards J110325-264515} 
\label{consistentvel}

The model here includes only two species, MgII and FeII.
In order to force a consistent velocity structure for both species, 
we followed the procedure described in Section \ref{interlopers}.
One interloper was found, illustrated as a green dashed vertical
line at approximately +125 km/s in Figure \ref{fig:j11_pre_one}.
In this particular case, it is obvious that the interloper is not
going to impact on the \da\ estimate so we did not bother to
follow the last part of the (more general) procedure outlined in
Section \ref{interlopers}, i.e. we did not enter the interloper
parameters as fixed and re-run {\sc GVPFIT}.  

Interestingly, for this particular absorption system, there did 
not appear to be any velocity components with MgII above the 
detection threshold and FeII below, otherwise the interloper 
procedure (Section \ref{interlopers}) would have identified them.  
The weak interloper that was found is detected only in MgII 2803 
and not in MgII2796 so cannot be MgII.  Looking at the detail 
in the MgII panels of Figure \ref{fig:j11_pre_one}, there is 
an apparent difference between the interloper strength and 
position for the 2000a and 2006 spectra.  We presume there are 
data problems in the 2000a spectrum at that point but made no
attempt to mask data as the feature is well offset from the
metal features relevant to \da.

\subsection{Results} \label{sect:results}

In the analysis described here we use {\sc VPFIT} Version 10, 
coupled with the new genetic algorithm {\sc GVPFIT}.

Figure \ref{fig:res_gro_one} shows how the model evolves at each
generation. Each sub-plot in these figures illustrates the ``most
successful'' model (i.e. minimum-$\chi^2$) out of
40 trials. Each generation up to around generation 15 (in
this particular case) results in a substantial improvement in the model,
seen clearly by looking at the residuals), $(D_i - M_i)/\sigma_i$ , 
where at pixel $i$, $D_i$ and $M_i$ are the flux of the
data and model respectively.

By generation 32 the models are already good, with a normalised 
$\chi^2 = 1.006$ at that generation. After generation 
32, occasionally 
new components are added, and consecutive $\chi^2$ values change only
slightly.  We allowed calculations to go on to generation 84
(by which time $\chi^2_{\nu}$ has fallen to 0.897 --- Section \ref{stop}).

\begin{figure*}
\begin{center}
\begin{tabular}{cc}
\includegraphics[trim=3.70cm 3.00cm 4.75cm 4.25cm, clip=true, width=19.00cm]{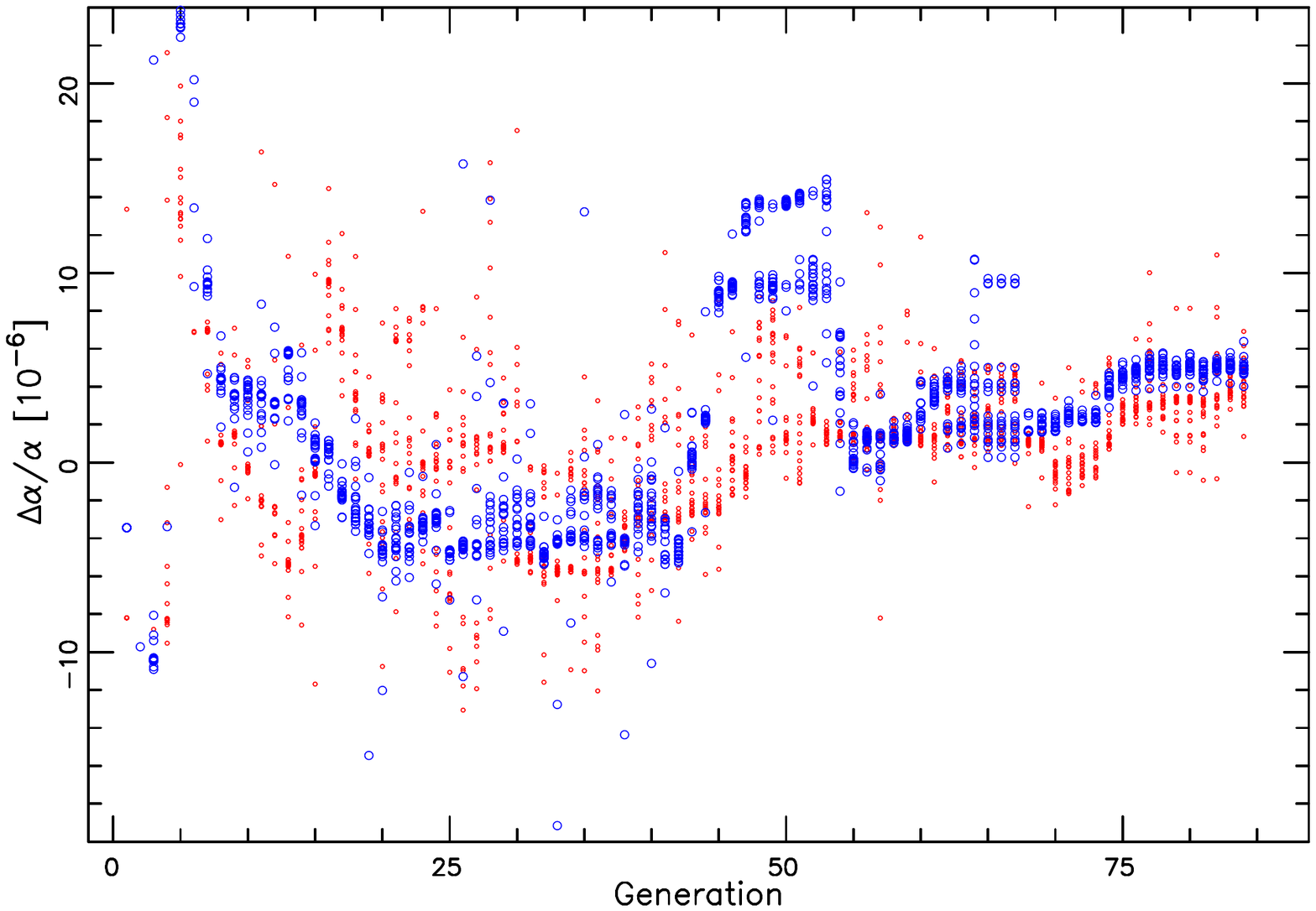} &
\end{tabular}
\end{center}
\caption[
\da\ vs generation. 
]{
\label{fig:da_fig0}\textbf{
\da\ vs generation} 
for all models generated during the analysis of the 
Fe II and Mg II absorption profiles at $z_{abs} = 1.839$ towards 
J$110325-264515$. The larger blue and smaller red circles show turbulent 
and thermal fits respectively. At increasing generation number, the models 
become increasingly complex. The data points suggest an 
underlying general 
trend, from which there is an abrupt departure over the generation range 
43 through 55.  Across this whole range, the goodness of fit (according 
to $\chi^2$) is acceptable.  This is clearly a local minimum in 
$\chi^2$-\da\ space, that would be impossible to identify without a 
thorough exploration of parameter space, but which is easily seen here.  
We discuss this point in detail in Section \ref{sect:results}.  Note 
that \da\ seems to stabilise to a ``plateau'' after 
approximately generation 53, where the scatter in \da\ is smaller than 
with the statistical uncertainty on \da\ in that region within thermal 
or turbulent models.  The statistical uncertainties are given in Table
\ref{tab:j11_sta} and illustrated in Figure \ref{fig:da_fig1}.
}
\end{figure*}

\begin{figure*}
\begin{center}
\begin{tabular}{cc}

\includegraphics[trim=3.00cm 3.99cm 4.75cm 4.25cm, clip=true, width=9.00cm]{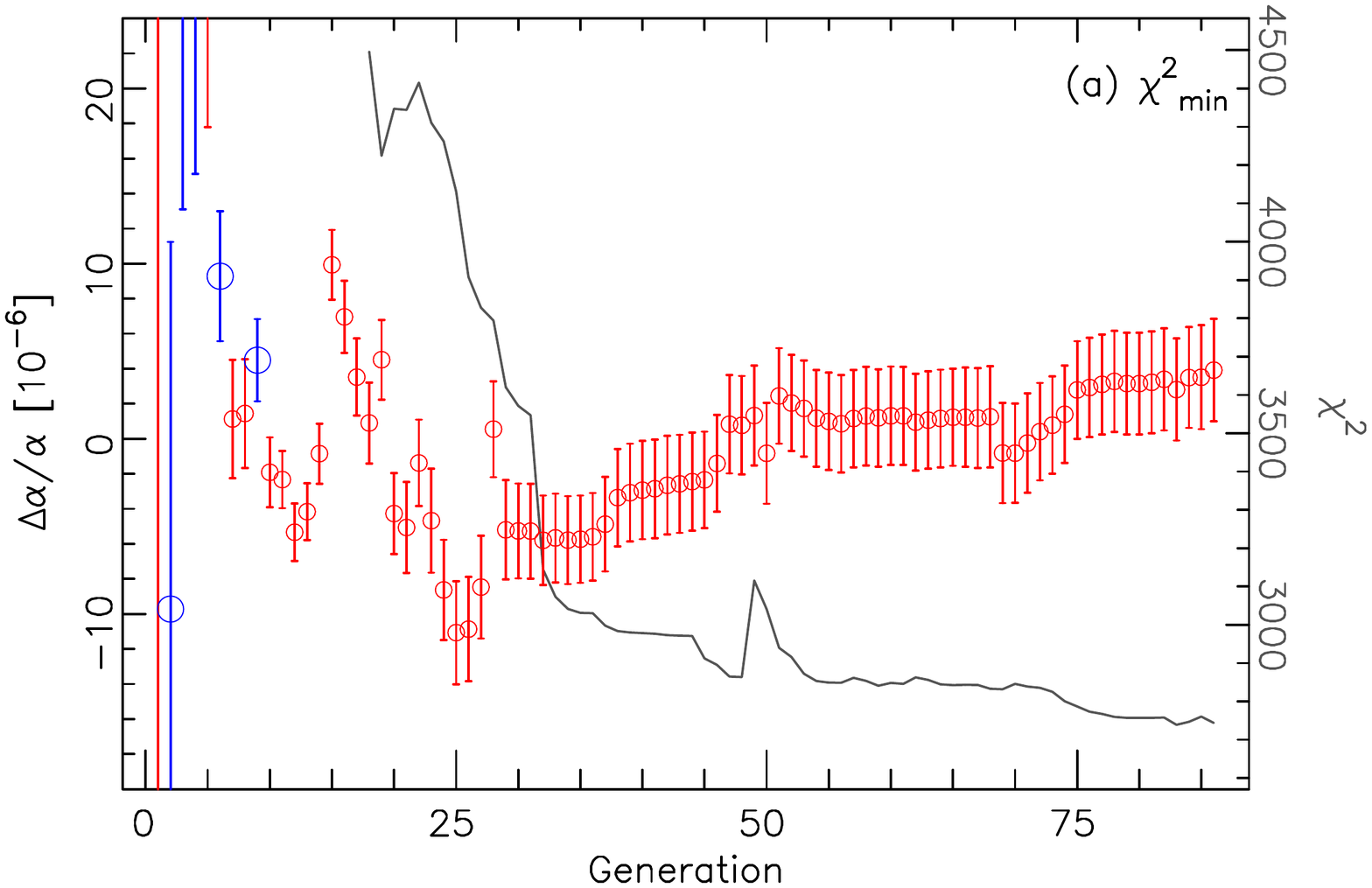} &
\includegraphics[trim=3.00cm 3.99cm 4.75cm 4.25cm, clip=true, width=9.00cm]{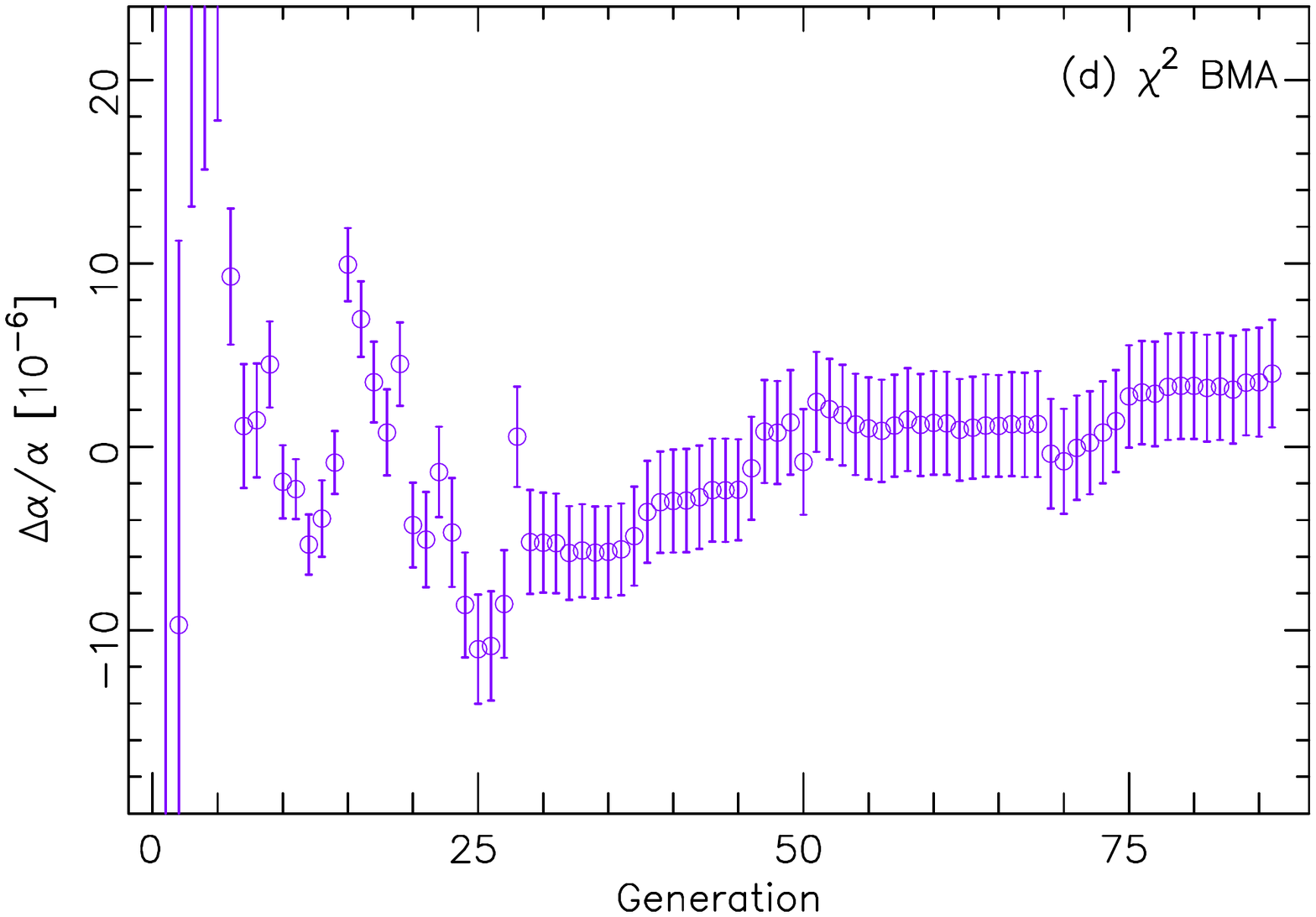} \\

\includegraphics[trim=3.00cm 3.99cm 4.75cm 4.25cm, clip=true, width=9.00cm]{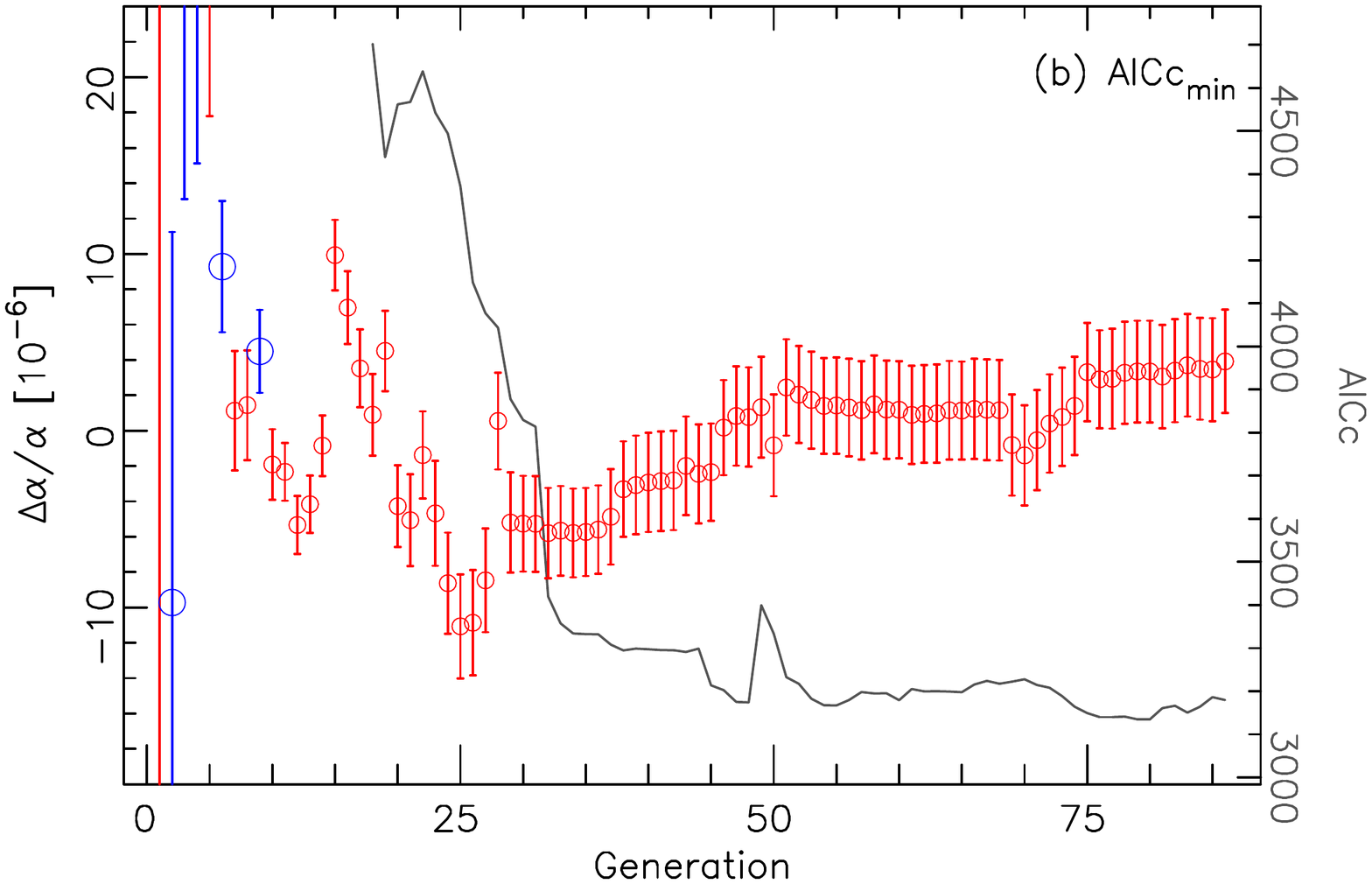} &
\includegraphics[trim=3.00cm 3.99cm 4.80cm 4.25cm, clip=true, width=9.00cm]{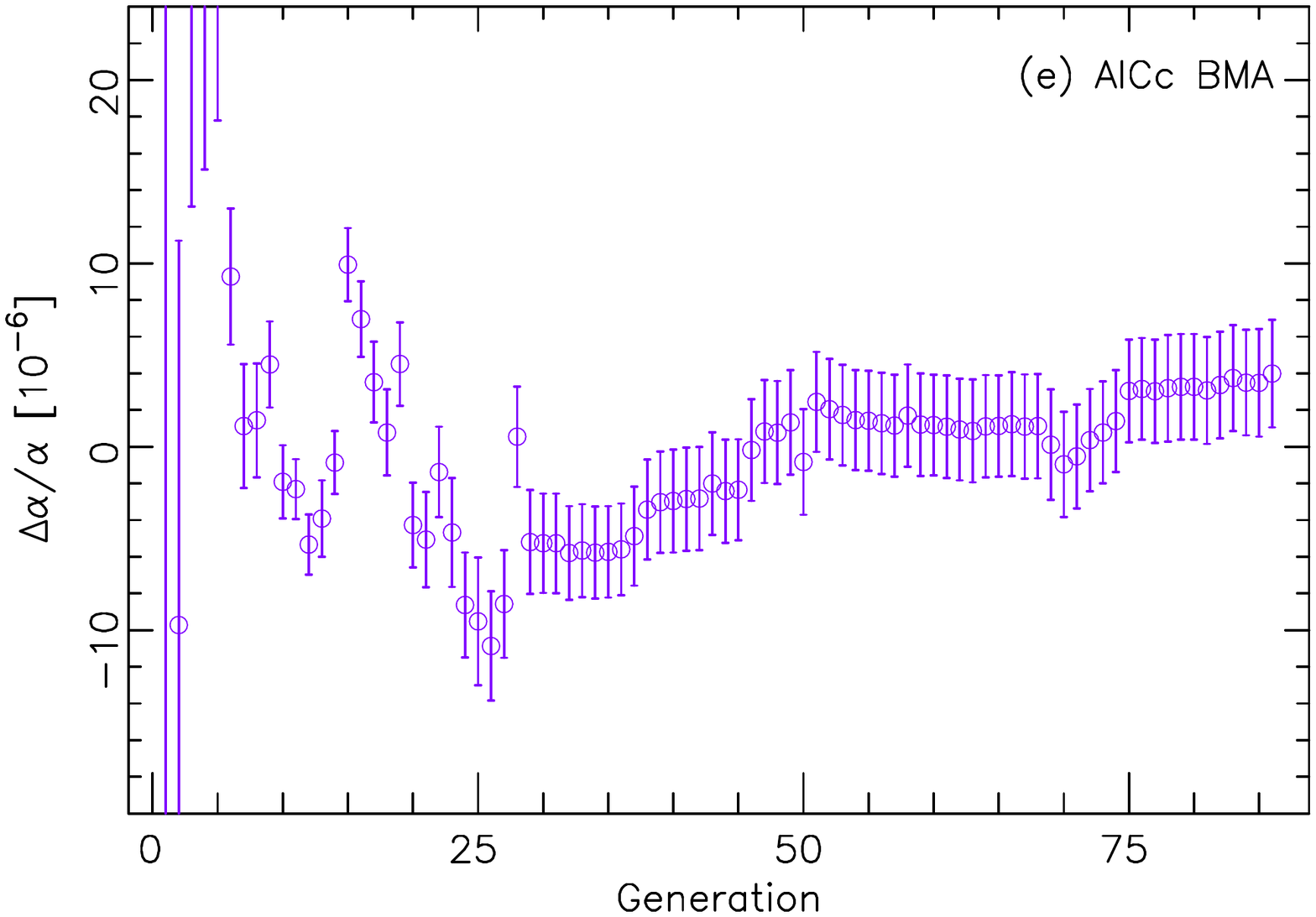} \\

\includegraphics[trim=3.00cm 2.50cm 4.75cm 4.25cm, clip=true, width=9.00cm]{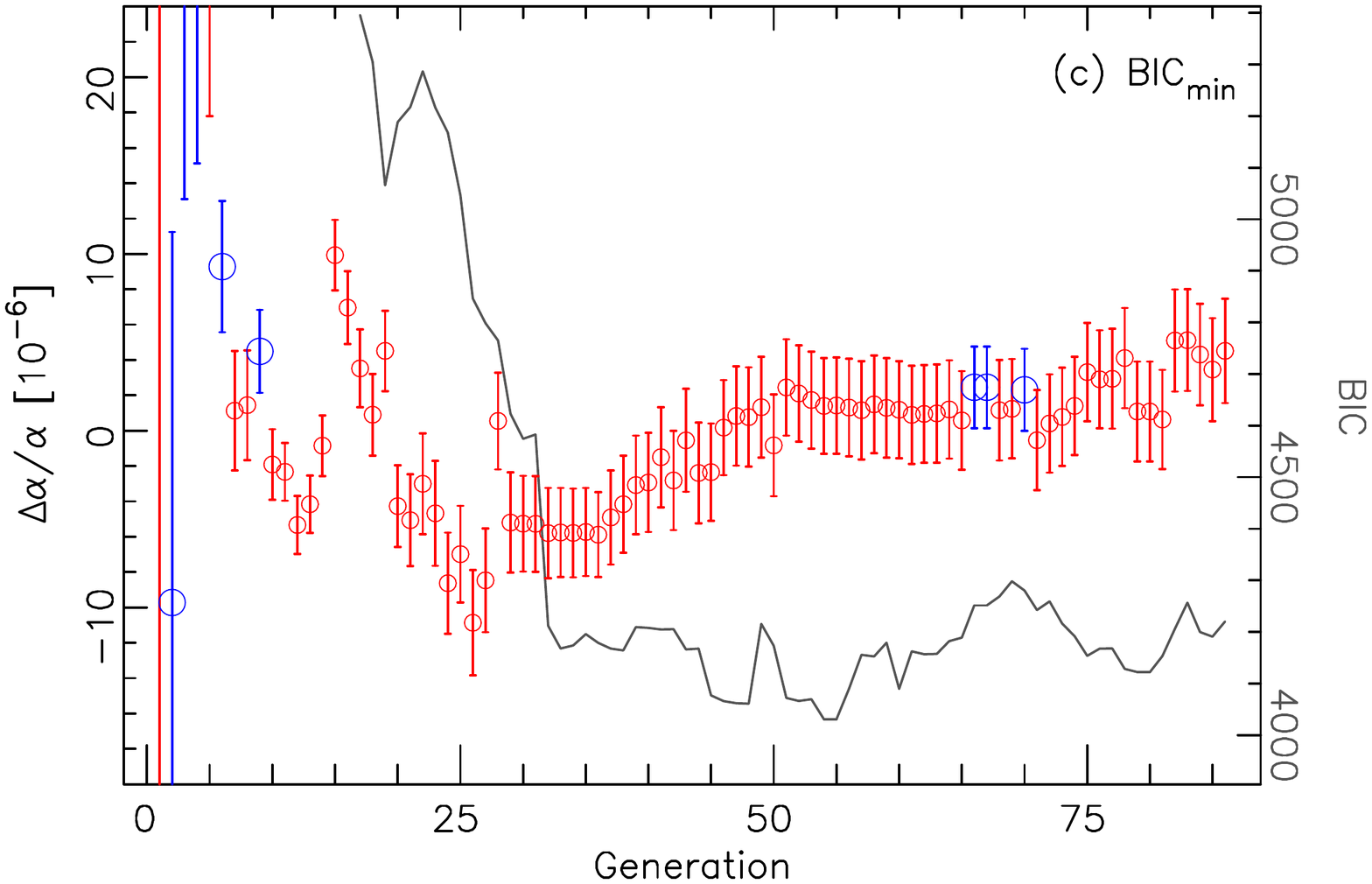} &
\includegraphics[trim=3.00cm 2.50cm 4.80cm 4.25cm, clip=true, width=9.00cm]{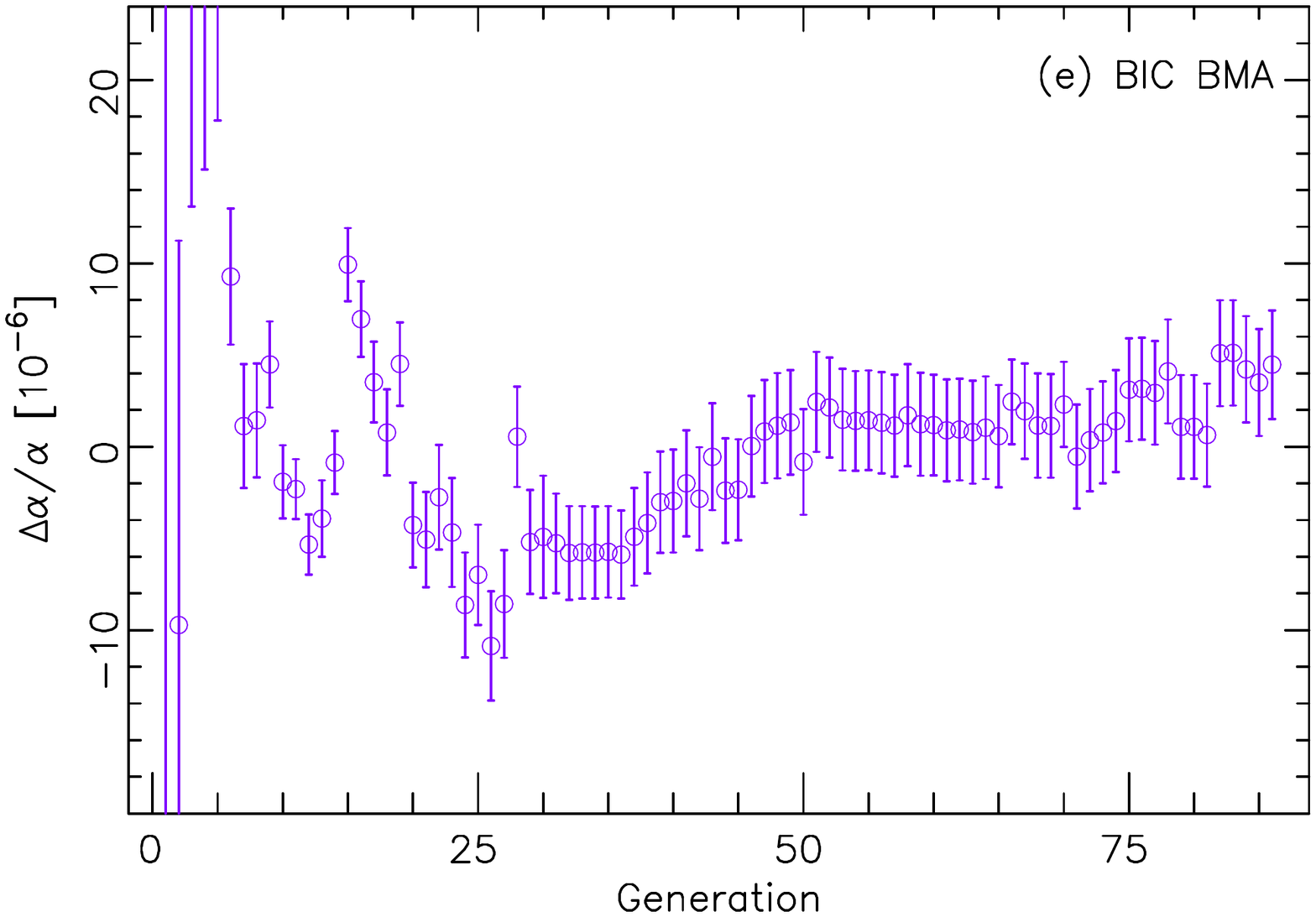} \\

\end{tabular}
\end{center}
\caption[
Comparison of \da\ evolution for different methods of analysis. 
]{
\label{fig:da_fig1}\textbf{Comparison of \da\ evolution for different 
methods of analysis.} 
Left-hand panels (a), (b), and (c):
Each point in these 3 panels corresponds to the \da\ estimate of the
model with a minimum in each statistic ($\chi^2$,
AICc and BIC respectively) 
from the combined set of thermal and turbulent models for that generation. 
Again, larger blue and
smaller red circles show turbulent and thermal fits respectively. 
After generation 10 the thermal models are preferred over turbulent models, with only a few exceptions, in all three statistics ($\chi^2$, AICc and BIC). 
The
continuous grey line illustrates the value of the statistic itself. The
features or spikes in the grey line are present because we plot the
absolute (as opposed to the normalised) value of the statistic, and the
number of free parameters in the fit is fluctuating slightly between
competing models.
Right-hand panels (d), (e), and (f):
Each point in these 3 panels corresponds to the BMA 
\da\ estimate for that generation, using the relevant statistic
($\chi^2$, $AICc$ and $BIC$ respectively). 
In all panels, error bars illustrate statistical uncertainties, 
derived from the diagonal of the covariance matrix in {\sc VPFIT}. 
}
\end{figure*}

Figures \ref{fig:da_fig0} and \ref{fig:da_fig1} show the \da\ estimates
for every model generated during the fitting process. In Figure
\ref{fig:da_fig0}, red and blue hollow circles illustrate the turbulent
and thermal models respectively.  Figure \ref{fig:da_fig1} (a), (b) and
(c) show the best-fit models according to $\chi^2$-test, $AICc$ and
$BIC$.  Figure \ref{fig:da_fig1} (d), (e) and (f) illustrate the
BMA estimates.

Table \ref{tab:j11_dal} gives the final estimates of \da\ for this
system.  The top three lines correspond to minimum values of the
three statistical goodness-of-fit measures, $\chi^2$-test, $AICc$ and
$BIC$, i.e. these correspond to specific models. The lower three lines
give the BMA results, as described in Section
\ref{postfit}.

Figures \ref{fig:da_fig0} and \ref{fig:da_fig1} and Table
\ref{tab:j11_dal} illustrate three important results:
\begin{itemize}

\item[(1)] The generation figure, Figure \ref{fig:da_fig0}, shows some
interesting characteristics. There is what appears to be an underlying
linear trend from approximately generation 20 onwards.  The most
conspicuous departure from that underlying trend happens between
generations 45 and 55, where \da\ rises abruptly but temporarily to a
much higher value. Despite the rather dramatic change in the value of
\da\ in this region, the models remain statistically good fits to the
data, as Table \ref{tab:j11_sta} shows. Given the procedure, this
feature therefore corresponds to a local minimum.  In the absence of the
broad picture provided by Figure 6, that is, if we were modelling the
spectrum following previous methods, where only one "best fit" to the
data is obtained, it would be easy to find a highly significant non-zero
value of \da. In Section \ref{emulate} we attempt to emulate human interactive
fitting and Table \ref{tab:int_comp} provides some interesting quantitative 
information. It turns out that the AICc and BIC ``preferred models''
(in this emulation) fall within this generation range and hence the 
results are clearly anomalous. This feature and its interpretation highlight 
the importance of a procedure such as GVPFIT.

\item[(2)] Beyond generation 55, where the normalised $\chi^2_{turb}$
and $\chi^2_{therm}$ values had dropped to $0.998$ and $0.918$
respectively, the scatter reduces and a greater degree of consistency
emerges for \da\ within each generation, i.e. the scatter within one
generation becomes smaller than the statistical error bar.

\item[(3)] Beyond generation 55, not only does the scatter within each
generation reduce, but the results become consistent across generations. 
\da\ is robust over a broad range of models (Table \ref{tab:j11_sta}).

\item[(4)] Variations between the values of \da\ given in Table
\ref{tab:j11_dal} are comfortably within the estimated statistical
errors.  The result for \da\ is thus insensitive to which statistic is
used. Figure \ref{fig:da_fig1} illustrates that in this case choice of
statistic is unimportant.

\end{itemize}

\subsubsection{An example model}

Clearly, each of the three BMA results are derived
from a large number of model spectra, so we cannot easily illustrate a
``best-fit'' spectrum.  Instead, for illustrative purposes, 
Figure \ref{fig:j11_pre_one} shows the minimum-AICc model.

Using AICc, we find a minimum value occurs at AICc$_{min} = 3035$ in
generation 67 (highlighted in Table \ref{tab:j11_sta}).  
As expected, 
since BIC has a stronger penalty for increased number of free
parameters and $\chi^2$ does not penalise at all, AICc$_{min}$ falls in
between the $\chi^2$ and BIC minima.

We note again here that this model (and all others described in this
paper), were derived without any human decision making other than
initially defining spectral fitting ranges and identifying species
detected. Initial parameter guesses were defined by the genetic
algorithm and all fitting procedures are completely automated.

Figure \ref{fig:j11_pre_one} shows that the genetic algorithm has
resulted in an excellent fit to the data, at least as good as could be
done interactively. The residuals are well behaved and there are no
discrepancies between the data and the model.  The only minor exceptions
are in one or two small areas where residual cosmic rays or other
problems remain in the extracted spectrum (which have an insignificant
impact on parameter estimation in this case).  The model parameters for
the fit illustrated in Figure \ref{fig:j11_pre_one} are given in Table
\ref{tab:preferred_model}.

\begin{figure*}
\begin{center}
\includegraphics[trim=1.45cm 0.95cm 2.59cm 3.23cm, clip=true, width=14.5cm]{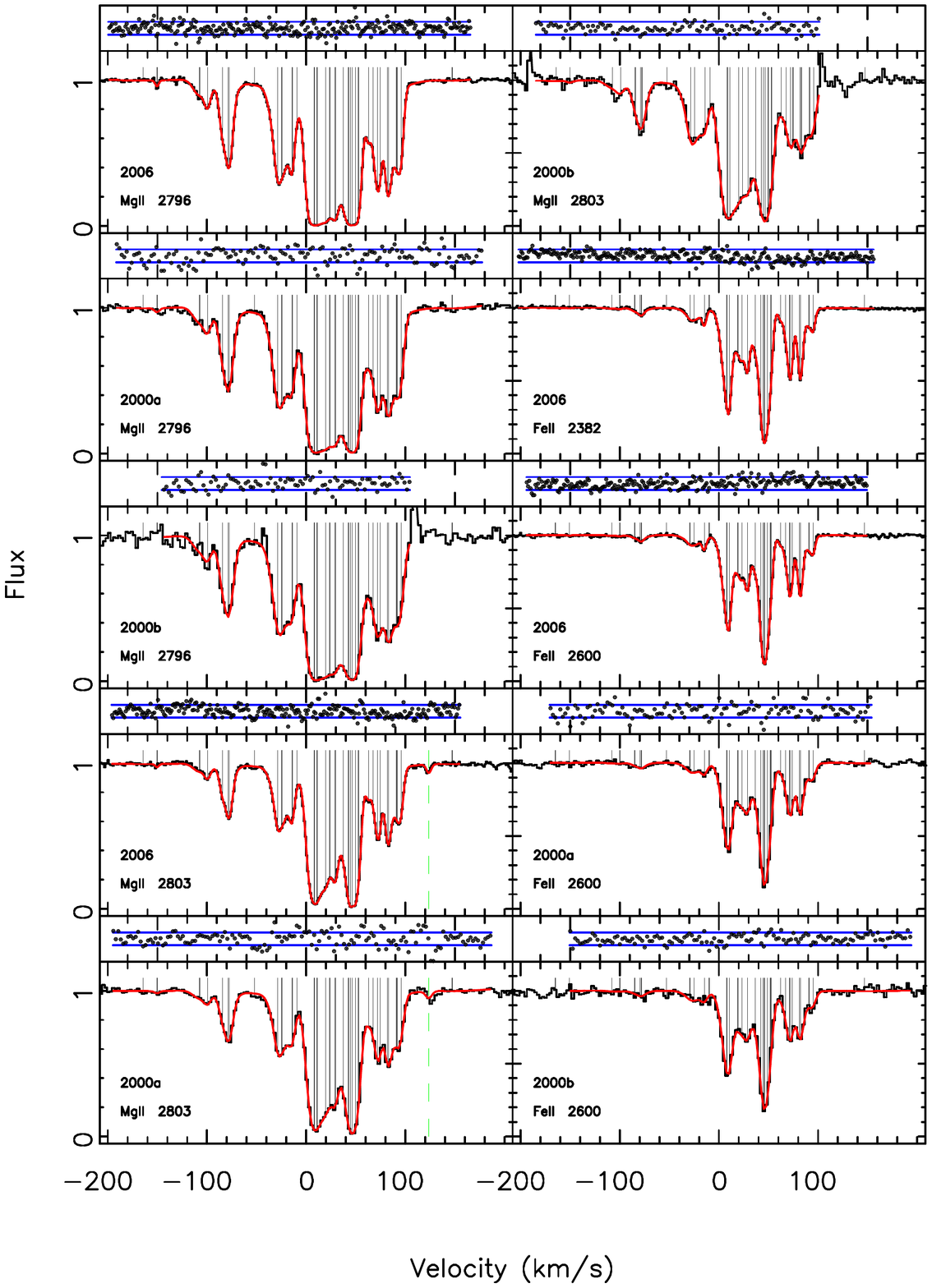}
\end{center}
\caption[
Lowest $AICc$ model. 
]{
\label{fig:j11_pre_one}
\textbf{
Comparison of the minimum-$AICc$ model and the spectral data.}
There are two panels on each row: the top panel compares the residuals
(black points) to the $1 \sigma$ expected fluctuations (horizontal blue
lines). The bottom panel compares the minimum-AICc model (smooth red line)
to the data (black). Each velocity component is shown as a grey
vertical line.  Vertical dashed (green) lines indicate interlopers (i.e.
additional components assumed to be from some other species at some
other redshifts).
}
\end{figure*}

\begin{figure*}
\begin{center}
\includegraphics[trim=1.45cm 0.95cm 2.59cm 5.23cm, clip=true, width=16.0cm]{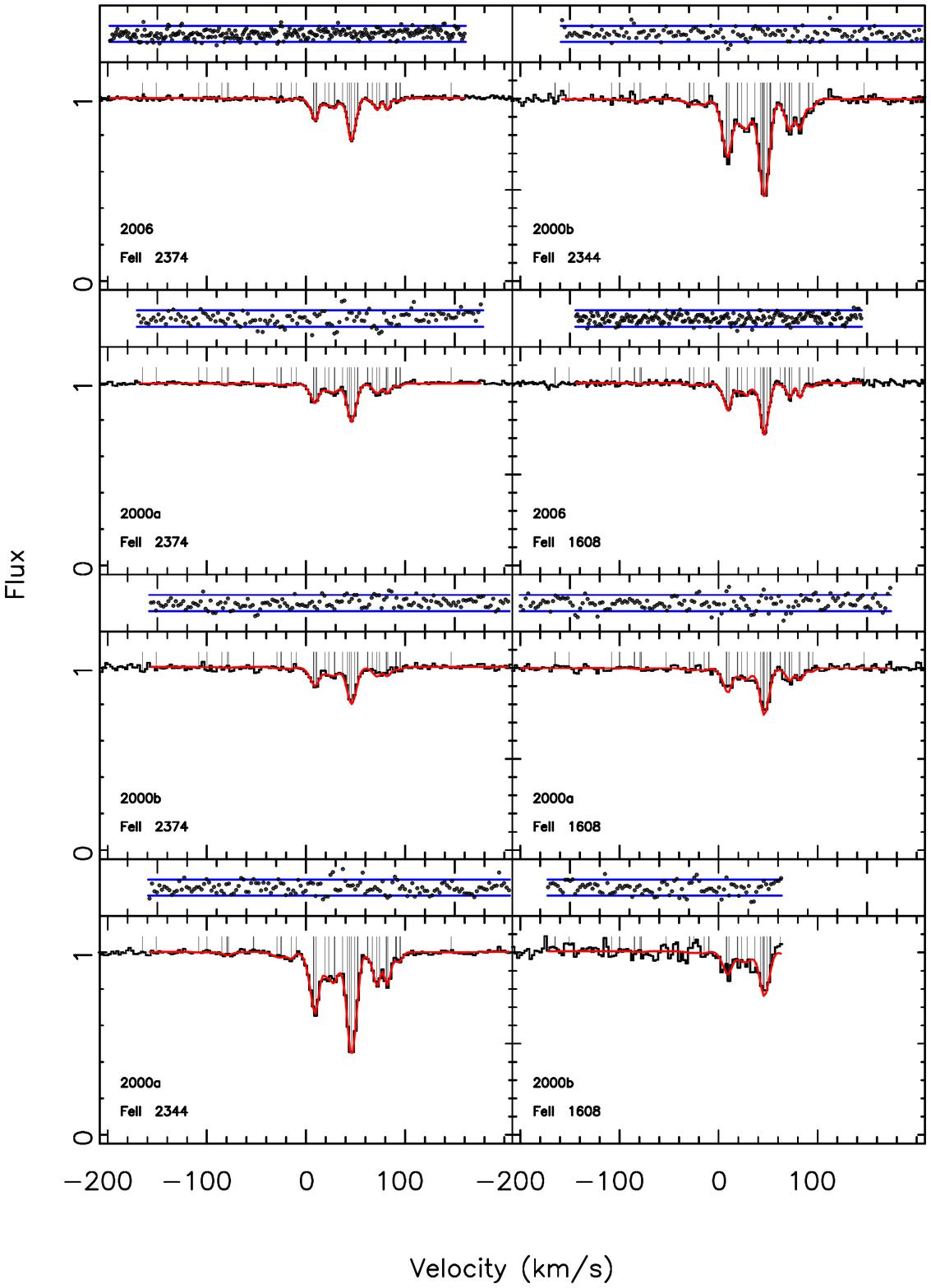}
\end{center}
\contcaption{
$\!\!.$ 
\textbf{
Comparison of the minimum-$AICc$ model and the spectral data. 
}
}
\end{figure*}

\setlength{\tabcolsep}{0.50em}
\begin{table}
\centering
\caption[
Comparison of \da\ estimates by statistical criteria
]{
\label{tab:j11_dal}
\textbf{Comparison of final \da\ estimates by statistical criteria.}
Each statistical criteria is used to analyse the set of candidate models 
and estimate a single \da\ value. 
Row 1: minimum $\chi^2$ model. 
Row 2: minimum Akaike Information Criteria corrected for small sample size ($AICc$) model. 
Row 3: minimum Bayesian Information Criteria ($BIC$) model. 
These three models correspond to the minimum in the respective statistic when the entire set of thermal and turbulent models are compared. The three models, in this case, are thermally broadened and highlighted in Table \ref{tab:j11_sta}; 
$\chi^2_{min}$ generation 83 ($k = 198$), $AICc_{min}$ 
generation 79 ($k = 176$), and $BIC_{min}$ generation 55 ($k = 144$), respectively. 
Row 4, 5 and 6: Bayesian Model Averaging (BMA), applied to the combined set of all thermal and turbulent models, using the $\chi^2$, $AICc$ and $BIC$ statistics, respectively. 
See Section \ref{postfit} for a discussion of BMA}. 
\begin{tabular}{lr}
\hline
Source                          & $\Delta\alpha/\alpha$ \\
                                & $[10^{-6}]$           \\
\hline
$\chi^2_{min}$              & 2.8 $\pm$ 2.9 \\
$AICc_{min}$                & 3.3 $\pm$ 2.9 \\
$BIC_{min}$                 & 1.4 $\pm$ 2.7 \\
\hline
$\chi^2$, BMA & 3.2 $\pm$ 3.0 \\
AICc, BMA     & 3.3 $\pm$ 2.9 \\
BIC, BMA      & 1.4 $\pm$ 2.7 \\
\hline
\end{tabular}
\vspace{1mm}
\end{table}
\setlength{\tabcolsep}{\oldtabcolsep}

\section{Comparison with Interactive Model Fitting}

\subsection{Emulating manual fitting}\label{emulate}

Table \ref{tab:j11_sta} gives the statistical details for a large number
of fits. For each generation, the minimum values of each of the 3
statistics used (i.e. $\chi^2$, AICc and BIC) is given, for fully
thermal or fully turbulent models.  The corresponding $\alpha$
measurements and number of free parameters $k$ is also shown. The total
number of data points in our analysis is $n = 3258$.

\setlength{\tabcolsep}{0.50em}
\begin{table}
\centering
\caption[
\da\ results from emulating human interactive fitting. 
]{
\label{tab:int_comp}
\textbf{
\da\ results from emulating human interactive fitting. 
}
Here we treat the thermal and turbulent cases separately, 
as has been done in many previous analyses of \da.  
To emulate a human interactive fit, we use three statistical 
measures and select a "preferred model" from Table \ref{tab:j11_sta}
in each case.
See Section \ref{emulate} for more information. 
Column 1: statistical criteria used in the fitting process. 
Column 3: model generation, from Table \ref{tab:j11_sta}. 
Column 4: number of free parameters in the model. }
\begin{tabular}{llccr}
\hline
Statistic      &             & Gen. & k & $\Delta\alpha/\alpha$ \\
&                            &      &   & $[10^{-6}]$           \\
\hline		        		       
$\chi^2_{min}$ & (thermal)   & 32   & 132 &         -5.8 $\pm$ 2.6 \\
$AICc_{min}$   & (thermal)   & 38   & 148 &         -3.3 $\pm$ 2.7 \\
$BIC_{min}$    & (thermal)   & 33   & 124 &         -5.8 $\pm$ 2.5 \\
\hline		        		       
$\chi^2_{min}$ & (turbulent) & 55   & 137 &         1.1 $\pm$ 1.8 \\
$AICc_{min}$   & (turbulent) & 52   & 140 &        14.1 $\pm$ 2.2 \\
$BIC_{min}$    & (turbulent) & 47   & 121 &        13.7 $\pm$ 1.9 \\
\hline
\end{tabular}
\vspace{1mm}
\end{table}
\setlength{\tabcolsep}{\oldtabcolsep}

We can use the data presented in Table \ref{tab:j11_sta} to emulate the
results that would be obtained using previous methodology i.e. a human
interactively fitting the data in real-time.  The emulation here is not
to be considered as entirely accurate, since the way in which a human
constructs models is not the same as the way in which {\sc GVPFIT}
constructs models. That aside, we can at least make some comparisons,
bearing in mind that the usual approach used in earlier (human) analyses
has been to increase the number of components in an absorption complex
until the overall fit reaches a normalised $\chi^2$ of about unity.

For the thermal model, we can see in Table \ref{tab:j11_sta} that
$\chi^2 / (n-k) \approx 1$ in generation 32.  Similarly, for a turbulent
model, generation 55 would be the most likely solution chosen in an
interactive fit.

If instead, an interactive modeller used AICc as the preferred
statistic, the likely approach would be to keep adding components to the 
model until both an adequate $\chi^2$ is reached (say $\chi^2_{\nu} < 1.1$) and 
AICc increased.  For the thermal models, Table
\ref{tab:j11_sta} shows that between generation 38 and generation 39, the
AICc statistic increases (the minimum-$AICc$ models from both generations 
have $\chi^2_{\nu} \approx 0.960$). 
Thus generation 38 would most likely be chosen for the thermal model.  
Similarly, generation 52 ($\chi^2_{\nu} = 1.038$) 
would most likely be chosen for the best turbulent model.

Using the BIC statistic (although to our knowledge BIC has not been used
in this context before), Table \ref{tab:j11_sta} shows that the chosen
thermal model would be generation 33 ($\chi^2_{\nu} = 0.992$) and the turbulent model would be generation 47 ($\chi^2_{\nu} = 1.068$).

Accepting the shortcomings of this comparison, a human (using AICc)
would therefore probably select the solutions in generations 52
(turbulent) or 38 (thermal).  In contrast, as can be seen in Table 
\ref{tab:j11_dal}, {\sc GVPFIT} would select generation 79. Table
\ref{tab:int_comp} summarises the values above, with the corresponding
values of \da. Emulating the ``usual'' human interactive fitting
procedure clearly results in selecting models with fewer free parameters 
($k$ = 132-137, 140-148 and 121-124 when using $\chi^2$, AICc and BIC respectively) 
compared to models selected using the same statistics after applying {\sc GVPFIT}. 
Table \ref{tab:j11_sta} shows that the three non-BMA values 
in Table \ref{tab:j11_dal} have global minima at much higher complexity,
generations 83, 79 and 55 for $\chi^2_{min}$ ($k=198$), AICc$_{min}$
($k=176$) and BIC$_{min}$ ($k=144$) respectively.

Of particular importance in this ``human emulation'' attempt, are the
\da\ results revealed in Table 5.  Using established statistical selection
criteria, AICc and BIC, we happen to end up in a local minimum in
parameter space that produces an anomalous result for \da\ yet provides
a statistically good fir to the data.  We discuss this local minimum
more detail in Section \ref{sect:results} and note here that the GVPFIT
procedure provides a ``global picture'', enabling the anomalous region
to be clearly seen.  We thus argue that GVPFIT out-performs any human
interactive modeller in this respect.

\subsection{Comparison with previous results for the $z_{abs}=
1.839$ absorption system towards J110325-264515}

Table \ref{tab:dis_com} compares our AICc BMA result
with the results from 5 previous studies of the same absorption system. 
Note that none of the results given in Table \ref{tab:dis_com} can be
considered independent since the data used have at least some
commonality in all analyses listed.

\setlength{\tabcolsep}{0.25em}
\begin{table*}
\begin{center}
\caption[
Comparison to Previously Published Models.
]{
\label{tab:dis_com}
\textbf{
Comparison to Previously Published Models.
}
Column 2: This work's $AICc$ BMA estimate, 
Column 3: This work's minimum-$AICc$ model, 
Column 4: \citet{levshakov2005}, 
Column 5: \citet{levshakov2007}, 
Column 6: \citet{molaro2008}, 
Column 7: \citet{levshakov2009}, 
Column 8: \citet{king2012}. 
Row 1: \da\ in units of $10^{-6}$, 
Row 2: normalised $\chi^2$, 
Row 3: number of data points, 
Row 5: number of velocity components between -50 km/s and 40 km/s (the common velocity range in the models). 
}
\begin{tabular}{c|cc|c|c|c|c|c}
\hline
&                                       This Work         & This Work         & Levshakov     & Levshakov     & Molaro        & Levshakov     & King             \\
&                                       $AICc$            & Minimum-$AICc$     &   2005        & 2007          &   2008        &   2009        &  2012            \\
&                                       BMA & Model             &               &               &               &               &                  \\
\hline
$\Delta\alpha/\alpha$                 & 3.3 $\pm$ 2.9    &                   & 2.4 $\pm$ 3.8 & 5.4 $\pm$ 2.5 & 5.7 $\pm$ 2.7 & 4.0 $\pm$ 2.8 &    6.1 $\pm$ 4.0 \\
$\chi^2_{\nu}$                        &                   & 0.896             &         1.097 &         0.901 &          0.90 &          0.87 &            0.935 \\
$n$                                   & 3258              & 3258              &           973 &           306 &           --- &           352 &              388 \\
$n_{\mbox{comp}}$                     &                   & 14                &             6 &            11 &            11 &            15 &                9 \\
\hline
\end{tabular}
\end{center}
\footnotesize{
--- information unavailable from publication. 
}
\end{table*}
\setlength{\tabcolsep}{\oldtabcolsep}

Our minimum-AICc model is also included in Table \ref{tab:dis_com}, for
completeness and so that we can compare $\chi^2$ values, number of
velocity components, and number of data points used in each analysis.

When comparing a spectral range common to all analyses including this
new one (see Table 6), our minimum-AICc model has 14 components.
Previous analyses have between 6 and 15 components in that same range.
We have used a somewhat larger dataset in the present analysis,
incorporating observations over many years (Table \ref{tab:dat_spe_obs})
so our overall signal-to-noise ratio is somewhat higher.  We have also
included MgII whereas none of the previous analyses did.  It is
therefore unsurprising that the number of components in our AICC model
is relatively high. We note that the \citet{king2012} model uses only 9
components yet also has used AICc to select the most appropriate model. 

\subsection{Apparent over-fitting}\label{overfit}

Careful inspection of Figure \ref{fig:j11_pre_one} may suggest too many
parameters have been used to model the data. For example, looking at the
top left panel for MgII 2796 (where the signal-to-noise is high), there
are 3 cases of very close blends.  An experienced ``manual-fitter'' may
intuitively regard this as over-fitting.  However, in fact, in
all 3 cases, the close pairs of blends have significantly different
$b$-parameters, so contribute quite differently to the profile shape. 
The model presented in Figure \ref{fig:j11_pre_one} is not intended to
represent a definitive model for this absorption system. Nevertheless,
it is the minimum AICc model, the procedure is objective and
reproducible, the normalised residuals obtained for the fit are good,
and BMA in any case downplays the importance of any
specific model.

\section{Testing GVPFIT using synthetic spectra} \label{synthetic}

\begin{figure}
\begin{center}
\begin{tabular}{cc}
\includegraphics[trim=0.00cm 0.00cm 0.00cm 0.00cm, clip=true, width=8.00cm]{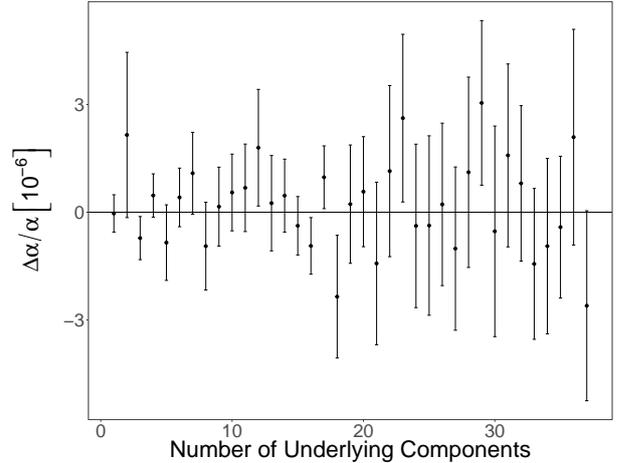} &
\end{tabular}
\end{center}
\caption[
\da\ vs number of underlying components for synthetic data. 
]{
\label{fig:da_vs_ncomp}\textbf{
\da\ vs number of velocity components for the set of synthetic spectra described in Section \ref{synthetic}. 
We created 37 synthetic spectra whose parameters were
drawn from the minimum-AICc models for the $z_{abs} = 1.839$
system towards J110325-264515 for models that contained 
1, 2, 3,...,37 velocity components.
\da\ is set to zero in these simulations.
The results illustrated here show the BMA 
results of applying GVPFIT to all 37 synthetic spectra.
The 1$\sigma$ statistical uncertainties are derived
in the same way as for the real data.
As can be seen {\sc GVPFIT} recovers the input \da\ value 
within the expected statistical uncertainty. 
}
}
\end{figure}

In Section \ref{simplesim} we described some relatively simple
simulations that served to illustrate how {\sc GVPFIT} works and also
as a basic test.  In this Section we describe simulations of
spectra with a broad range of velocity structures, ranging from
extremely simple up to highly complex.  The simulations described
in this Section are based on the characteristics of the real absorption
system analysed in Section \ref{applic}.

From the whole database created when running {\sc GVPFIT} on the real 
spectra, we selected the minimum-AICc model with 1 through 37 
velocity components.  The best-fit parameters from these 37 
models were then used to create 37 model spectra.  We go up to 37 
because a 37-component fit corresponds to the minimum-AICc 
model for the real data.

Synthetic models were created directly from the appropriate 
{\sc VPFIT} output, hence were convolved with the same instrumental
profile as used in fitting the real spectra.  Gaussian noise was
added to each model using the Box-Muller transform approach
\citep{box1958}.  We used the actual error arrays from the real 
spectra to assign Gaussian standard deviation at each pixel
in each synthetic spectrum created, thus emulating the 
characteristics of the real data closely.  The real spectra have
multiple observations from different epochs.  Synthetic
spectra corresponding to each of the real spectra were created.
The synthetic spectra were created and fitted using turbulent 
b-parameters only.  The calculations were not repeated for thermal 
models.  This is a minor difference between the treatment of the 
simulations compared to the real data but this has no relevance in 
terms the usefulness of these simulations as a stringent test of GVPFIT.

We then treated the synthetic spectra described above as if they 
were real data, applied GVPFIT to each of these 37 models and 
AICc BMA values for \da, as described in Section \ref{sect:results}.
The 37 results are illustrated in Figure \ref{fig:da_vs_ncomp}.  
Error bars grow as the absorption system complexity increases,
as would be expected.  The weighted mean over the 37 values of
\da\ is $0.04 \pm 0.20 \times 10^{-6}$ and no bias for the 
\da\ estimate is seen.

The simulation results provide additional informative
performance details.  The set of input models comprises
$\sum_{i=1}^{37} n_i = 703$ components, where $n_i$ is the
number of velocity components for the $i^{th}$ model.
In other words, if {\sc GVPFIT} was a perfect procedure, it
should have detected exactly 703 components in total for
the entire set of simulations.

As a test of this, we compared the 37 minimum-AICc models from
the simulations with the input models.  We found that
653 of the 703 real velocity components were
successfully identified.  Also, 4 additional velocity
components were introduced in the {\sc GVPFIT} fitting process
that were not present in the original models.  GVPFIT therefore
successfully identified 92.9\% of the real components present
and only 0.6\% of spurious velocity components were introduced. 

We summarise the outcome of these simulations with two conclusions: 
(1) Using realistic simulated spectra covering a broad range in 
complexity, {\sc GVPFIT} returns results that are accurate and 
unbiased; (2) A second interesting implication is that selecting 
the minimum-AICc does a good job in selecting the 
``right'' model, with no evidence for over- or under-fitting.

\section{Conclusions}\label{conclusions}

\begin{itemize}

\item[(1)] We have introduced Genetic Voigt Profile FIT (GVPFIT), 
a new ``artificial intelligence''
genetic algorithm for the analysis of absorption spectra 
(in quasars, the interstellar medium, or stellar astrophysics).

\item[(2)] Using this genetic algorithm we present a new method for the analysis of absorption spectra. The method is memetic, 
unifying a genetic algorithm ({\sc GVPFIT}),
non-linear least-squares optimisation ({\sc VPFIT}), and Bayesian Model
Averaging (BMA). In contrast to previous methodologies, BMA derives
results using a large set of models, providing a more robust final
answer compared to picking a single preferred model since it avoids the
systematic uncertainties associated with model choice. This new method
eliminates most of the human decision-making processes and hence
minimises or removes any possible bias.

\item[(3)] Robust and automated \da\ estimation permits a more complete
analysis (i.e. a large set of models rather than just a single model).
Since GVPFIT provides a ``global view'' over many generations
(i.e. complexities) and we can identify local minima.  
This may apply to other ``interesting'' parameters, not only \da.
GVPFIT out-performs human interactive fitting in this respect.

\item[(4)] We have applied GVPFIT to the measurement of the fine
structure constant in a high-redshift quasar absorption system, at
$z_{abs}$ = 1.839, towards the quasar J110325-264515. The result is
consistent both with zero and with the tentative spatial variation of
$\alpha$ reported in \citet{webb2011} and \citet{king2012}.

\item[(5)] We find that sets of statistically acceptable models
can be found which nonetheless are anomalous in terms of \da. 
A human interactive fitting process could not easily identify these
regions, whereas GVPFIT can.

\item[(6)] Comparing GVPFIT with emulations of previous interactive
fitting methods, which require far more human decision-making, we see
that GVPFIT provides greater consistency and stability (Tables
\ref{tab:j11_dal}, \ref{tab:int_comp}, \ref{tab:dis_com} and associated
discussion).

\item[(7)] Since the human-work in modelling absorption spectra is now
removed by GVPFIT, future projects that were previously unrealistic are
now feasible. One example of this is modelling both thermal and
turbulent models for each absorption system independently, allowing a
more reliable comparison between models and data.  This analysis has
been done for the first time here (applied to the $z_{abs}=1.839$
absorption system towards the quasar J110325-264515).

\end{itemize}

\section*{Acknowledgements}
We would like to thank Bob Carswell.  Much of the content of this paper
is built upon our collaborative work over many years. We thank
Signe Riemer-S\o rensen, Michael Wilczynska and Evgeny Zavarygin for 
detailed comments, Vincent Dumont for discussions
about atomic data, and Michael Murphy for communications regarding initial
data reduction and for comments on an early draft.  We also are grateful
to Adrian Raftery and Jennifer Hoeting for input concerning BMA, and
to referee, Ewan Cameron, for his detailed criticisms, which we think
resulted in a much improved paper.  MBB is grateful 
for postgraduate scholarship support from UNSW.

\bibliographystyle{mn2e.bst}
\bibliography{jkw,mbb}

\appendix

\section{Atomic data}\label{app:atomic_data}

Here we provide details for the MgII and FeII atomic data used in our
analysis. 
The transitions used in this work are identified in Table \ref{tab:app_ato_ide}. The characteristics and Voigt profile parameters for these transitions are given in Table \ref{tab:app_ato_dat}. 

\subsection{MgII 2796/2803\AA\, hyperfine transitions}

The MgII hyperfine structure results from the interaction of the nuclear
electric and magnetic fields with the electron angular momentum and
spin.  It impacts those isotopes with a non-zero nuclear spin which are
thus themselves split further.  Hyperfine splitting occurs to both the
lower and upper levels of a transition, although the effect on the lower
level dominates and the upper level splitting is often not considered.

The $^{25}$MgII isotope is split into the twelve hyperfine transitions 
(illustrated in Figure \ref{mgii_hfel}),
whose wavenumbers can be calculated from the magnetic dipole hyperfine
constants, $A_i$ and $A_j$, and the electric quadrupole hyperfine constant, $B$
\be
\omega_{25}^{i,j} = \omega_{24} + \Delta \omega_{24,25} + \frac{1}{2} 
\left[ A^i K^i + A^j K^j + B C^j \right]
\ee
where $\omega_{25}^{i,j}$ is the wavenumber for the hyperfine transition
from the $i^{th}$ lower level to the to $j^{th}$ upper level,
$\Delta\omega_{24,25}$ is the wavenumber shift between $^{24}$MgII and
the observed centroid for $^{25}$MgII, and $K$ is given by
\be
K = F(F + 1) - J(J + 1) - I(I + 1)
\ee
and $C$ is
\be
C = \frac{3K(K+1)-4(I+1)J(J+1)}{2I(2I-1)2J(2J-1)}
\ee
where $F = |I + J|$, $I$ is the nuclear angular momentum and $J$ is
the electron angular momentum. The uncertainty on the $^{25}$MgII
hyperfine wavenumbers are given by the centroid uncertainties i.e.
\be
\sigma \omega^{i,j}_{25} \approx \Delta \omega_{24}
\ee
The $^{25}$MgII hyperfine structures abundances are taken to be equally 
populated, under the assumption of local thermodynamic equilibrium at
the cosmic microwave background temperature (i.e. $\frac{N_{i}}{N_{j}} =
\frac{g_{i}}{g_{j}} e^{-\Delta E_{i,j}/kT} \approx 1)$. 

The details of the MgII hyperfine structure are given in Table \ref{tab:app_ato_mgii}. 
The MgII vacuum wavelengths, $\lambda_{vac}$ and wavenumbers, $\omega =
f/c$, where $f$ is the frequency and $c$ is the speed of light in a
vacuum, and their associated uncertainties, are given in Table
\ref{tab:app_ato_wav}. 

\begin{tikzpicture}[
  level/.style   = { ultra thick, blue },
  connect/.style = { dashed, red },
  label/.style   = { text width=2cm },
  transition/.style={black,->,>=stealth',shorten >=1pt}
 ]
\begin{scope}[scale=0.75]
  \draw[level] (0,0) -- node[color=black,above] {3s$_{1/2}$} (2,0);
  \draw[level] (0,5) -- node[color=black,above] {3p} (2,5);
  \draw[connect] (2,0) -- (3,0);
  \draw[connect] (2,5) -- (3,4);
  \draw[connect] (2,5) -- (3,6);
  \draw[level] (3,0) -- node[color=black,above] {3s$_{1/2}$} (5,0);
  \draw[level] (3,4) -- node[color=black,above] {3p$_{1/2}$} (5,4);
  \draw[level] (3,6) -- node[color=black,above] {3p$_{3/2}$} (5,6);
  \draw[connect] (5,0) -- (6,-0.3);
  \draw[connect] (5,0) -- (6,0.3);
  \draw[connect] (5,4) -- (6,3.7);
  \draw[connect] (5,4) -- (6,4.3);
  \draw[connect] (5,6) -- (6,5.4);
  \draw[connect] (5,6) -- (6,5.8);
  \draw[connect] (5,6) -- (6,6.2);
  \draw[connect] (5,6) -- (6,6.6);
  \draw[level] (6,-0.3) -- (9.25,-0.3) node[color=black,right] {F=3};
  \draw[level]  (6,0.3) -- (9.25,0.3)  node[color=black,right] {F=2};
  \draw[level]  (6,3.7) -- (9.25,3.7)  node[color=black,right] {F=3};
  \draw[level]  (6,4.3) -- (9.25,4.3)  node[color=black,right] {F=2};
  \draw[level]  (6,5.4) -- (9.25,5.4)  node[color=black,right] {F=4};
  \draw[level]  (6,5.8) -- (9.25,5.8)  node[color=black,right] {F=3};  
  \draw[level]  (6,6.2) -- (9.25,6.2)  node[color=black,right] {F=2};
  \draw[level]  (6,6.6) -- (9.25,6.6)  node[color=black,right] {F=1};
  \draw[transition] (6.25,-0.3) -- (6.25,6.6);
  \draw[transition] (6.5,-0.3) -- (6.5,6.2);
  \draw[transition] (6.75,-0.3) -- (6.75,5.8);
  \draw[transition] (7.0,-0.3) -- (7.0,5.4);
  \draw[transition] (7.25,0.3) -- (7.25,6.6);
  \draw[transition] (7.5,0.3) -- (7.5,6.2);
  \draw[transition] (7.75,0.3) -- (7.75,5.8);
  \draw[transition] (8.0,0.3) -- (8.0,5.4);
  \draw[transition] (8.25,-0.3) -- (8.25,4.3);
  \draw[transition] (8.5,-0.3) -- (8.5,3.7);
  \draw[transition] (8.75,0.3) -- (8.75,4.3);
  \draw[transition] (9.0,0.3) -- (9.0,3.7);
  \draw[color=black] (4,-1.5) node{FINE};
  \draw[color=black] (8,-1.5) node{HYPERFINE};
  \draw[color=black] (4,-2) node{STRUCTURE};
  \draw[color=black] (8,-2) node{(OBSERVED)};
\end{scope}
\end{tikzpicture}
\captionof{figure}{\label{mgii_hfel} Energy level diagram for observed $^{25}$MgII
hyperfine transitions. Ordering of transitions, from left to right, is
in ascending wavelength, as in Table \ref{tab:app_ato_mgii}.}

\subsection{FeII}

We have used slightly different wavelengths to those in \citet{king2012}
since newer FeII laboratory (centroid) wavelengths are available from \citet{nave2012}.

The frequency shift, $\delta \omega^{A,A'}$, between the transitions of two isotopes with mass numbers $A$ and $A'$ can be approximated, by neglecting the field shift, as shown in \citep{porsev2009}
\be
\label{feii_domega}
\delta \omega^{A,A'} = \omega^{A} - \omega^{A'} \approx k_{\mbox{MS}} \left\{ \frac{1}{A} - \frac{1}{A'} \right\}
\ee
where $k_{\mbox{MS}}$ is the total mass shift given in Table \ref{tab:app_ato_feii2} and the mass numbers are given in \ref{tab:app_ato_feii1}. 

In addition we can relate the frequencies of the isotopes of FeII to the observed centroid frequency, $\omega$, assuming terrestrial isotopic abundance, 
\be
\label{feii_omega}
\omega = \omega^{54} P^{54} + \omega^{56} P^{56} + \omega^{57} P^{57} + \omega^{58} P^{58}
\ee
where $P^{A}$ is the terrestrial isotopic abundance of the isotope with mass number $A$, shown in Table \ref{tab:app_ato_wav} as Relative Strength. 

Combining Equations \ref{feii_domega} and \ref{feii_omega} we solved for the frequency of the FeII$^{54}$ transitions, and from that frequency the remaining isotopic structure as shown in Table \ref{tab:app_ato_wav}. 

\setlength{\tabcolsep}{0.25em}
\begin{table}
\centering
\caption[
Atomic Data -- Transition Classification.
]{
\label{tab:app_ato_ide}
\textbf{
Atomic Data -- Transition Classification.
}
Column 1 and 2: atomic species$^{[2]}$ and transition label$^{[2]}$, 
Column 3 and 4: lower and upper atomic state electronic configurations.  
The superscripts, e.g. $^{[3]}$, correspond to the list of references given in Table \ref{tab:app_ato_ref}. 
}
\begin{tabular}{llll}
\hline
Species & Tran. & Lower State                          & Upper State                                \\
        &       &                                      &                                            \\
\hline          
Fe II   & 1608  &    $3d^6(^5D)4s a^6D_{9/2}$$^{[3]}$  & $3d^5(^6S)4s4p(^3P) y^6P^o_{7/2}$$^{[3]}$  \\
Fe II   & 2344  &    $3d^6(^5D)4s a^6D_{9/2}$$^{[3]}$  &        $3d^6(^5D)4p z^6P^o_{7/2}$$^{[3]}$  \\
Fe II   & 2374  &    $3d^6(^5D)4s a^6D_{9/2}$$^{[3]}$  &        $3d^6(^5D)4p z^6F^o_{9/2}$$^{[3]}$  \\
Fe II   & 2382  &    $3d^6(^5D)4s a^6D_{9/2}$$^{[3]}$  &       $3d^6(^5D)4p z^6F^o_{11/2}$$^{[3]}$  \\
Fe II   & 2586  &    $3d^6(^5D)4s a^6D_{9/2}$$^{[3]}$  &        $3d^6(^5D)4p z^6D^o_{7/2}$$^{[3]}$  \\
Fe II   & 2600  &    $3d^6(^5D)4s a^6D_{9/2}$$^{[3]}$  &        $3d^6(^5D)4p z^6D^o_{9/2}$$^{[3]}$  \\
\hline
Mg II   & 2796  &               $3s^2S_{1/2}$$^{[16]}$ &                    $3p ^2P_{3/2}$$^{[16]}$ \\
Mg II   & 2803  &               $3s^2S_{1/2}$$^{[16]}$ &                    $3p ^2P_{1/2}$$^{[16]}$ \\
\hline
\end{tabular}
\end{table}
\setlength{\tabcolsep}{\oldtabcolsep}

\setlength{\tabcolsep}{0.25em}
\begin{table*}
\centering
\caption[
Atomic Data -- Voigt Profile Parameters.
]{
\label{tab:app_ato_dat}
\textbf{Atomic Data -- Voigt Profile Parameters}. 
The rest wavelengths are shown in Table \ref{tab:app_ato_wav}. 
Column 1 and 2: atomic species and transition$^{[2]}$. 
Column 3: atomic mass. 
Column 4 and 5: lower and upper ionisation potentials. 
Column 6: oscillator strength of the centroid. 
Column 7: energy level lifetime. 
Column 8: centroid transition dampening constant. 
Column 9: $q$-coefficient (uncertainties as in \citet{berengut2011}). 
The superscripts, e.g. $^{[3]}$, correspond to the list of references given in Table \ref{tab:app_ato_ref}. 
}
\begin{tabular}{lllllllll}
\hline
Species & Tran. & Atomic Mass             & Lower IP             & Upper IP              & \hspace{0.1mm} $f$   & \hspace{0.1mm} $\tau$ & \hspace{0.1mm} $\Gamma$           & \hspace{0.1mm} $q$               \\
        &       & \hspace{0.1mm} [amu]    & \hspace{0.1mm} [eV]  & \hspace{0.1mm} [eV]   &                      & \hspace{0.1mm} [ns]   & \hspace{0.1mm} [$10^{8}$s$^{-1}$] & \hspace{0.1mm} [cm$^{-1}$]       \\
\hline
Fe II   & 1608  & 55.845(19)$^{[4]}$      & 7.9024(1)$^{[7]}$    & 16.19920(5)$^{[3]}$   & 0.058(5)$^{[8]}$     & 3.65(20)$^{[10]}$     &  2.74(15)$^{[26]}$                &  -1165(300)$^{[12,13]}$          \\
Fe II   & 2344  &                         &                      &                       & 0.114(2)$^{[9]}$     & 3.73(5)$^{[11]}$      &  2.68(4)$^{[26]}$                 & \ 1375(300)$^{[12,13]}$          \\
Fe II   & 2374  &                         &                      &                       & 0.0313(14)$^{[9]}$   & 3.24(6)$^{[11]}$      &  3.09(6)$^{[26]}$                 & \ 1625(100)$^{[12,13]}$          \\
Fe II   & 2382  &                         &                      &                       & 0.320(4)$^{[9]}$     & 3.19(4)$^{[11]}$      &  3.13(4)$^{[26]}$                 & \ 1505(100)$^{[12,13]}$          \\
Fe II   & 2586  &                         &                      &                       & 0.0691(25)$^{[9]}$   & 3.68(7)$^{[11]}$      &  2.72(5)$^{[26]}$                 & \ 1515(100)$^{[12,13]}$          \\
Fe II   & 2600  &                         &                      &                       & 0.239(4)$^{[9]}$     & 3.70(6)$^{[11]}$      &  2.70(4)$^{[26]}$                 & \ 1370(100)$^{[12,13]}$          \\
Mg II   & 2796  & 24.3051(53)$^{[4]}$     & 7.646238(5)$^{[19]}$ & 15.035277(8)$^{[19]}$ & 0.611(2)$^{[20]}$    & 3.854(30)$^{[21]}$    &  2.595(20)$^{[26]}$               & \hspace{1.5mm} 212(1)$^{[22]}$   \\
Mg II   & 2803  &                         &                      &                       & 0.306(5)$^{[20]}$    & 3.810(40)$^{[21]}$    &  2.625(28)$^{[26]}$               & \hspace{1.625mm} 121(1)$^{[22]}$ \\
\hline
\end{tabular}
\end{table*}
\setlength{\tabcolsep}{\oldtabcolsep}

\setlength{\tabcolsep}{0.25em}
\begin{table*}
\centering
\caption[
Atomic Data -- MgII 25 Hyperfine Structure. 
]{
\label{tab:app_ato_mgii}
\textbf{
Atomic Data -- MgII 25 Hyperfine Structure. 
}
For this atom I = 5/2. 
Columns 2 and 5: atomic configuration. 
Columns 3 and 6: electron angular momentum.  
Columns 4 and 7: magnitude of the F quantum number, which is the vector sum of the total angular momentum for the nuclear state and electron state. 
Column 8: isotopic shift from the $^{24}$MgII centroid to the $^{25}$MgII centroid energy levels, 
Column 9 and 10: magnetic dipole hyperfine constant, 
for the $i^{th}$ and $j^{th}$ state, respectively. 
Column 11: electric dipole hyperfine constant for the $j^{th}$ state. 
Column 12: $i^{th}$ energy level. 
Column 13: $j^{th}$ energy level. 
The superscripts, e.g. $^{[3]}$, correspond to the list of references given in Table \ref{tab:app_ato_ref}. 
}
\begin{tabular}{ccccccccccccc}
\hline
\# & Lower          & $\mathbf{J}^{i}$   & $F^{i}$ & Upper            & $\mathbf{J}^{j}$   & $F^{j}$ & $\Delta \omega_{24,25}$\ $^{[16]}$ & $A^{i}$\ $^{[25]}$  & $A^{j}$\ $^{[28]}$ & $B^{j}$\ $^{[28]}$ & $E^{i}$           & $E^{j}$           \\
&  State\ $^{[16]}$ &                    &         & State\ $^{[16]}$ &                    &         & $(10^{-2})$                        & $(10^{-2})$         & $(10^{-4})$        & $(10^{-4})$        & $(10^{-2})$       & [cm$^{-1}$]       \\
&                   &                    &         &                  &                    &         & [cm$^{-1}$]                        & [cm$^{-1}$]           & [cm$^{-1}$]      & [cm$^{-1}$]        & [cm$^{-1}$]       &                   \\
                                                                                                       
\hline                                                                                                 
1  & $3s^2S_{1/2}$  & \hspace{0.1mm} 1/2 & 3       & $3p ^2P_{3/2}$ &               -3/2 & 1       & 5.407(63)                        & -1.98889051(18)   & -0.6301          & -0.7642          &                -2.48611314(23)   & 35760.8940400(53) \\
2  & $3s^2S_{1/2}$  & \hspace{0.1mm} 1/2 & 3       & $3p ^2P_{3/2}$ &               -1/2 & 2       & 5.407(63)                        & -1.98889051(18)   & -0.6301          & -0.7642          &                -2.48611314(23)   & 35760.8925701(53) \\
3  & $3s^2S_{1/2}$  & \hspace{0.1mm} 1/2 & 3       & $3p ^2P_{3/2}$ & \hspace{0.1mm} 1/2 & 3       & 5.407(63)                        & -1.98889051(18)   & -0.6301          & -0.7642          &                -2.48611314(23)   & 35760.8906798(53) \\
4  & $3s^2S_{1/2}$  & \hspace{0.1mm} 1/2 & 3       & $3p ^2P_{3/2}$ & \hspace{0.1mm} 3/2 & 4       & 5.407(63)                        & -1.98889051(18)   & -0.6301          & -0.7642          &                -2.48611314(23)   & 35760.8889135(53) \\
5  & $3s^2S_{1/2}$  &               -1/2 & 2       & $3p ^2P_{3/2}$ &               -3/2 & 1       & 5.407(63)                        & -1.98889051(18)   & -0.6301          & -0.7642          & \hspace{0.00mm} 3.48055840(32)   & 35760.8940400(53) \\
6  & $3s^2S_{1/2}$  &               -1/2 & 2       & $3p ^2P_{3/2}$ &               -1/2 & 2       & 5.407(63)                        & -1.98889051(18)   & -0.6301          & -0.7642          & \hspace{0.00mm} 3.48055840(32)   & 35760.8925701(53) \\
7  & $3s^2S_{1/2}$  &               -1/2 & 2       & $3p ^2P_{3/2}$ & \hspace{0.1mm} 1/2 & 3       & 5.407(63)                        & -1.98889051(18)   & -0.6301          & -0.7642          & \hspace{0.00mm} 3.48055840(32)   & 35760.8906798(53) \\
8  & $3s^2S_{1/2}$  &               -1/2 & 2       & $3p ^2P_{3/2}$ & \hspace{0.1mm} 3/2 & 4       & 5.407(63)                        & -1.98889051(18)   & -0.6301          & -0.7642          & \hspace{0.00mm} 3.48055840(32)   & 35760.8889135(53) \\
\hline
9  & $3s^2S_{1/2}$  & \hspace{0.1mm} 1/2 & 3       & $3p ^2P_{1/2}$ &               -1/2 & 2       & 5.404(63)                        & -1.98889051(18)   & -3.3923          & N/A              &                -2.48611314(23)   & 35669.3476437(53) \\
10 & $3s^2S_{1/2}$  & \hspace{0.1mm} 1/2 & 3       & $3p ^2P_{1/2}$ & \hspace{0.1mm} 1/2 & 3       & 5.404(63)                        & -1.98889051(18)   & -3.3923          & N/A              &                -2.48611314(23)   & 35669.3374666(53) \\
11 & $3s^2S_{1/2}$  &               -1/2 & 2       & $3p ^2P_{1/2}$ &               -1/2 & 2       & 5.404(63)                        & -1.98889051(18)   & -3.3923          & N/A              & \hspace{0.00mm} 3.48055840(32)   & 35669.3476437(53) \\
12 & $3s^2S_{1/2}$  &               -1/2 & 2       & $3p ^2P_{1/2}$ & \hspace{0.1mm} 1/2 & 3       & 5.404(63)                        & -1.98889051(18)   & -3.3923          & N/A              & \hspace{0.00mm} 3.48055840(32)   & 35669.3374666(53) \\
\hline
\end{tabular}
\end{table*}
\setlength{\tabcolsep}{\oldtabcolsep}

\setlength{\tabcolsep}{0.25em}
\begin{table}
\centering
\caption[
Atomic Data -- FeII Isotopic Atomic Mass. 
]{
\label{tab:app_ato_feii1}
\textbf{
Atomic Data -- FeII Isotopic Atomic Mass. 
}
The superscript corresponds to the list of references given in Table \ref{tab:app_ato_ref}. 
}
\begin{tabular}{cc}
\hline
A  & Atomic Mass           \\
&    [amu]                 \\
\hline                                                                                                 
54 & 53.9396105(7)$^{[5]}$ \\
56 & 55.9349375(7)$^{[5]}$ \\
57 & 56.9353940(7)$^{[5]}$ \\
58 & 57.9332756(8)$^{[5]}$ \\
\hline
\end{tabular}
\end{table}
\setlength{\tabcolsep}{\oldtabcolsep}

\setlength{\tabcolsep}{0.25em}
\begin{table}
\centering
\caption[
Atomic Data -- FeII Isotopic Mass Shift. 
]{
\label{tab:app_ato_feii2}
\textbf{
Atomic Data -- FeII Isotopic Mass Shift. 
}
The superscript corresponds to the list of references given in Table \ref{tab:app_ato_ref}. 
}
\begin{tabular}{cc}
\hline
Trans & Mass Shift                      \\
&       [cm$^{-1}$]                     \\
\hline
1608  &                 -68.45$^{[15]}$ \\
2433  & \hspace{0.25mm}  39.03$^{[15]}$ \\
2374  & \hspace{0.25mm}  48.43$^{[15]}$ \\
2382  & \hspace{0.25mm}  48.13$^{[15]}$ \\
2600  & \hspace{0.25mm}  40.16$^{[15]}$ \\
\hline
\end{tabular}
\end{table}
\setlength{\tabcolsep}{\oldtabcolsep}

\setlength{\tabcolsep}{0.25em}
\begin{table*}
\caption[
Atomic Data -- Wavelength Table.
]{
\label{tab:app_ato_wav}
\textbf{Atomic Data -- Wavelength Table}.
Note that the FeII 2586 transition from Tables \ref{tab:app_ato_ide} and \ref{tab:app_ato_dat} is not used in the 
estimation of \da\ for this system, and so has been omitted from this table. 
Column 1 and 2: atomic species and transition labels$^{[2]}$. 
Column 3: mass number. 
Columns 4 and 5: magnitude of the F quantum number, which is the vector sum of the total angular momentum for the nuclear state and electron state. 
Column 6: relative (percentage) strength assuming terrestrial isotopic abundance 
and equal population in the $^{25}$MgII hyperfine energy levels. 
Column 7: centroid wavenumber. 
Column 8: vacuum wavelength. 
Column 9: effective oscillator strength ($f_{eff}$ = oscillator strength multiplied by relative strength. Needed by VPFIT to compute composite Voigt profiles). 
The superscripts, e.g. $^{[3]}$, correspond to the list of references given in Table \ref{tab:app_ato_ref}. 
}
\begin{center}
\begin{tabular}{cccccclll}
\hline
Species & Trans & A                 & $F^{i}$ & $F^{j}$ & Relative                            & \hspace{0.1mm} Wavenumber  & \hspace{4.5mm} $\lambda_{\mbox{vac}}$ & \hspace{6.5mm} $f_{eff}$ \\
&               &                   &         &         & Strength [\%]                       & \hspace{3.0mm} [cm$^{-1}$] & \hspace{6.0mm} [\AA]                  &                          \\   
\hline
FeII    & 1608  &  ---              &         &         &              ---                    & 62171.623(3)$^{[3]}$       & 1608.45085(7)$^{[3]}$                 &  \hspace{5mm} ---        \\
&               &  54$^{[5]}$       &         &         &  \hspace{0.9mm} 5.845(23)$^{[6]}$   & 62171.580(3)$^{[14]}$      & 1608.45197(7)$^{[27]}$                &  0.00339(29)$^{[18]}$    \\
&               &  56$^{[5]}$       &         &         &              91.754(24)$^{[6]}$     & 62171.625(3)$^{[14]}$      & 1608.45080(7)$^{[27]}$                &  0.053(5))$^{[18]}$      \\
&               &  57$^{[5]}$       &         &         &  \hspace{2.5mm} 2.1191(65)$^{[6]}$  & 62171.647(3)$^{[14]}$      & 1608.45024(7)$^{[27]}$                &  0.00123(11)$^{[18]}$    \\
&               &  58$^{[5]}$       &         &         &  \hspace{2.5mm} 0.2819(27)$^{[6]}$  & 62171.667(3)$^{[14]}$      & 1608.44971(7)$^{[27]}$                &  0.000164(14)$^{[18]}$   \\
&         2344  &  ---              &         &         &              ---                    & 42658.2440(19)$^{[3]}$     & 2344.21276(11)$^{[27]}$               &  \hspace{5mm} ---        \\
&               &  54$^{[5]}$       &         &         &  \hspace{0.9mm} 5.845(23)$^{[6]}$   & 42658.2686(19)$^{[14]}$    & 2344.21141(11)$^{[27]}$               &  0.00666(12)$^{[18]}$    \\
&               &  56$^{[5]}$       &         &         &              91.754(24)$^{[6]}$     & 42658.2428(19)$^{[14]}$    & 2344.21283(11)$^{[27]}$               &  0.1046(18)$^{[18]}$     \\
&               &  57$^{[5]}$       &         &         &  \hspace{2.5mm} 2.1191(65)$^{[6]}$  & 42658.2306(19)$^{[14]}$    & 2344.21350(11)$^{[27]}$               &  0.00242(4)$^{[18]}$     \\
&               &  58$^{[5]}$       &         &         &  \hspace{2.5mm} 0.2819(27)$^{[6]}$  & 42658.2188(19)$^{[14]}$    & 2344.21415(11)$^{[27]}$               &  0.000321(6)$^{[18]}$    \\
&         2374  &  ---              &         &         &              ---                    & 42114.8372(19)$^{[3]}$     & 2374.46009(11)$^{[27]}$               &  \hspace{5mm} ---        \\
&               &  54$^{[5]}$       &         &         &  \hspace{0.9mm} 5.845(23)$^{[6]}$   & 42114.8678(19)$^{[14]}$    & 2374.45836(11)$^{[27]}$               &  0.00183(8)$^{[18]}$     \\
&               &  56$^{[5]}$       &         &         &              91.754(24)$^{[6]}$     & 42114.8357(19)$^{[14]}$    & 2374.46017(11)$^{[27]}$               &  0.0287(13)$^{[18]}$     \\
&               &  57$^{[5]}$       &         &         &  \hspace{2.5mm} 2.1191(65)$^{[6]}$  & 42114.8205(19)$^{[14]}$    & 2374.46103(11)$^{[27]}$               &  0.00066(3)$^{[18]}$     \\
&               &  58$^{[5]}$       &         &         &  \hspace{2.5mm} 0.2819(27)$^{[6]}$  & 42114.8059(19)$^{[14]}$    & 2374.46185(11)$^{[27]}$               &  0.000088(4)$^{[18]}$    \\
&         2382  &  ---              &         &         &              ---                    & 41968.0679(20)$^{[3]}$     & 2382.76397(11)$^{[27]}$               &  \hspace{5mm} ---        \\
&               &  54$^{[5]}$       &         &         &  \hspace{0.9mm} 5.845(23)$^{[6]}$   & 41968.0983(20)$^{[14]}$    & 2382.76224(11)$^{[27]}$               &  0.01870(25)$^{[18]}$    \\
&               &  56$^{[5]}$       &         &         &              91.754(24)$^{[6]}$     & 41968.0664(20)$^{[14]}$    & 2382.76405(11)$^{[27]}$               &  0.294(4)$^{[18]}$       \\
&               &  57$^{[5]}$       &         &         &  \hspace{2.5mm} 2.1191(65)$^{[6]}$  & 41968.0513(20)$^{[14]}$    & 2382.76491(11)$^{[27]}$               &  0.00678(9)$^{[18]}$     \\
&               &  58$^{[5]}$       &         &         &  \hspace{2.5mm} 0.2819(27)$^{[6]}$  & 41968.0368(20)$^{[14]}$    & 2382.76574(11)$^{[27]}$               &  0.000902(14)$^{[18]}$   \\
&         2600  &  ---              &         &         &              ---                    & 38458.9928(17)$^{[3]}$     & 2600.17210(11)$^{[27]}$               &  \hspace{5mm} ---        \\
&               &  54$^{[5]}$       &         &         &  \hspace{0.9mm} 5.845(23)$^{[6]}$   & 38458.0181(17)$^{[14]}$    & 2600.17038(11)$^{[27]}$               &  0.01397(24)$^{[18]}$    \\
&               &  56$^{[5]}$       &         &         &              91.754(24)$^{[6]}$     & 38458.9916(17)$^{[14]}$    & 2600.17218(11)$^{[27]}$               &  0.219(4)$^{[18]}$       \\
&               &  57$^{[5]}$       &         &         &  \hspace{2.5mm} 2.1191(65)$^{[6]}$  & 38459.9790(17)$^{[14]}$    & 2600.17303(11)$^{[27]}$               &  0.00506(9)$^{[18]}$     \\
&               &  58$^{[5]}$       &         &         &  \hspace{2.5mm} 0.2819(27)$^{[6]}$  & 38459.9668(17)$^{[14]}$    & 2600.17385(11)$^{[27]}$               &  0.000674(13)$^{[18]}$   \\
\hline
MgII    & 2796  &  ---              &         &         &              ---                    & 35760.85414(6)$^{[16]}$    & 2796.353789(5)$^{[27]}$               &  \hspace{5mm} ---        \\
&               & 26$^{[5]}$        &         &         &  \hspace{-2mm} 11.01(4)$^{[17]}$    & 35760.9403866(53)$^{[16]}$ & 2796.34704565(42)$^{[27]}$            &  0.0673(6)$^{[18]}$      \\
&               & 25$^{[5]}$        &  3      & 1       &  \hspace{4.0mm} 1.2500(12)$^{[18]}$ & 35760.9189011(53)$^{[23]}$ & 2796.34872573(42)$^{[27]}$            &  0.00764(3)$^{[18]}$     \\
&               & 25$^{[5]}$        &  3      & 2       &  \hspace{4.0mm} 1.2500(12)$^{[18]}$ & 35760.9174313(53)$^{[23]}$ & 2796.34884066(42)$^{[27]}$            &  0.00764(3)$^{[18]}$     \\
&               & 25$^{[5]}$        &  3      & 3       &  \hspace{4.0mm} 1.2500(12)$^{[18]}$ & 35760.9155410(53)$^{[23]}$ & 2796.34898848(42)$^{[27]}$            &  0.00764(3)$^{[18]}$     \\
&               & 25$^{[5]}$        &  3      & 4       &  \hspace{4.0mm} 1.2500(12)$^{[18]}$ & 35760.9137747(53)$^{[23]}$ & 2796.34912660(42)$^{[27]}$            &  0.00764(3)$^{[18]}$     \\
&               & 25$^{[5]}$        &  2      & 1       &  \hspace{4.0mm} 1.2500(12)$^{[18]}$ & 35760.8592344(53)$^{[23]}$ & 2796.35339142(42)$^{[27]}$            &  0.00764(3)$^{[18]}$     \\
&               & 25$^{[5]}$        &  2      & 2       &  \hspace{4.0mm} 1.2500(12)$^{[18]}$ & 35760.8577646(53)$^{[23]}$ & 2796.35350635(42)$^{[27]}$            &  0.00764(3)$^{[18]}$     \\
&               & 25$^{[5]}$        &  2      & 3       &  \hspace{4.0mm} 1.2500(12)$^{[18]}$ & 35760.8558742(53)$^{[23]}$ & 2796.35365416(42)$^{[27]}$            &  0.00764(3)$^{[18]}$     \\
&               & 25$^{[5]}$        &  2      & 4       &  \hspace{4.0mm} 1.2500(12)$^{[18]}$ & 35760.8541079(53)$^{[23]}$ & 2796.35379228(42)$^{[27]}$            &  0.00764(3)$^{[18]}$     \\
&               & 24$^{[5]}$        &         &         &  \hspace{-2mm} 78.99(3)$^{[17]}$    & 35760.8373967(53)$^{[16]}$ & 2796.35509903(42)$^{[27]}$            &  0.483(4)$^{[18]}$       \\
&         2803  &  ---              &         &         &              ---                    & 35669.30440(6)$^{[16]}$    & 2803.530982(5)$^{[27]}$               &  \hspace{5mm} ---        \\
&               & 26$^{[5]}$        &         &         &  \hspace{-2.0mm} 11.01(4)$^{[17]}$  & 35669.3905712(53)$^{[16]}$ & 2803.52420938(42)$^{[27]}$            &  0.03369(24)$^{[18]}$    \\
&               & 25$^{[5]}$        &  3      & 2       &  \hspace{-0.5mm}  2.50(2)$^{[18]}$  & 35669.3725048(53)$^{[23]}$ & 2803.52642924(42)$^{[27]}$            &  0.00765(13)$^{[18]}$    \\
&               & 25$^{[5]}$        &  3      & 3       &  \hspace{-0.5mm}  2.50(2)$^{[18]}$  & 35669.3623278(53)$^{[23]}$ & 2803.52562935(42)$^{[27]}$            &  0.00765(13)$^{[18]}$    \\
&               & 25$^{[5]}$        &  2      & 2       &  \hspace{-0.5mm} 2.50(2)$^{[18]}$   & 35669.3128381(53)$^{[23]}$ & 2803.53111891(42)$^{[27]}$            &  0.00765(13)$^{[18]}$    \\
&               & 25$^{[5]}$        &  2      & 3       &  \hspace{-0.5mm} 2.50(2)$^{[18]}$   & 35669.3026610(53)$^{[23]}$ & 2803.53031902(42)$^{[27]}$            &  0.00765(13)$^{[18]}$    \\
&               & 24$^{[5]}$        &         &         &  \hspace{-2.0mm} 78.99(3)$^{[17]}$  & 35669.2876697(53)$^{[16]}$ & 2803.53229720(42)$^{[27]}$            &  0.2417(16)$^{[18]}$     \\
\hline
\end{tabular}
\end{center}
\end{table*}
\setlength{\tabcolsep}{\oldtabcolsep}

\setlength{\tabcolsep}{0.25em}
\begin{table*}
\centering
\caption[
Atomic Data -- List of References and Derivations.
]{
\label{tab:app_ato_ref}
\textbf{Atomic Data -- List of References and Derivations}.
}
\begin{tabular}{cll}
\hline
Label & Reference             & Comment \\
\hline
    1 & This Work             & Labeling convention \\
    2 & \citet{king2012}      & Labeling convention \\
    3 & \citet{nave2012}      & FeII Transition Electronic Configurations, \\
      &                       & \ \ \ FeII Centroid Wavenumber, \\
      &                       & \ \ \ and FeII 1608 Vacuum Wavelength \\
    4 &                       & Derived from Isotope Mass$^{[5]}$ \\
      &                       & \ \ \ and Isotope Abundance$^{[6]}$ \\
    5 & \citet{audi2003}      & FeII Isotope Mass and FeII Mass Numbers \\
    6 & \citet{taylor1992}    & FeII Isotope Abundance \\
    7 & \citet{worden1984}    & FeI Ionisation Potential \\
    8 & \citet{bergeson1996b} & Fe II 1608 Oscillator Strength \\
    9 & \citet{bergeson1996a} & Fe II Oscillator Strengths \\
   10 & \citet{li2000}        & Fe II 1608 Transition Lifetime \\
   11 & \citet{biemont1991}   & Fe II Transition Lifetimes \\
   12 & \citet{dzuba2002}     & Fe II $q$-coefficients \\
   13 & \citet{porsev2007}    & Fe II $q$-coefficients \\
   14 &                       & Derived from Atomic Mass$^{[4]}$, \\
      &                       & \ \ \ FeII Isotope Mass$^{[5]}$, \\
      &                       & \ \ \ FeII Centroid Wavenumber$^{[3]}$, \\
      &                       & \ \ \ and FeII Mass Shift$^{[15]}$ \\
   15 & \citet{porsev2009}    & FeII Isotope Mass Shift \\
   16 & \citet{batteiger2009} & MgII Transition Electronic Classifications, \\
      &                       & \ \ \ MgII Centroid Wavenumber, \\
      &                       & \ \ \ MgII Mass Numbers, $^{24}$MgII Wavenumber, \\
      &                       & \ \ \ and $^{25}$MgII/$^{24}$MgII + $^{26}$MgII/$^{24}$MgII \\
      &                       & \ \ \ Frequency Shifts \\
   17 & \citet{rosman1998}    & $^{26}$MgII + $^{24}$MgII Isotopic Abundancies \\
   18 &                       & Derived from \\
      &                       & \ \ \ FeII Isotope Abundance$^{[6]}$, \\
      &                       & \ \ \ FeII Oscillator Strengths$^{[9]}$, \\
      &                       & \ \ \ $^{26}$MgII + $^{24}$MgII Isotope Abundance$^{[17]}$, \\
      &                       & \ \ \ MgII Oscillator Strengths$^{[20]}$, \\
      &                       & \ \ \ and Assumption of Local Thermodynamic Equilibrium. \\
   19 & \citet{kaufman1991}   & MgII Ionisation Potentials \\
   20 & \citet{godefroid1999} & MgII Oscillator Strengths (Theory) \\
   21 & \citet{ansbacher1989} & MgII Transition Lifetimes \\
   22 & \citet{dzuba2007}     & MgII $q$-coefficients \\
   23 &                       & Derived from $^{24}$MgII Wavenumber$^{[16]}$, \\
      &                       & \ \ \ $^{25}$MgII/$^{24}$MgII Frequency Shifts$^{[16]}$, \\
      &                       & \ \ \ $^{25}$MgII Magnetic Dipole Hyperfine Constants$^{[25,28]}$, \\
      &                       & \ \ \ and $^{25}$MgII Electric Quadrupole Hyperfine Constants$^{[28]}$ \\
   24 &                       & Derived from $^{25}$MgII Centroid Wavenumber$^{[23]}$, \\
      &                       & \ \ \ MgII Transition Electronic Configurations$^{[16]}$, \\
      &                       & \ \ \ and $^{25}$MgII Magnetic Dipole Hyperfine Constants$^{[25]}$ \\
   25 & \citet{itano1981}     & $^{25}$MgII Magnetic Dipole Hyperfine Constants \\
   26 &                       & Derived from Lifetimes$^{[10,11,21]}$ \\
   27 &                       & Derived from Wavenumber$^{[3,14,16,23,24]}$ \\
   28 & \citet{sur2005}       & $^{25}$MgII Magnetic Dipole Hyperfine Constants, \\
      &                       & \ \ \ and $^{25}$MgII Electric Quadrupole Hyperfine Constants \\
\hline
\end{tabular}
\end{table*}
\setlength{\tabcolsep}{\oldtabcolsep}

\clearpage

\section{Supplementary Results}\label{app:supp_data}

\begin{table*}
\footnotesize{
\caption[
Model Parameters for the minimum-$AICc$ model. 
]{
\label{tab:preferred_model}
\textbf{
Model Parameters for the minimum-$AICc$ model. 
The thermally broadened model from Generation 79, as shown in Table \ref{tab:j11_sta}. 
This model estimates \da\ $= 3.3 \pm 2.9$ with $\chi^2 = 2762.3$, 101 free parameters and $AICc = 3134$ and $BIC = 4185$. 
}
Column 2: redshift of the velocity component. 
Column 3 and 4: Fe II column density and Doppler b-parameters. 
Column 5 and 6: Mg II column density and Doppler b-parameters. 
We found one unidentified interloping absorption component and the wavelength, column density and b-parameter is given below the main table. 
b-parameter and column densities are modelled using H I 1215.67 as the species (i.e. $\Gamma=6.265\times10^{8}s^{-1}$ and $f=0.416400$). 
The interloping absorption component is does not contribute to estimation of \da. 
}
\begin{center}
\begin{tabular}{cc|rr|rr}
\hline
\#   & z                                    & FeII \hspace{10.0mm}                 &                                            & MgII \hspace{10.0mm}                 &                             \\
&                                           & N \hspace{6.8mm}                     & b \hspace{5.5mm}                           & N \hspace{6.8mm}                     & b \hspace{5.5mm}            \\
&                                           & $\log_{10} [$cm$^{-2}]$ \hspace{0mm} & $[$kms$^{-1}]$ \hspace{1mm}                & $\log_{10} [$cm$^{-2}]$ \hspace{0mm} & $[$kms$^{-1}]$ \hspace{1mm} \\
\hline                                      
 1 & 1.8369044 $\pm$ 0.0000580 & 10.91 $\pm$ 1.84 &  1.72 $\pm$ \hspace{0.5mm}  3.57 & 12.15 $\pm$ 1.18 &  2.48 $\pm$ \hspace{0.5mm}  5.15 \\
 2 & 1.8370372 $\pm$ 0.0000039 &  9.85 $\pm$ 0.75 &  0.20 $\pm$ \hspace{0.5mm}  1.70 & 10.64 $\pm$ 0.25 &  0.29 $\pm$ \hspace{0.5mm}  2.45 \\
 3 & 1.8374427 $\pm$ 0.0000159 & 10.54 $\pm$ 0.30 &  5.79 $\pm$ \hspace{0.5mm}  1.20 & 11.60 $\pm$ 0.12 &  8.34 $\pm$ \hspace{0.5mm}  1.73 \\
 4 & 1.8375204 $\pm$ 0.0000027 & 10.42 $\pm$ 0.28 &  2.58 $\pm$ \hspace{0.5mm}  0.48 & 11.46 $\pm$ 0.15 &  3.72 $\pm$ \hspace{0.5mm}  0.69 \\
 5 & 1.8376617 $\pm$ 0.0000042 & 10.23 $\pm$ 0.94 &  2.30 $\pm$ \hspace{0.5mm}  0.32 & 11.71 $\pm$ 0.09 &  3.31 $\pm$ \hspace{0.5mm}  0.47 \\
 6 & 1.8377127 $\pm$ 0.0000115 & 11.38 $\pm$ 0.18 &  9.69 $\pm$ \hspace{0.5mm}  2.92 & 11.95 $\pm$ 0.09 & 13.96 $\pm$ \hspace{0.5mm}  4.21 \\
 7 & 1.8377247 $\pm$ 0.0000021 & 10.97 $\pm$ 0.26 &  2.79 $\pm$ \hspace{0.5mm}  0.21 & 12.15 $\pm$ 0.06 &  4.01 $\pm$ \hspace{0.5mm}  0.30 \\
 8 & 1.8379646 $\pm$ 0.0000279 & 10.08 $\pm$ 1.38 &  7.06 $\pm$ \hspace{0.5mm}  4.01 & 11.30 $\pm$ 0.45 & 10.17 $\pm$ \hspace{0.5mm}  5.77 \\
 9 & 1.8381890 $\pm$ 0.0000022 & 10.44 $\pm$ 0.47 &  1.97 $\pm$ \hspace{0.5mm}  0.39 & 11.66 $\pm$ 0.13 &  2.84 $\pm$ \hspace{0.5mm}  0.56 \\
10 & 1.8382263 $\pm$ 0.0000098 & 11.86 $\pm$ 0.12 & 10.76 $\pm$ \hspace{0.5mm}  2.87 & 12.22 $\pm$ 0.30 & 15.51 $\pm$ \hspace{0.5mm}  4.14 \\
11 & 1.8382286 $\pm$ 0.0000061 & 10.95 $\pm$ 1.08 &  6.37 $\pm$ \hspace{0.5mm}  0.80 & 12.49 $\pm$ 0.14 &  9.18 $\pm$ \hspace{0.5mm}  1.16 \\
12 & 1.8383266 $\pm$ 0.0000026 & 11.43 $\pm$ 0.06 &  1.79 $\pm$ \hspace{0.5mm}  0.30 & 11.94 $\pm$ 0.11 &  2.58 $\pm$ \hspace{0.5mm}  0.43 \\
13 & 1.8383724 $\pm$ 0.0000303 &  9.99 $\pm$ 1.57 &  2.19 $\pm$ \hspace{0.5mm}  3.18 & 11.05 $\pm$ 0.92 &  3.15 $\pm$ \hspace{0.5mm}  4.58 \\
14 & 1.8385374 $\pm$ 0.0000163 & 12.40 $\pm$ 0.21 &  4.60 $\pm$ \hspace{0.5mm}  0.93 & 13.20 $\pm$ 0.21 &  6.63 $\pm$ \hspace{0.5mm}  1.34 \\
15 & 1.8385412 $\pm$ 0.0000275 & 12.37 $\pm$ 0.25 &  8.56 $\pm$ \hspace{0.5mm}  1.42 & 12.43 $\pm$ 0.41 & 12.34 $\pm$ \hspace{0.5mm}  2.04 \\
16 & 1.8385637 $\pm$ 0.0000018 & 12.23 $\pm$ 0.34 &  2.14 $\pm$ \hspace{0.5mm}  0.81 & 11.97 $\pm$ 1.04 &  3.08 $\pm$ \hspace{0.5mm}  1.17 \\
17 & 1.8386464 $\pm$ 0.0000498 & 12.05 $\pm$ 0.60 &  4.07 $\pm$ \hspace{0.5mm}  2.91 & 12.90 $\pm$ 0.79 &  5.86 $\pm$ \hspace{0.5mm}  4.20 \\
18 & 1.8386842 $\pm$ 0.0000149 & 11.80 $\pm$ 1.94 &  2.13 $\pm$ \hspace{0.5mm}  2.74 & 11.18 $\pm$ 9.39 &  3.06 $\pm$ \hspace{0.5mm}  3.95 \\
19 & 1.8387369 $\pm$ 0.0000124 & 12.13 $\pm$ 3.55 &  2.66 $\pm$ \hspace{0.5mm}  3.78 & 12.66 $\pm$ 3.11 &  3.83 $\pm$ \hspace{0.5mm}  5.44 \\
20 & 1.8387383 $\pm$ 0.0000892 & 11.82 $\pm$ 5.02 &  4.63 $\pm$                10.26 & 11.99 $\pm$ 9.12 &  6.68 $\pm$                14.80 \\
21 & 1.8388127 $\pm$ 0.0000843 & 12.03 $\pm$ 2.06 &  3.28 $\pm$ \hspace{0.5mm}  7.51 & 12.44 $\pm$ 3.42 &  4.73 $\pm$                10.82 \\
22 & 1.8388635 $\pm$ 0.0001241 & 11.42 $\pm$ 9.43 &  2.72 $\pm$ \hspace{0.5mm}  9.52 & 12.72 $\pm$ 3.37 &  3.91 $\pm$                13.72 \\
23 & 1.8388837 $\pm$ 0.0000071 & 10.35 $\pm$ 9.29 &  0.65 $\pm$ \hspace{0.5mm}  2.72 & 14.41 $\pm$ 9.79 &  0.93 $\pm$ \hspace{0.5mm}  3.92 \\
24 & 1.8388991 $\pm$ 0.0000455 & 13.04 $\pm$ 1.13 &  4.09 $\pm$ \hspace{0.5mm}  8.51 & 11.33 $\pm$ 9.71 &  5.89 $\pm$                12.27 \\
25 & 1.8389290 $\pm$ 0.0000058 & 11.32 $\pm$ 2.05 &  0.46 $\pm$ \hspace{0.5mm}  1.30 & 14.43 $\pm$ 8.11 &  0.66 $\pm$ \hspace{0.5mm}  1.87 \\
26 & 1.8389573 $\pm$ 0.0000154 & 11.85 $\pm$ 0.84 &  1.60 $\pm$ \hspace{0.5mm}  0.84 & 12.38 $\pm$ 0.74 &  2.31 $\pm$ \hspace{0.5mm}  1.22 \\
27 & 1.8389619 $\pm$ 0.0001910 & 11.87 $\pm$ 5.04 &  3.36 $\pm$ \hspace{0.5mm}  7.19 & 12.34 $\pm$ 3.38 &  4.83 $\pm$                10.37 \\
28 & 1.8390541 $\pm$ 0.0000063 & 11.37 $\pm$ 0.11 &  2.00 $\pm$ \hspace{0.5mm}  0.54 & 11.86 $\pm$ 0.21 &  2.88 $\pm$ \hspace{0.5mm}  0.77 \\
29 & 1.8390994 $\pm$ 0.0000090 & 11.59 $\pm$ 0.33 &  1.60 $\pm$ \hspace{0.5mm}  1.36 & 11.55 $\pm$ 0.74 &  2.31 $\pm$ \hspace{0.5mm}  1.96 \\
30 & 1.8391411 $\pm$ 0.0000048 & 12.23 $\pm$ 0.10 &  2.20 $\pm$ \hspace{0.5mm}  0.74 & 12.27 $\pm$ 0.17 &  3.16 $\pm$ \hspace{0.5mm}  1.06 \\
31 & 1.8391659 $\pm$ 0.0000048 & 11.59 $\pm$ 0.33 &  0.54 $\pm$ \hspace{0.5mm}  0.70 & 11.37 $\pm$ 0.42 &  0.78 $\pm$ \hspace{0.5mm}  1.02 \\
32 & 1.8392328 $\pm$ 0.0000166 & 12.04 $\pm$ 0.32 &  5.12 $\pm$ \hspace{0.5mm}  2.78 & 12.24 $\pm$ 0.38 &  7.37 $\pm$ \hspace{0.5mm}  4.01 \\
33 & 1.8392446 $\pm$ 0.0000020 & 12.13 $\pm$ 0.29 &  2.19 $\pm$ \hspace{0.5mm}  0.68 & 12.20 $\pm$ 0.37 &  3.16 $\pm$ \hspace{0.5mm}  0.98 \\
34 & 1.8393218 $\pm$ 0.0000127 & 11.50 $\pm$ 0.12 &  2.73 $\pm$ \hspace{0.5mm}  0.79 & 12.05 $\pm$ 0.14 &  3.94 $\pm$ \hspace{0.5mm}  1.14 \\
35 & 1.8393281 $\pm$ 0.0002711 & 11.07 $\pm$ 1.14 & 16.57 $\pm$                15.47 & 11.84 $\pm$ 0.90 & 23.88 $\pm$                22.29 \\
36 & 1.8393643 $\pm$ 0.0000049 & 11.62 $\pm$ 0.12 &  2.05 $\pm$ \hspace{0.5mm}  0.24 & 11.98 $\pm$ 0.22 &  2.96 $\pm$ \hspace{0.5mm}  0.34 \\
37 & 1.8398527 $\pm$ 0.0000914 & 11.11 $\pm$ 0.76 &  1.24 $\pm$ \hspace{0.5mm}  3.17 & 11.82 $\pm$ 1.20 &  1.79 $\pm$ \hspace{0.5mm}  4.58 \\
\hline
&                           &  &                            &                             &                                      \\ 
& Unidentified Interloper                          &   &                            &                             &                                      \\ 
& $\lambda$                            & N \hspace{6.8mm}                     & b \hspace{5.5mm}                           &                             &                                      \\ 
& [\AA]                                          & $\log_{10} [$cm$^{-2}]$ \hspace{0mm} & $[$kms$^{-1}]$ \hspace{1mm}                &                             &                                      \\ 
\hline
1 & 7960.991 $\pm$ 0.011 & 11.45 $\pm$ 0.06 & 2.44 $\pm$ 0.70 \\
\hline
\end{tabular}
\end{center}
}
\end{table*}
\setlength{\tabcolsep}{\oldtabcolsep}

\begin{table*}
\footnotesize{
\caption[
Details of the spectral wavelength regions for the minimum-$AICc$ model. 
]{
\label{tab:preferred_model_data_regions}
\textbf{
Details of the spectral wavelength regions for the minimum-$AICc$ model. 
}
The model parameters of the minimum-$AICc$ model are shown in Table \ref{tab:preferred_model}. 
Column 1: spectrum label (see Table \ref{tab:dat_spe_obs}). 
Columns 2 and 3: species and transition labels. 
Column 4 and 5: observed start and end wavelength of absorption complex. 
Column 6: correction applied to the continuum in this region (original continuum is multiplied by this offset). 
Column 7: correction applied to the zero level in this region (offset added to original zero flux level). 
}
\begin{center}
\begin{tabular}{ccccccc}
\hline
Spectrum & Spec.        & Tran.        & $\lambda_{\mbox{start}}$  & $\lambda_{\mbox{end}}$    & Continuum           & Zero Level                          \\
&        &              & [\AA]                     & [\AA]                     & Correction          & Correction                                    \\
\hline
2000a & Fe II & 1608 & 4563.332 & 4567.731 & 1.0026 $\pm$ 0.0013 & \hspace{0.5mm}   \\
2000a & Fe II & 2344 & 6735.368 & 6743.444 & 1.0041 $\pm$ 0.0009 & \hspace{0.5mm}   \\
2000a & Fe II & 2374 & 6758.838 & 6766.909 & 1.0010 $\pm$ 0.0049 & \hspace{0.5mm}   \\
2000a & Fe II & 2600 & 7375.726 & 7384.171 & 1.0032 $\pm$ 0.0040 & \hspace{0.5mm}   \\
2000a & Mg II & 2796 & 7932.058 & 7941.719 & 1.0395 $\pm$ 0.0765 & \hspace{0.5mm}  0.0028 $\pm$ 0.0007 \\
2000a & Mg II & 2803 & 7952.504 & 7961.837 & 1.0145 $\pm$ 0.0375 & \hspace{0.5mm}  0.0004 $\pm$ 0.0028 \\
\hline                                                                                                
2000b & Fe II & 1608 & 4562.478 & 4568.180 & 0.9990 $\pm$ 0.0015 & \hspace{0.5mm}   \\
2000b & Fe II & 2344 & 6650.434 & 6658.533 & 1.0035 $\pm$ 0.0020 & \hspace{0.5mm}   \\
2000b & Fe II & 2374 & 6735.988 & 6743.839 & 1.0002 $\pm$ 0.0009 & \hspace{0.5mm}   \\
2000b & Fe II & 2600 & 7376.299 & 7384.288 & 1.0039 $\pm$ 0.0041 & \hspace{0.5mm}   \\
2000b & Mg II & 2796 & 7932.272 & 7942.033 & 1.0395 $\pm$ 0.0765 & \hspace{0.5mm}  0.0035 $\pm$ 0.0010 \\
2000b & Mg II & 2803 & 7952.542 & 7962.670 & 1.0191 $\pm$ 0.0375 &                -0.0003 $\pm$ 0.0031 \\
\hline                                                                                                
2006  & Fe II & 1608 & 4562.905 & 4566.501 & 1.0082 $\pm$ 0.0043 & \hspace{0.5mm}   \\
2006  & Fe II & 2374 & 6650.434 & 6658.533 & 0.9990 $\pm$ 0.0027 & \hspace{0.5mm}   \\
2006  & Fe II & 2382 & 6736.266 & 6744.437 & 1.0055 $\pm$ 0.0016 & \hspace{0.5mm}   \\
2006  & Fe II & 2600 & 7376.791 & 7385.271 & 1.0015 $\pm$ 0.0050 & \hspace{0.5mm}   \\
2006  & Mg II & 2796 & 7933.485 & 7940.107 & 1.0308 $\pm$ 0.0761 & \hspace{0.5mm}  0.0036 $\pm$ 0.0017 \\
2006  & Mg II & 2803 & 7952.814 & 7960.398 & 1.0165 $\pm$ 0.0378 & \hspace{0.5mm}  0.0030 $\pm$ 0.0043 \\
\hline
\end{tabular}
\end{center}
}
\end{table*}
\setlength{\tabcolsep}{\oldtabcolsep}

\definecolor{band1}{rgb}{1.0,0.5,0.5} 
\definecolor{band2}{rgb}{1.0,0.7,0.4} 
\definecolor{band3}{rgb}{1.0,1.0,0.5} 
\definecolor{band0}{rgb}{0.6,1.0,0.5} 
\definecolor{white}{rgb}{1.0,1.0,1.0}

\setlength{\tabcolsep}{0.10em}
\begin{table*}
\begin{center}
{\tiny
\caption{
\label{tab:j11_sta}
\textbf{
Comparison of the best-fit models from each generation. 
}
Column 1 gives the generation number. Columns 2 to 10 show the thermally
broadened models, columns 11 to 19 show the turbulently broadened
models. For each generation, the \da\ estimate and number of free
parameters $k$ are shown for the three best-fit models, according to
each of the three statistics, $\chi^2$, $AICc$ and $BIC$. 
Columns 2,5,8,11,14 and 17 show
the \da\ estimates. Columns 3,6,9,12,15 and 18 show the number of free
parameters contributing to each model. 
Columns 4,7,10,13,16 and 19 show respectively 
the $\chi^2$, $AICc$ and $BIC$ statistics. 
The models with a minimum 
in each statistic are highlighted in yellow. 
The large jump in the number of free parameters between generation 
5 and 6 (from $k$ = 21 to $k$ = 52) is due to the addition of parameters 
to correct the continuum and zero levels in the models. 
When these correction parameters are included in generations 1--5 
the bad fit to the data (due to the models lacking 
the complexity required) leads to obviously bad corrections. 
Hence these correction parameters are not included until after generation 5. 
The number of data points is 3258. 
}
\begin{tabular}{|c|rrr|rrr|rrr|rrr|rrr|rrr|}\hline
\multicolumn{1}{|c|}{\#}&
\multicolumn{9}{c|}{\textbf{THERMAL}}&
\multicolumn{9}{c|}{\textbf{TURBULENT}}\\
\cline{2-19}

\multicolumn{1}{|c|}{\rule{0pt}{4ex}} &
\multicolumn{3}{c|}{$\chi^2$}&
\multicolumn{3}{c|}{$AICc$}&
\multicolumn{3}{c|}{$BIC$}&
\multicolumn{3}{c|}{$\chi^2$}&
\multicolumn{3}{c|}{$AICc$}&
\multicolumn{3}{c|}{$BIC$}\\
\cline{2-19}

\multicolumn{1}{|c|}{\rule{0pt}{4ex}}&
\multicolumn{1}{c}{$\Delta\alpha/\alpha$}&
\multicolumn{1}{c}{$k$}&
\multicolumn{1}{c|}{$\chi^2$}&
\multicolumn{1}{c}{$\Delta\alpha/\alpha$}&
\multicolumn{1}{c}{$k$}&
\multicolumn{1}{c|}{$AICc$}&
\multicolumn{1}{c}{$\Delta\alpha/\alpha$}&
\multicolumn{1}{c}{$k$}&
\multicolumn{1}{c|}{$BIC$}&
\multicolumn{1}{c}{$\Delta\alpha/\alpha$}&
\multicolumn{1}{c}{$k$}&
\multicolumn{1}{c|}{$\chi^2$}&
\multicolumn{1}{c}{$\Delta\alpha/\alpha$}&
\multicolumn{1}{c}{$k$}&
\multicolumn{1}{c|}{$AICc$}&
\multicolumn{1}{c}{$\Delta\alpha/\alpha$}&
\multicolumn{1}{c}{$k$}&
\multicolumn{1}{c|}{$BIC$}\\

\multicolumn{1}{|c|}{}&
\multicolumn{1}{c}{$[1\times10^{-6}]$}&
\multicolumn{1}{c}{}&
\multicolumn{1}{c|}{}&
\multicolumn{1}{c}{$[1\times10^{-6}]$}&
\multicolumn{1}{c}{}&
\multicolumn{1}{c|}{}&
\multicolumn{1}{c}{$[1\times10^{-6}]$}&
\multicolumn{1}{c}{}&
\multicolumn{1}{c|}{}&
\multicolumn{1}{c}{$[1\times10^{-6}]$}&
\multicolumn{1}{c}{}&
\multicolumn{1}{c|}{}&
\multicolumn{1}{c}{$[1\times10^{-6}]$}&
\multicolumn{1}{c}{}&
\multicolumn{1}{c|}{}&
\multicolumn{1}{c}{$[1\times10^{-6}]$}&
\multicolumn{1}{c}{}&
\multicolumn{1}{c|}{}\\
\hline

1  & -465.6 $\pm$                 47.5 &   5 & 712137.8                    & -465.6 $\pm$                 47.5 &   5 & 712147.8                   & -465.6 $\pm$                 47.5 &   5 & 712178.2                   & -510.3 $\pm$                 60.4 &   5 & 723245.3 & -510.3 $\pm$                 60.4 &   5 & 723255.3 & -510.3 $\pm$                 60.4 &   5 & 723285.7 \\
2  & -444.4 $\pm$                 42.1 &   9 & 585493.0                    & -444.4 $\pm$                 42.1 &   9 & 585511.0                   & -444.4 $\pm$                 42.1 &   9 & 585565.8                   &   -9.7 $\pm$                 21.0 &   9 & 520557.9 &   -9.7 $\pm$                 21.0 &   9 & 520575.9 &   -9.7 $\pm$                 21.0 &   9 & 520630.7 \\
3  &   -8.8 $\pm$                 19.4 &  13 & 421947.5                    &   -8.8 $\pm$                 19.4 &  13 & 421973.6                   &   -8.8 $\pm$                 19.4 &  13 & 422052.6                   &   28.2 $\pm$                 10.9 &  13 & 354211.6 &   28.2 $\pm$                 10.9 &  13 & 354237.7 &   28.2 $\pm$                 10.9 &  13 & 354316.8 \\
4  &   21.6 $\pm$                 10.3 &  17 & 242770.8                    &   21.6 $\pm$                 10.3 &  17 & 242805.0                   &   21.6 $\pm$                 10.3 &  17 & 242908.3                   &   26.2 $\pm$ \hspace{0.45mm}  8.9 &  17 & 240147.3 &   26.2 $\pm$ \hspace{0.45mm}  8.9 &  17 & 240181.5 &   26.2 $\pm$ \hspace{0.45mm}  8.9 &  17 & 240284.8 \\
5  &   27.2 $\pm$ \hspace{0.45mm}  6.2 &  21 & 115307.0                    &   27.2 $\pm$ \hspace{0.45mm}  6.2 &  21 & 115349.3                   &   27.2 $\pm$ \hspace{0.45mm}  6.2 &  21 & 115476.8                   &   27.0 $\pm$ \hspace{0.45mm}  7.3 &  21 & 160988.6 &   27.0 $\pm$ \hspace{0.45mm}  7.3 &  21 & 161030.9 &   27.0 $\pm$ \hspace{0.45mm}  7.3 &  21 & 161158.5 \\
6  &    6.9 $\pm$ \hspace{0.45mm}  3.8 &  52 &  42820.4                    &    6.9 $\pm$ \hspace{0.45mm}  3.8 &  52 &  42926.1                   &    6.9 $\pm$ \hspace{0.45mm}  3.8 &  52 &  43241.0                   &    9.3 $\pm$ \hspace{0.45mm}  3.7 &  52 &  41024.2 &    9.3 $\pm$ \hspace{0.45mm}  3.7 &  52 &  41129.9 &    9.3 $\pm$ \hspace{0.45mm}  3.7 &  52 &  41444.8 \\
7  &    1.1 $\pm$ \hspace{0.45mm}  3.4 &  56 &  34460.0                    &    1.1 $\pm$ \hspace{0.45mm}  3.4 &  56 &  34574.0                   &    1.1 $\pm$ \hspace{0.45mm}  3.4 &  56 &  34913.0                   &    4.7 $\pm$ \hspace{0.45mm}  3.3 &  56 &  34737.7 &    4.7 $\pm$ \hspace{0.45mm}  3.3 &  56 &  34851.7 &    4.7 $\pm$ \hspace{0.45mm}  3.3 &  56 &  35190.7 \\
8  &    1.4 $\pm$ \hspace{0.45mm}  3.1 &  60 &  29019.1                    &    1.4 $\pm$ \hspace{0.45mm}  3.1 &  60 &  29141.4                   &    1.4 $\pm$ \hspace{0.45mm}  3.1 &  60 &  29504.4                   &    4.0 $\pm$ \hspace{0.45mm}  3.0 &  60 &  29153.3 &    4.0 $\pm$ \hspace{0.45mm}  3.0 &  60 &  29275.6 &    4.0 $\pm$ \hspace{0.45mm}  3.0 &  60 &  29638.6 \\
9  &   -0.3 $\pm$ \hspace{0.45mm}  2.8 &  64 &  24340.2                    &   -0.3 $\pm$ \hspace{0.45mm}  2.8 &  64 &  24470.8                   &   -0.3 $\pm$ \hspace{0.45mm}  2.8 &  64 &  24857.8                   &    4.5 $\pm$ \hspace{0.45mm}  2.3 &  64 &  19076.5 &    4.5 $\pm$ \hspace{0.45mm}  2.3 &  64 &  19207.1 &    4.5 $\pm$ \hspace{0.45mm}  2.3 &  64 &  19594.2 \\
10 &   -1.9 $\pm$ \hspace{0.45mm}  2.0 &  68 &  13748.8                    &   -1.9 $\pm$ \hspace{0.45mm}  2.0 &  68 &  13887.7                   &   -1.9 $\pm$ \hspace{0.45mm}  2.0 &  68 &  14298.8                   &    3.6 $\pm$ \hspace{0.45mm}  2.1 &  68 &  14472.1 &    3.6 $\pm$ \hspace{0.45mm}  2.1 &  68 &  14611.0 &    3.6 $\pm$ \hspace{0.45mm}  2.1 &  68 &  15022.1 \\
11 &   -2.3 $\pm$ \hspace{0.45mm}  1.6 &  72 &   9349.1                    &   -2.3 $\pm$ \hspace{0.45mm}  1.6 &  72 &   9496.4                   &   -2.3 $\pm$ \hspace{0.45mm}  1.6 &  72 &   9931.5                   &    3.1 $\pm$ \hspace{0.45mm}  1.8 &  72 &  11182.5 &    3.1 $\pm$ \hspace{0.45mm}  1.8 &  72 &  11329.8 &    3.1 $\pm$ \hspace{0.45mm}  1.8 &  72 &  11764.9 \\
12 &   -5.3 $\pm$ \hspace{0.45mm}  1.6 &  76 &   8445.2                    &   -5.3 $\pm$ \hspace{0.45mm}  1.6 &  76 &   8600.9                   &   -5.3 $\pm$ \hspace{0.45mm}  1.6 &  76 &   9060.0                   &    5.8 $\pm$ \hspace{0.45mm}  1.7 &  76 &   9256.3 &    5.8 $\pm$ \hspace{0.45mm}  1.7 &  76 &   9412.0 &    5.8 $\pm$ \hspace{0.45mm}  1.7 &  76 &   9871.0 \\
13 &   -4.2 $\pm$ \hspace{0.45mm}  1.6 &  80 &   7311.5                    &   -4.2 $\pm$ \hspace{0.45mm}  1.6 &  80 &   7475.5                   &   -4.2 $\pm$ \hspace{0.45mm}  1.6 &  80 &   7958.6                   &    3.4 $\pm$ \hspace{0.45mm}  1.7 &  80 &   8860.4 &    3.4 $\pm$ \hspace{0.45mm}  1.7 &  80 &   9024.5 &    3.4 $\pm$ \hspace{0.45mm}  1.7 &  80 &   9507.5 \\
14 &   -0.8 $\pm$ \hspace{0.45mm}  1.7 &  84 &   5974.8                    &   -0.8 $\pm$ \hspace{0.45mm}  1.7 &  84 &   6147.3                   &   -0.8 $\pm$ \hspace{0.45mm}  1.7 &  84 &   6654.3                   &    1.3 $\pm$ \hspace{0.45mm}  1.9 &  84 &   8710.2 &    1.3 $\pm$ \hspace{0.45mm}  1.9 &  84 &   8882.7 &    1.3 $\pm$ \hspace{0.45mm}  1.9 &  84 &   9389.7 \\
15 &    9.9 $\pm$ \hspace{0.45mm}  2.0 &  88 &   5215.0                    &    9.9 $\pm$ \hspace{0.45mm}  2.0 &  88 &   5396.0                   &    9.9 $\pm$ \hspace{0.45mm}  2.0 &  88 &   5926.8                   &    0.9 $\pm$ \hspace{0.45mm}  2.0 &  84 &   8387.0 &    0.9 $\pm$ \hspace{0.45mm}  2.0 &  84 &   8559.5 &    0.9 $\pm$ \hspace{0.45mm}  2.0 &  84 &   9066.5 \\
16 &    7.0 $\pm$ \hspace{0.45mm}  2.1 &  92 &   4834.2                    &    7.0 $\pm$ \hspace{0.45mm}  2.1 &  92 &   5023.7                   &    7.0 $\pm$ \hspace{0.45mm}  2.1 &  92 &   5578.4                   &   -0.9 $\pm$ \hspace{0.45mm}  2.0 &  88 &   7779.9 &   -0.9 $\pm$ \hspace{0.45mm}  2.0 &  88 &   7960.9 &   -0.9 $\pm$ \hspace{0.45mm}  2.0 &  88 &   8491.8 \\
17 &    3.5 $\pm$ \hspace{0.45mm}  2.2 &  96 &   4619.4                    &    3.5 $\pm$ \hspace{0.45mm}  2.2 &  96 &   4817.3                   &    3.5 $\pm$ \hspace{0.45mm}  2.2 &  96 &   5395.9                   &   -0.1 $\pm$ \hspace{0.45mm}  2.0 &  92 &   7382.1 &   -0.1 $\pm$ \hspace{0.45mm}  2.0 &  92 &   7571.5 &   -0.1 $\pm$ \hspace{0.45mm}  2.0 &  92 &   8126.3 \\
18 &    0.9 $\pm$ \hspace{0.45mm}  2.3 & 100 &   4495.8                    &    0.9 $\pm$ \hspace{0.45mm}  2.3 & 100 &   4702.2                   &    0.9 $\pm$ \hspace{0.45mm}  2.3 & 100 &   5304.7                   &   -0.2 $\pm$ \hspace{0.45mm}  2.0 &  92 &   7574.7 &   -0.2 $\pm$ \hspace{0.45mm}  2.0 &  92 &   7764.1 &   -0.2 $\pm$ \hspace{0.45mm}  2.0 &  92 &   8318.8 \\
19 &    4.5 $\pm$ \hspace{0.45mm}  2.3 & 104 &   4224.5                    &    4.5 $\pm$ \hspace{0.45mm}  2.3 & 104 &   4439.4                   &    4.5 $\pm$ \hspace{0.45mm}  2.3 & 104 &   5065.8                   &   -4.8 $\pm$ \hspace{0.45mm}  1.9 &  92 &   6269.3 &   -4.8 $\pm$ \hspace{0.45mm}  1.9 &  92 &   6458.7 &   -4.8 $\pm$ \hspace{0.45mm}  1.9 &  92 &   7013.5 \\
20 &   -4.3 $\pm$ \hspace{0.45mm}  2.3 & 104 &   4347.0                    &   -4.3 $\pm$ \hspace{0.45mm}  2.3 & 104 &   4561.9                   &   -4.3 $\pm$ \hspace{0.45mm}  2.3 & 104 &   5188.3                   &   -4.5 $\pm$ \hspace{0.45mm}  1.9 &  96 &   5877.8 &   -4.5 $\pm$ \hspace{0.45mm}  1.9 &  96 &   6075.7 &   -4.5 $\pm$ \hspace{0.45mm}  1.9 &  96 &   6654.3 \\
21 &   -5.1 $\pm$ \hspace{0.45mm}  2.6 & 108 &   4343.9                    &   -5.1 $\pm$ \hspace{0.45mm}  2.6 & 108 &   4567.4                   &   -5.1 $\pm$ \hspace{0.45mm}  2.6 & 108 &   5217.5                   &   -5.7 $\pm$ \hspace{0.45mm}  1.9 &  92 &   5934.5 &   -5.7 $\pm$ \hspace{0.45mm}  1.9 &  92 &   6123.9 &   -5.7 $\pm$ \hspace{0.45mm}  1.9 &  92 &   6678.7 \\
22 &   -1.4 $\pm$ \hspace{0.45mm}  2.5 & 108 &   4415.2                    &   -1.4 $\pm$ \hspace{0.45mm}  2.5 & 108 &   4638.7                   &   -3.0 $\pm$ \hspace{0.45mm}  2.9 & 104 &   5286.9                   &   -3.7 $\pm$ \hspace{0.45mm}  1.9 &  92 &   6850.9 &   -3.7 $\pm$ \hspace{0.45mm}  1.9 &  92 &   7040.3 &   -3.5 $\pm$ \hspace{0.45mm}  1.9 &  88 &   7584.5 \\
23 &   -4.7 $\pm$ \hspace{0.45mm}  3.0 & 112 &   4310.3                    &   -4.7 $\pm$ \hspace{0.45mm}  3.0 & 112 &   4542.4                   &   -4.7 $\pm$ \hspace{0.45mm}  3.0 & 112 &   5216.3                   &   -2.9 $\pm$ \hspace{0.45mm}  1.9 &  96 &   6755.0 &   -2.9 $\pm$ \hspace{0.45mm}  1.9 &  96 &   6952.9 &   -2.9 $\pm$ \hspace{0.45mm}  1.9 &  96 &   7531.5 \\
24 &   -8.6 $\pm$ \hspace{0.45mm}  2.9 & 112 &   4262.1                    &   -8.6 $\pm$ \hspace{0.45mm}  2.9 & 112 &   4494.1                   &   -8.6 $\pm$ \hspace{0.45mm}  2.9 & 112 &   5168.1                   &   -4.6 $\pm$ \hspace{0.45mm}  1.9 &  97 &   6572.5 &   -4.6 $\pm$ \hspace{0.45mm}  1.9 &  97 &   6772.5 &   -4.6 $\pm$ \hspace{0.45mm}  1.9 &  97 &   7357.1 \\
25 &  -11.1 $\pm$ \hspace{0.45mm}  2.9 & 116 &   4130.9                    &  -11.1 $\pm$ \hspace{0.45mm}  2.9 & 116 &   4371.5                   &   -7.0 $\pm$ \hspace{0.45mm}  2.7 & 112 &   5046.4                   &   -5.1 $\pm$ \hspace{0.45mm}  1.8 & 101 &   5878.9 &   -5.1 $\pm$ \hspace{0.45mm}  1.8 & 101 &   6087.4 &   -5.1 $\pm$ \hspace{0.45mm}  1.8 & 101 &   6695.8 \\
26 &  -10.9 $\pm$ \hspace{0.45mm}  3.0 & 116 &   3907.9                    &  -10.9 $\pm$ \hspace{0.45mm}  3.0 & 116 &   4148.5                   &  -10.9 $\pm$ \hspace{0.45mm}  3.0 & 116 &   4846.2                   &   -4.5 $\pm$ \hspace{0.45mm}  1.9 & 105 &   5739.3 &   -4.5 $\pm$ \hspace{0.45mm}  1.9 & 105 &   5956.4 &   -4.5 $\pm$ \hspace{0.45mm}  1.9 & 105 &   6588.6 \\
27 &   -8.5 $\pm$ \hspace{0.45mm}  2.9 & 120 &   3827.9                    &   -8.5 $\pm$ \hspace{0.45mm}  2.9 & 120 &   4077.1                   &   -8.5 $\pm$ \hspace{0.45mm}  2.9 & 120 &   4798.5                   &   -4.3 $\pm$ \hspace{0.45mm}  1.9 & 109 &   5729.6 &   -4.4 $\pm$ \hspace{0.45mm}  1.9 & 105 &   5947.9 &   -4.4 $\pm$ \hspace{0.45mm}  1.9 & 105 &   6580.2 \\
28 &    0.5 $\pm$ \hspace{0.45mm}  2.7 & 120 &   3794.1                    &    0.5 $\pm$ \hspace{0.45mm}  2.7 & 120 &   4043.4                   &    0.5 $\pm$ \hspace{0.45mm}  2.7 & 120 &   4764.8                   &   -4.3 $\pm$ \hspace{0.45mm}  1.9 & 109 &   5731.3 &   -4.3 $\pm$ \hspace{0.45mm}  1.9 & 109 &   5956.9 &   -4.3 $\pm$ \hspace{0.45mm}  1.9 & 109 &   6613.0 \\
29 &   -5.2 $\pm$ \hspace{0.45mm}  2.8 & 124 &   3620.0                    &   -5.2 $\pm$ \hspace{0.45mm}  2.8 & 124 &   3877.9                   &   -5.2 $\pm$ \hspace{0.45mm}  2.8 & 124 &   4623.0                   &   -4.2 $\pm$ \hspace{0.45mm}  1.9 & 105 &   5733.2 &   -4.2 $\pm$ \hspace{0.45mm}  1.9 & 105 &   5950.3 &   -4.2 $\pm$ \hspace{0.45mm}  1.9 & 105 &   6582.5 \\
30 &   -5.3 $\pm$ \hspace{0.45mm}  2.7 & 124 &   3571.3                    &   -5.3 $\pm$ \hspace{0.45mm}  2.7 & 124 &   3829.2                   &   -5.3 $\pm$ \hspace{0.45mm}  2.7 & 124 &   4574.4                   &   -4.2 $\pm$ \hspace{0.45mm}  1.9 & 109 &   5740.3 &   -4.2 $\pm$ \hspace{0.45mm}  1.9 & 109 &   5965.9 &   -4.2 $\pm$ \hspace{0.45mm}  1.9 & 109 &   6622.0 \\
31 &   -5.3 $\pm$ \hspace{0.45mm}  2.7 & 128 &   3547.2                    &   -5.3 $\pm$ \hspace{0.45mm}  2.7 & 128 &   3813.8                   &   -5.3 $\pm$ \hspace{0.45mm}  2.7 & 128 &   4582.6                   &   -4.4 $\pm$ \hspace{0.45mm}  1.9 & 105 &   6490.1 &   -4.4 $\pm$ \hspace{0.45mm}  1.9 & 105 &   6707.1 &   -4.4 $\pm$ \hspace{0.45mm}  1.9 & 105 &   7339.4 \\
32 &   -5.8 $\pm$ \hspace{0.45mm}  2.6 & 132 &   3144.1                    &   -5.8 $\pm$ \hspace{0.45mm}  2.6 & 132 &   3419.4                   &   -5.8 $\pm$ \hspace{0.45mm}  2.6 & 132 &   4211.9                   &   -4.5 $\pm$ \hspace{0.45mm}  1.8 & 109 &   5702.8 &   -4.5 $\pm$ \hspace{0.45mm}  1.8 & 109 &   5928.5 &   -4.5 $\pm$ \hspace{0.45mm}  1.8 & 109 &   6584.5 \\
33 &   -5.7 $\pm$ \hspace{0.45mm}  2.5 & 136 &   3073.3                    &   -5.7 $\pm$ \hspace{0.45mm}  2.5 & 136 &   3357.3                   &   -5.8 $\pm$ \hspace{0.45mm}  2.5 & 132 &   4167.8                   &   -3.8 $\pm$ \hspace{0.45mm}  1.8 & 109 &   5625.5 &   -3.8 $\pm$ \hspace{0.45mm}  1.8 & 109 &   5851.1 &   -3.8 $\pm$ \hspace{0.45mm}  1.8 & 109 &   6507.2 \\
34 &   -5.8 $\pm$ \hspace{0.45mm}  2.5 & 140 &   3041.4                    &   -5.8 $\pm$ \hspace{0.45mm}  2.5 & 140 &   3334.1                   &   -5.8 $\pm$ \hspace{0.45mm}  2.5 & 140 &   4173.8                   &   -3.9 $\pm$ \hspace{0.45mm}  1.8 & 109 &   5598.8 &   -3.9 $\pm$ \hspace{0.45mm}  1.8 & 109 &   5824.4 &   -3.9 $\pm$ \hspace{0.45mm}  1.8 & 109 &   6480.5 \\
35 &   -5.7 $\pm$ \hspace{0.45mm}  2.5 & 144 &   3031.0                    &   -5.7 $\pm$ \hspace{0.45mm}  2.5 & 144 &   3332.4                   &   -5.7 $\pm$ \hspace{0.45mm}  2.5 & 144 &   4195.8                   &   -2.8 $\pm$ \hspace{0.45mm}  1.8 & 105 &   6152.6 &   -2.8 $\pm$ \hspace{0.45mm}  1.8 & 105 &   6369.6 &   -2.8 $\pm$ \hspace{0.45mm}  1.8 & 105 &   7001.9 \\
36 &   -5.6 $\pm$ \hspace{0.45mm}  2.5 & 144 &   3030.5                    &   -5.6 $\pm$ \hspace{0.45mm}  2.5 & 144 &   3331.9                   &   -5.9 $\pm$ \hspace{0.45mm}  2.4 & 136 &   4178.7                   &   -4.2 $\pm$ \hspace{0.45mm}  1.9 &  97 &   6467.4 &   -4.2 $\pm$ \hspace{0.45mm}  1.9 &  97 &   6667.4 &   -4.2 $\pm$ \hspace{0.45mm}  1.9 &  97 &   7252.0 \\
37 &   -4.9 $\pm$ \hspace{0.45mm}  2.7 & 148 &   2997.8                    &   -4.9 $\pm$ \hspace{0.45mm}  2.7 & 148 &   3308.0                   &   -4.9 $\pm$ \hspace{0.45mm}  2.7 & 140 &   4167.7                   &   -4.4 $\pm$ \hspace{0.45mm}  1.8 & 101 &   5739.2 &   -4.4 $\pm$ \hspace{0.45mm}  1.8 & 101 &   5947.7 &   -4.4 $\pm$ \hspace{0.45mm}  1.8 & 101 &   6556.2 \\
38 &   -3.4 $\pm$ \hspace{0.45mm}  2.8 & 152 &   2983.5                    &   -3.3 $\pm$ \hspace{0.45mm}  2.7 & 148 &   3294.7                   &   -4.2 $\pm$ \hspace{0.45mm}  2.8 & 144 &   4164.1                   &   -4.0 $\pm$ \hspace{0.45mm}  1.8 & 105 &   5335.0 &   -4.0 $\pm$ \hspace{0.45mm}  1.8 & 105 &   5552.1 &   -4.0 $\pm$ \hspace{0.45mm}  1.8 & 105 &   6184.4 \\
39 &   -3.1 $\pm$ \hspace{0.45mm}  2.8 & 152 &   2979.9                    &   -3.1 $\pm$ \hspace{0.45mm}  2.8 & 152 &   3298.8                   &   -3.1 $\pm$ \hspace{0.45mm}  2.8 & 152 &   4209.4                   &   -2.5 $\pm$ \hspace{0.45mm}  1.8 & 105 &   5745.0 &   -2.5 $\pm$ \hspace{0.45mm}  1.8 & 105 &   5962.1 &   -2.5 $\pm$ \hspace{0.45mm}  1.8 & 105 &   6594.4 \\
40 &   -2.9 $\pm$ \hspace{0.45mm}  2.8 & 152 &   2978.4                    &   -2.9 $\pm$ \hspace{0.45mm}  2.8 & 152 &   3297.4                   &   -2.9 $\pm$ \hspace{0.45mm}  2.8 & 152 &   4207.9                   &   -2.5 $\pm$ \hspace{0.45mm}  1.8 & 105 &   6262.9 &   -2.5 $\pm$ \hspace{0.45mm}  1.8 & 105 &   6479.9 &   -2.5 $\pm$ \hspace{0.45mm}  1.8 & 105 &   7112.2 \\
41 &   -2.9 $\pm$ \hspace{0.45mm}  2.8 & 152 &   2976.4                    &   -2.9 $\pm$ \hspace{0.45mm}  2.8 & 152 &   3295.4                   &   -1.5 $\pm$ \hspace{0.45mm}  2.8 & 144 &   4204.8                   &   -5.4 $\pm$ \hspace{0.45mm}  1.8 & 109 &   5999.2 &   -5.4 $\pm$ \hspace{0.45mm}  1.8 & 109 &   6224.9 &   -4.2 $\pm$ \hspace{0.45mm}  1.8 & 105 &   6860.5 \\
42 &   -2.6 $\pm$ \hspace{0.45mm}  2.8 & 156 &   2973.0                    &   -2.8 $\pm$ \hspace{0.45mm}  2.8 & 152 &   3294.8                   &   -2.8 $\pm$ \hspace{0.45mm}  2.8 & 152 &   4205.4                   &   -0.5 $\pm$ \hspace{0.45mm}  1.5 & 109 &   5196.4 &   -0.5 $\pm$ \hspace{0.45mm}  1.5 & 109 &   5422.0 &   -0.5 $\pm$ \hspace{0.45mm}  1.5 & 109 &   6078.1 \\
43 &   -2.6 $\pm$ \hspace{0.45mm}  2.8 & 156 &   2971.6                    &   -2.0 $\pm$ \hspace{0.45mm}  2.8 & 152 &   3290.9                   &   -0.5 $\pm$ \hspace{0.45mm}  2.9 & 144 &   4166.0                   &    2.6 $\pm$ \hspace{0.45mm}  1.5 & 109 &   4220.9 &    2.6 $\pm$ \hspace{0.45mm}  1.5 & 109 &   4446.5 &    2.6 $\pm$ \hspace{0.45mm}  1.5 & 109 &   5102.5 \\
44 &   -2.4 $\pm$ \hspace{0.45mm}  2.8 & 156 &   2970.9                    &   -2.4 $\pm$ \hspace{0.45mm}  2.8 & 156 &   3298.7                   &   -2.4 $\pm$ \hspace{0.45mm}  2.8 & 144 &   4167.7                   &    8.0 $\pm$ \hspace{0.45mm}  1.6 & 113 &   3554.7 &    8.0 $\pm$ \hspace{0.45mm}  1.6 & 113 &   3788.9 &    8.0 $\pm$ \hspace{0.45mm}  1.6 & 113 &   4468.8 \\
45 &   -2.3 $\pm$ \hspace{0.45mm}  2.8 & 144 &   2912.0                    &   -2.3 $\pm$ \hspace{0.45mm}  2.8 & 144 &   3213.5                   &   -2.3 $\pm$ \hspace{0.45mm}  2.8 & 144 &   4076.8                   &    8.5 $\pm$ \hspace{0.45mm}  1.6 & 117 &   3468.9 &    8.5 $\pm$ \hspace{0.45mm}  1.6 & 117 &   3711.7 &    8.5 $\pm$ \hspace{0.45mm}  1.6 & 117 &   4415.3 \\
46 &   -1.4 $\pm$ \hspace{0.45mm}  2.8 & 148 &   2895.2                    &    0.2 $\pm$ \hspace{0.45mm}  2.7 & 144 &   3202.3                   &    0.2 $\pm$ \hspace{0.45mm}  2.7 & 144 &   4065.7                   &   12.1 $\pm$ \hspace{0.45mm}  1.9 & 121 &   3393.8 &   12.1 $\pm$ \hspace{0.45mm}  1.9 & 121 &   3645.2 &   12.1 $\pm$ \hspace{0.45mm}  1.9 & 121 &   4372.5 \\
47 &    0.8 $\pm$ \hspace{0.45mm}  2.8 & 148 &   2864.8                    &    0.8 $\pm$ \hspace{0.45mm}  2.8 & 148 &   3175.0                   &    0.8 $\pm$ \hspace{0.45mm}  2.8 & 148 &   4061.9                   &   13.7 $\pm$ \hspace{0.45mm}  1.9 & 125 &   3341.5 &   13.7 $\pm$ \hspace{0.45mm}  1.9 & 125 &   3601.6 &   13.7 $\pm$ \hspace{0.45mm}  1.9 & 121 &   4329.2 \\
48 &    0.8 $\pm$ \hspace{0.45mm}  2.8 & 148 &   2863.7                    &    0.8 $\pm$ \hspace{0.45mm}  2.8 & 148 &   3173.9                   &    0.8 $\pm$ \hspace{0.45mm}  2.8 & 148 &   4060.9                   &   13.4 $\pm$ \hspace{0.45mm}  1.9 & 129 &   3310.7 &   13.4 $\pm$ \hspace{0.45mm}  1.9 & 129 &   3579.4 &   13.7 $\pm$ \hspace{0.45mm}  1.9 & 125 &   4348.9 \\
49 &    1.3 $\pm$ \hspace{0.45mm}  2.8 & 136 &   3115.4                    &    1.3 $\pm$ \hspace{0.45mm}  2.8 & 136 &   3399.3                   &    1.3 $\pm$ \hspace{0.45mm}  2.8 & 136 &   4215.4                   &   13.6 $\pm$ \hspace{0.45mm}  1.9 & 129 &   3310.1 &   13.6 $\pm$ \hspace{0.45mm}  1.9 & 129 &   3578.8 &    9.3 $\pm$ \hspace{0.45mm}  1.5 & 121 &   4352.7 \\
50 &   -0.8 $\pm$ \hspace{0.45mm}  2.9 & 140 &   3040.8                    &   -0.8 $\pm$ \hspace{0.45mm}  2.9 & 140 &   3333.5                   &   -0.8 $\pm$ \hspace{0.45mm}  2.9 & 140 &   4173.3                   &   13.9 $\pm$ \hspace{0.45mm}  2.2 & 133 &   3285.6 &   13.9 $\pm$ \hspace{0.45mm}  2.2 & 133 &   3563.0 &   13.8 $\pm$ \hspace{0.45mm}  1.9 & 129 &   4347.0 \\
51 &    2.4 $\pm$ \hspace{0.45mm}  2.7 & 140 &   2939.7                    &    2.4 $\pm$ \hspace{0.45mm}  2.7 & 140 &   3232.4                   &    2.4 $\pm$ \hspace{0.45mm}  2.7 & 140 &   4072.2                   &   14.0 $\pm$ \hspace{0.45mm}  2.2 & 137 &   3262.0 &   14.0 $\pm$ \hspace{0.45mm}  2.2 & 137 &   3548.1 &   13.8 $\pm$ \hspace{0.45mm}  2.0 & 133 &   4347.3 \\
52 &    2.0 $\pm$ \hspace{0.45mm}  2.7 & 144 &   2915.8                    &    2.0 $\pm$ \hspace{0.45mm}  2.7 & 144 &   3217.2                   &    2.1 $\pm$ \hspace{0.45mm}  2.7 & 140 &   4066.0                   &   14.1 $\pm$ \hspace{0.45mm}  2.2 & 137 &   3239.5 &   14.1 $\pm$ \hspace{0.45mm}  2.2 & 137 &   3525.6 &    9.2 $\pm$ \hspace{0.45mm}  1.8 & 129 &   4346.0 \\
53 &    1.7 $\pm$ \hspace{0.45mm}  2.7 & 148 &   2872.4                    &    1.7 $\pm$ \hspace{0.45mm}  2.7 & 148 &   3182.5                   &    1.7 $\pm$ \hspace{0.45mm}  2.7 & 148 &   4069.5                   &    6.8 $\pm$ \hspace{0.45mm}  1.7 & 133 &   3264.1 &    6.8 $\pm$ \hspace{0.45mm}  1.7 & 133 &   3541.5 &    6.8 $\pm$ \hspace{0.45mm}  1.7 & 133 &   4339.9 \\
54 &    1.2 $\pm$ \hspace{0.45mm}  2.8 & 152 &   2853.3                    &    1.4 $\pm$ \hspace{0.45mm}  2.7 & 144 &   3167.1                   &    1.4 $\pm$ \hspace{0.45mm}  2.7 & 144 &   4030.5                   &    2.1 $\pm$ \hspace{0.45mm}  1.6 & 133 &   3202.6 &    2.1 $\pm$ \hspace{0.45mm}  1.6 & 133 &   3480.0 &    2.1 $\pm$ \hspace{0.45mm}  1.6 & 133 &   4278.4 \\
55 &    1.0 $\pm$ \hspace{0.45mm}  2.8 & 156 &   2849.2                    &    1.4 $\pm$ \hspace{0.45mm}  2.7 & 144 &   3167.0                   &    1.4 $\pm$ \hspace{0.45mm}  2.7 & 144 &   4030.4\cellcolor{band3}  &    1.1 $\pm$ \hspace{0.45mm}  1.8 & 137 &   3114.8 &    1.1 $\pm$ \hspace{0.45mm}  1.8 & 137 &   3401.0 &    1.1 $\pm$ \hspace{0.45mm}  1.8 & 137 &   4223.0 \\
56 &    0.9 $\pm$ \hspace{0.45mm}  2.8 & 160 &   2848.9                    &    1.3 $\pm$ \hspace{0.45mm}  2.8 & 152 &   3179.4                   &    1.3 $\pm$ \hspace{0.45mm}  2.8 & 152 &   4089.9                   &    1.4 $\pm$ \hspace{0.45mm}  1.8 & 141 &   3070.0 &    1.4 $\pm$ \hspace{0.45mm}  1.8 & 141 &   3364.9 &    1.4 $\pm$ \hspace{0.45mm}  1.8 & 141 &   4210.6 \\
57 &    1.2 $\pm$ \hspace{0.45mm}  2.8 & 160 &   2861.4                    &    1.2 $\pm$ \hspace{0.45mm}  2.8 & 160 &   3198.0                   &    1.2 $\pm$ \hspace{0.45mm}  2.8 & 160 &   4155.6                   &    1.0 $\pm$ \hspace{0.45mm}  1.9 & 145 &   3059.4 &    1.4 $\pm$ \hspace{0.45mm}  1.8 & 141 &   3362.4 &    1.4 $\pm$ \hspace{0.45mm}  1.8 & 141 &   4208.1 \\
58 &    1.3 $\pm$ \hspace{0.45mm}  2.8 & 164 &   2853.7                    &    1.5 $\pm$ \hspace{0.45mm}  2.8 & 160 &   3194.8                   &    1.5 $\pm$ \hspace{0.45mm}  2.8 & 160 &   4152.4                   &    1.3 $\pm$ \hspace{0.45mm}  1.9 & 145 &   3053.0 &    1.3 $\pm$ \hspace{0.45mm}  1.9 & 145 &   3356.6 &    2.2 $\pm$ \hspace{0.45mm}  2.0 & 137 &   4198.9 \\
59 &    1.2 $\pm$ \hspace{0.45mm}  2.8 & 168 &   2840.7                    &    1.2 $\pm$ \hspace{0.45mm}  2.8 & 168 &   3195.0                   &    1.3 $\pm$ \hspace{0.45mm}  2.8 & 164 &   4179.2                   &    2.4 $\pm$ \hspace{0.45mm}  2.1 & 145 &   3038.6 &    2.4 $\pm$ \hspace{0.45mm}  2.1 & 145 &   3342.2 &    2.4 $\pm$ \hspace{0.45mm}  2.1 & 145 &   4211.5 \\
60 &    1.3 $\pm$ \hspace{0.45mm}  2.8 & 172 &   2848.2                    &    1.2 $\pm$ \hspace{0.45mm}  2.7 & 152 &   3179.1                   &    1.2 $\pm$ \hspace{0.45mm}  2.7 & 152 &   4089.6                   &    3.1 $\pm$ \hspace{0.45mm}  2.2 & 149 &   3020.7 &    4.3 $\pm$ \hspace{0.45mm}  2.2 & 141 &   3324.2 &    4.3 $\pm$ \hspace{0.45mm}  2.2 & 141 &   4169.9 \\
61 &    1.3 $\pm$ \hspace{0.45mm}  2.8 & 172 &   2845.7                    &    0.9 $\pm$ \hspace{0.45mm}  2.8 & 160 &   3204.9                   &    0.9 $\pm$ \hspace{0.45mm}  2.8 & 160 &   4162.4                   &    3.5 $\pm$ \hspace{0.45mm}  2.2 & 153 &   3006.4 &    3.6 $\pm$ \hspace{0.45mm}  2.2 & 145 &   3320.4 &    3.6 $\pm$ \hspace{0.45mm}  2.2 & 145 &   4189.7 \\
62 &    0.9 $\pm$ \hspace{0.45mm}  2.8 & 160 &   2862.7                    &    0.9 $\pm$ \hspace{0.45mm}  2.8 & 160 &   3199.3                   &    0.9 $\pm$ \hspace{0.45mm}  2.8 & 160 &   4156.9                   &    4.1 $\pm$ \hspace{0.45mm}  2.2 & 153 &   3003.0 &    3.8 $\pm$ \hspace{0.45mm}  2.3 & 149 &   3316.4 &    4.3 $\pm$ \hspace{0.45mm}  2.3 & 145 &   4194.0 \\
63 &    1.1 $\pm$ \hspace{0.45mm}  2.8 & 164 &   2856.1                    &    1.0 $\pm$ \hspace{0.45mm}  2.8 & 160 &   3199.9                   &    1.0 $\pm$ \hspace{0.45mm}  2.8 & 160 &   4157.5                   &    4.0 $\pm$ \hspace{0.45mm}  2.3 & 149 &   3000.5 &    4.0 $\pm$ \hspace{0.45mm}  2.3 & 149 &   3312.9 &    4.0 $\pm$ \hspace{0.45mm}  2.3 & 149 &   4205.8 \\
64 &    1.2 $\pm$ \hspace{0.45mm}  2.8 & 168 &   2844.7                    &    1.2 $\pm$ \hspace{0.45mm}  2.8 & 168 &   3199.1                   &    1.2 $\pm$ \hspace{0.45mm}  2.8 & 164 &   4182.3                   &    3.7 $\pm$ \hspace{0.45mm}  2.3 & 153 &   3001.2 &    3.7 $\pm$ \hspace{0.45mm}  2.3 & 153 &   3322.4 &    3.4 $\pm$ \hspace{0.45mm}  2.3 & 149 &   4225.8 \\
65 &    1.2 $\pm$ \hspace{0.45mm}  2.8 & 172 &   2842.7                    &    1.1 $\pm$ \hspace{0.45mm}  2.8 & 168 &   3197.7                   &    0.6 $\pm$ \hspace{0.45mm}  2.8 & 164 &   4189.0                   &    1.8 $\pm$ \hspace{0.45mm}  2.3 & 153 &   3066.2 &    1.8 $\pm$ \hspace{0.45mm}  2.3 & 153 &   3387.4 &    2.5 $\pm$ \hspace{0.45mm}  2.3 & 141 &   4251.3 \\
66 &    1.2 $\pm$ \hspace{0.45mm}  2.8 & 176 &   2843.3                    &    1.2 $\pm$ \hspace{0.45mm}  2.8 & 176 &   3215.5                   &    1.2 $\pm$ \hspace{0.45mm}  2.8 & 176 &   4266.9                   &    1.8 $\pm$ \hspace{0.45mm}  2.3 & 153 &   3066.2 &    1.8 $\pm$ \hspace{0.45mm}  2.3 & 153 &   3387.4 &    2.5 $\pm$ \hspace{0.45mm}  2.3 & 141 &   4251.3 \\
67 &    1.2 $\pm$ \hspace{0.45mm}  2.8 & 180 &   2842.8                    &    1.2 $\pm$ \hspace{0.45mm}  2.8 & 180 &   3224.0                   &    1.0 $\pm$ \hspace{0.45mm}  2.8 & 172 &   4252.6                   &    1.8 $\pm$ \hspace{0.45mm}  2.3 & 153 &   3066.2 &    1.8 $\pm$ \hspace{0.45mm}  2.3 & 153 &   3387.4 &    2.5 $\pm$ \hspace{0.45mm}  2.3 & 141 &   4251.3 \\
68 &    1.3 $\pm$ \hspace{0.45mm}  2.9 & 184 &   2832.9                    &    1.2 $\pm$ \hspace{0.45mm}  2.8 & 176 &   3217.4                   &    1.2 $\pm$ \hspace{0.45mm}  2.8 & 176 &   4268.8                   &    1.7 $\pm$ \hspace{0.45mm}  2.3 & 157 &   3045.5 &    1.7 $\pm$ \hspace{0.45mm}  2.3 & 157 &   3375.5 &    1.7 $\pm$ \hspace{0.45mm}  2.3 & 153 &   4294.0 \\
69 &   -0.8 $\pm$ \hspace{0.45mm}  2.9 & 184 &   2831.9                    &   -0.8 $\pm$ \hspace{0.45mm}  2.9 & 184 &   3222.1                   &    1.2 $\pm$ \hspace{0.45mm}  2.8 & 180 &   4298.2                   &    1.7 $\pm$ \hspace{0.45mm}  2.3 & 157 &   3037.4 &    1.7 $\pm$ \hspace{0.45mm}  2.3 & 157 &   3367.4 &    1.7 $\pm$ \hspace{0.45mm}  2.3 & 157 &   4307.3 \\
70 &   -0.8 $\pm$ \hspace{0.45mm}  2.8 & 184 &   2845.8                    &   -1.4 $\pm$ \hspace{0.45mm}  2.8 & 180 &   3227.5                   &   -1.4 $\pm$ \hspace{0.45mm}  2.8 & 180 &   4302.3                   &    2.1 $\pm$ \hspace{0.45mm}  2.3 & 157 &   3033.6 &    2.1 $\pm$ \hspace{0.45mm}  2.3 & 157 &   3363.6 &    2.3 $\pm$ \hspace{0.45mm}  2.3 & 149 &   4280.2 \\
71 &   -0.2 $\pm$ \hspace{0.45mm}  2.8 & 188 &   2838.7                    &   -0.5 $\pm$ \hspace{0.45mm}  2.8 & 172 &   3214.4                   &   -0.5 $\pm$ \hspace{0.45mm}  2.8 & 172 &   4242.4                   &    2.3 $\pm$ \hspace{0.45mm}  2.3 & 160 &   3026.9 &    2.3 $\pm$ \hspace{0.45mm}  2.3 & 160 &   3363.6 &    2.7 $\pm$ \hspace{0.45mm}  2.4 & 157 &   4303.5 \\
72 &    0.4 $\pm$ \hspace{0.45mm}  2.8 & 176 &   2835.6                    &    0.4 $\pm$ \hspace{0.45mm}  2.8 & 176 &   3207.9                   &    0.4 $\pm$ \hspace{0.45mm}  2.8 & 176 &   4259.3                   &    2.2 $\pm$ \hspace{0.45mm}  2.4 & 164 &   3025.5 &    2.2 $\pm$ \hspace{0.45mm}  2.4 & 164 &   3371.0 &    2.2 $\pm$ \hspace{0.45mm}  2.4 & 164 &   4352.1 \\
73 &    0.8 $\pm$ \hspace{0.45mm}  2.8 & 172 &   2824.9                    &    0.8 $\pm$ \hspace{0.45mm}  2.8 & 172 &   3188.2                   &    0.8 $\pm$ \hspace{0.45mm}  2.8 & 172 &   4216.2                   &    3.5 $\pm$ \hspace{0.45mm}  2.3 & 163 &   2981.4 &    3.5 $\pm$ \hspace{0.45mm}  2.3 & 163 &   3324.6 &    3.5 $\pm$ \hspace{0.45mm}  2.3 & 163 &   4299.8 \\
74 &    1.4 $\pm$ \hspace{0.45mm}  2.8 & 172 &   2800.5                    &    1.4 $\pm$ \hspace{0.45mm}  2.8 & 172 &   3163.8                   &    1.4 $\pm$ \hspace{0.45mm}  2.8 & 172 &   4191.8                   &    4.3 $\pm$ \hspace{0.45mm}  2.3 & 163 &   2972.8 &    4.3 $\pm$ \hspace{0.45mm}  2.3 & 163 &   3316.1 &    4.4 $\pm$ \hspace{0.45mm}  2.3 & 159 &   4270.2 \\
75 &    2.8 $\pm$ \hspace{0.45mm}  2.8 & 176 &   2786.7                    &    3.3 $\pm$ \hspace{0.45mm}  2.8 & 168 &   3149.3                   &    3.3 $\pm$ \hspace{0.45mm}  2.8 & 168 &   4153.8                   &    4.7 $\pm$ \hspace{0.45mm}  2.3 & 167 &   2968.7 &    4.6 $\pm$ \hspace{0.45mm}  2.3 & 159 &   3317.5 &    4.6 $\pm$ \hspace{0.45mm}  2.3 & 159 &   4269.2 \\
76 &    2.9 $\pm$ \hspace{0.45mm}  2.8 & 176 &   2772.9                    &    2.9 $\pm$ \hspace{0.45mm}  2.8 & 172 &   3139.7                   &    2.9 $\pm$ \hspace{0.45mm}  2.8 & 172 &   4167.7                   &    4.9 $\pm$ \hspace{0.45mm}  2.3 & 161 &   2958.8 &    4.9 $\pm$ \hspace{0.45mm}  2.3 & 161 &   3297.6 &    4.9 $\pm$ \hspace{0.45mm}  2.3 & 161 &   4261.1 \\
77 &    3.1 $\pm$ \hspace{0.45mm}  2.9 & 180 &   2767.3                    &    2.9 $\pm$ \hspace{0.45mm}  2.8 & 172 &   3139.8                   &    2.9 $\pm$ \hspace{0.45mm}  2.8 & 172 &   4167.8                   &    4.9 $\pm$ \hspace{0.45mm}  2.3 & 165 &   2957.9 &    4.9 $\pm$ \hspace{0.45mm}  2.3 & 161 &   3299.7 &    5.3 $\pm$ \hspace{0.45mm}  2.3 & 154 &   4258.1 \\
78 &    3.3 $\pm$ \hspace{0.45mm}  2.9 & 180 &   2759.6                    &    3.3 $\pm$ \hspace{0.45mm}  2.9 & 180 &   3140.8                   &    4.1 $\pm$ \hspace{0.45mm}  2.8 & 164 &   4128.2                   &    4.9 $\pm$ \hspace{0.45mm}  2.3 & 165 &   2954.0 &    4.9 $\pm$ \hspace{0.45mm}  2.3 & 165 &   3301.7 &    5.7 $\pm$ \hspace{0.45mm}  2.3 & 162 &   4284.7 \\
79 &    3.2 $\pm$ \hspace{0.45mm}  2.9 & 184 &   2757.0                    &    3.3 $\pm$ \hspace{0.45mm}  2.9 & 176 &   3134.5\cellcolor{band3}  &    1.1 $\pm$ \hspace{0.45mm}  2.8 & 156 &   4122.0                   &    4.9 $\pm$ \hspace{0.45mm}  2.3 & 161 &   2958.8 &    4.9 $\pm$ \hspace{0.45mm}  2.3 & 161 &   3297.7 &    4.9 $\pm$ \hspace{0.45mm}  2.3 & 161 &   4261.1 \\
80 &    3.2 $\pm$ \hspace{0.45mm}  2.9 & 184 &   2757.0                    &    3.3 $\pm$ \hspace{0.45mm}  2.9 & 176 &   3134.5                   &    1.1 $\pm$ \hspace{0.45mm}  2.8 & 156 &   4122.0                   &    4.6 $\pm$ \hspace{0.45mm}  2.3 & 165 &   2960.1 &    4.6 $\pm$ \hspace{0.45mm}  2.3 & 165 &   3307.9 &    5.3 $\pm$ \hspace{0.45mm}  2.3 & 158 &   4254.7 \\
81 &    3.2 $\pm$ \hspace{0.45mm}  2.9 & 192 &   2757.1                    &    3.1 $\pm$ \hspace{0.45mm}  2.9 & 188 &   3160.5                   &    0.6 $\pm$ \hspace{0.45mm}  2.8 & 160 &   4153.0                   &    5.1 $\pm$ \hspace{0.45mm}  2.3 & 169 &   2960.5 &    5.2 $\pm$ \hspace{0.45mm}  2.3 & 162 &   3315.8 &    5.2 $\pm$ \hspace{0.45mm}  2.3 & 162 &   4285.2 \\
82 &    3.4 $\pm$ \hspace{0.45mm}  2.9 & 192 &   2757.9                    &    3.4 $\pm$ \hspace{0.45mm}  2.9 & 192 &   3166.1                   &    5.1 $\pm$ \hspace{0.45mm}  2.9 & 172 &   4206.2                   &    5.1 $\pm$ \hspace{0.45mm}  2.3 & 166 &   2957.7 &    5.1 $\pm$ \hspace{0.45mm}  2.3 & 166 &   3307.6 &    5.1 $\pm$ \hspace{0.45mm}  2.3 & 166 &   4300.4 \\
83 &    2.8 $\pm$ \hspace{0.45mm}  2.9 & 196 &   2738.7\cellcolor{band3}   &    3.7 $\pm$ \hspace{0.45mm}  2.9 & 192 &   3150.4                   &    5.1 $\pm$ \hspace{0.45mm}  2.9 & 180 &   4256.8                   &    5.0 $\pm$ \hspace{0.45mm}  2.4 & 165 &   2953.6 &    5.0 $\pm$ \hspace{0.45mm}  2.4 & 165 &   3301.3 &    5.0 $\pm$ \hspace{0.45mm}  2.4 & 165 &   4288.2 \\
84 &    3.5 $\pm$ \hspace{0.45mm}  2.9 & 196 &   2746.7                    &    3.5 $\pm$ \hspace{0.45mm}  2.9 & 196 &   3163.9                   &    4.3 $\pm$ \hspace{0.45mm}  2.9 & 164 &   4199.9                   &    4.9 $\pm$ \hspace{0.45mm}  2.5 & 169 &   2952.1 &    4.9 $\pm$ \hspace{0.45mm}  2.4 & 165 &   3301.4 &    5.0 $\pm$ \hspace{0.45mm}  2.3 & 162 &   4274.6 \\
\hline
\end{tabular}
}
\end{center}
\end{table*}
\setlength{\tabcolsep}{\oldtabcolsep}

\label{lastpage}

\end{document}